\documentclass[12pt,a4paper]{article}

\usepackage{ifthen} 
\newboolean{pdflatex}
\setboolean{pdflatex}{true} 

\newboolean{articletitles}
\setboolean{articletitles}{true} 

\newboolean{uprightparticles}
\setboolean{uprightparticles}{false} 

\newboolean{inbibliography}
\setboolean{inbibliography}{false} 

\usepackage[top=1in, bottom=1.25in, left=1in, right=1in]{geometry}

\usepackage{microtype}
\usepackage{lineno}  
\usepackage{xspace} 
\usepackage{caption} 

\usepackage{graphicx}  
\usepackage{color}
\usepackage{colortbl}
\graphicspath{{./figs/}} 

\usepackage{amsmath} 
\usepackage{amssymb}
\usepackage{amsfonts}
\usepackage{upgreek} 

\newcommand*\patchAmsMathEnvironmentForLineno[1]{%
\expandafter\let\csname old#1\expandafter\endcsname\csname #1\endcsname
\expandafter\let\csname oldend#1\expandafter\endcsname\csname
end#1\endcsname
 \renewenvironment{#1}%
   {\linenomath\csname old#1\endcsname}%
   {\csname oldend#1\endcsname\endlinenomath}%
}
\newcommand*\patchBothAmsMathEnvironmentsForLineno[1]{%
  \patchAmsMathEnvironmentForLineno{#1}%
  \patchAmsMathEnvironmentForLineno{#1*}%
}
\AtBeginDocument{%
\patchBothAmsMathEnvironmentsForLineno{equation}%
\patchBothAmsMathEnvironmentsForLineno{align}%
\patchBothAmsMathEnvironmentsForLineno{flalign}%
\patchBothAmsMathEnvironmentsForLineno{alignat}%
\patchBothAmsMathEnvironmentsForLineno{gather}%
\patchBothAmsMathEnvironmentsForLineno{multline}%
\patchBothAmsMathEnvironmentsForLineno{eqnarray}%
}

\usepackage{hyperref}    
\usepackage[all]{hypcap} 

\usepackage{xspace} 
\usepackage{upgreek}

\def\lhcb {\mbox{LHCb}\xspace}

\def\MagUp {\mbox{\em Mag\kern -0.05em Up}\xspace}

\ifthenelse{\boolean{uprightparticles}}%
{
 
 \def\Pgamma      {\ensuremath{\upgamma}\xspace}

 \def\Ppi         {\ensuremath{\uppi}\xspace}

 \def\PDelta      {\ensuremath{\Delta}\xspace}                 
 \def\PXi      {\ensuremath{\Xi}\xspace}                 
 \def\PLambda      {\ensuremath{\Lambda}\xspace}                 
 \def\PSigma      {\ensuremath{\Sigma}\xspace}                 
 \def\POmega      {\ensuremath{\Omega}\xspace}                 
 \def\PUpsilon      {\ensuremath{\Upsilon}\xspace}                 
 
 \def\PB      {\ensuremath{\mathrm{B}}\xspace}                 
                  
 \def\PD      {\ensuremath{\mathrm{D}}\xspace}

 \def\PK      {\ensuremath{\mathrm{K}}\xspace}

 \def\Pb      {\ensuremath{\mathrm{b}}\xspace}                 
 \def\Pc      {\ensuremath{\mathrm{c}}\xspace}

 \def\Pi      {\ensuremath{\mathrm{i}}\xspace}

 \def\Pp      {\ensuremath{\mathrm{p}}\xspace}

}
{
 
 \def\Pgamma      {\ensuremath{\gamma}\xspace}

 \def\Ppi         {\ensuremath{\pi}\xspace}

 \mathchardef\PDelta="7101
 \mathchardef\PXi="7104
 \mathchardef\PLambda="7103
 \mathchardef\PSigma="7106
 \mathchardef\POmega="710A
 \mathchardef\PUpsilon="7107
                  
 \def\PB      {\ensuremath{B}\xspace}                 
                  
 \def\PD      {\ensuremath{D}\xspace}

 \def\PK      {\ensuremath{K}\xspace}

 \def\Pb      {\ensuremath{b}\xspace}                 
 \def\Pc      {\ensuremath{c}\xspace}

 \def\Pi      {\ensuremath{i}\xspace}

 \def\Pp      {\ensuremath{p}\xspace}

}

\makeatletter
\ifcase \@ptsize \relax
  \newcommand{\miniscule}{\@setfontsize\miniscule{4}{5}}
\or
  \newcommand{\miniscule}{\@setfontsize\miniscule{5}{6}}
\or
  \newcommand{\miniscule}{\@setfontsize\miniscule{5}{6}}
\fi
\makeatother

\DeclareRobustCommand{\optbar}[1]{\shortstack{{\miniscule (\rule[.5ex]{1.25em}{.18mm})}
  \\ [-.7ex] $#1$}}

\def\g      {{\ensuremath{\Pgamma}}\xspace}

\def\cquark    {{\ensuremath{\Pc}}\xspace}

\def\bquark    {{\ensuremath{\Pb}}\xspace}

\def\pion   {{\ensuremath{\Ppi}}\xspace}
\def\piz    {{\ensuremath{\pion^0}}\xspace}

\def\pip    {{\ensuremath{\pion^+}}\xspace}
\def\pim    {{\ensuremath{\pion^-}}\xspace}

\def\kaon    {{\ensuremath{\PK}}\xspace}
  \def\Kbar    {{\kern 0.2em\overline{\kern -0.2em \PK}{}}\xspace}

\def\KorKbar    {\kern 0.18em\optbar{\kern -0.18em K}{}\xspace}

\def\Kp      {{\ensuremath{\kaon^+}}\xspace}
\def\Km      {{\ensuremath{\kaon^-}}\xspace}

  \def\Dbar    {{\kern 0.2em\overline{\kern -0.2em \PD}{}}\xspace}
\def\D       {{\ensuremath{\PD}}\xspace}

\def\DorDbar    {\kern 0.18em\optbar{\kern -0.18em D}{}\xspace}
\def\Dz      {{\ensuremath{\D^0}}\xspace}
\def\Dzb     {{\ensuremath{\Dbar{}^0}}\xspace}

\def\Dstarz  {{\ensuremath{\D^{*0}}}\xspace}

\def\Dstarp  {{\ensuremath{\D^{*+}}}\xspace}

\def\B       {{\ensuremath{\PB}}\xspace}
\def\Bbar    {{\ensuremath{\kern 0.18em\overline{\kern -0.18em \PB}{}}}\xspace}

\def\BorBbar    {\kern 0.18em\optbar{\kern -0.18em B}{}\xspace}

  \def\Y#1S{\ensuremath{\PUpsilon{(#1S)}}\xspace}

\def\proton      {{\ensuremath{\Pp}}\xspace}

\def\Lz          {{\ensuremath{\PLambda}}\xspace}
\def\Lbar        {{\ensuremath{\kern 0.1em\overline{\kern -0.1em\PLambda}}}\xspace}
\def\LorLbar    {\kern 0.18em\optbar{\kern -0.18em \PLambda}{}\xspace}
\def\Lambdares   {{\ensuremath{\PLambda}}\xspace}

\def\Lb      {{\ensuremath{\Lz^0_\bquark}}\xspace}

\def\Lc      {{\ensuremath{\Lz^+_\cquark}}\xspace}

\def\BF         {{\ensuremath{\mathcal{B}}}\xspace}

\def\BR         {\BF}

\def\to                 {\ensuremath{\rightarrow}\xspace}

\def\CP                {{\ensuremath{C\!P}}\xspace}

\def\AT#1     {\ensuremath{A_{\mathrm{T}}^{#1}}\xspace}           

\def\C#1      {\ensuremath{\mathcal{C}_{#1}}\xspace}                       
\def\Cp#1     {\ensuremath{\mathcal{C}_{#1}^{'}}\xspace}                    
\def\Ceff#1   {\ensuremath{\mathcal{C}_{#1}^{\mathrm{(eff)}}}\xspace}        
\def\Cpeff#1  {\ensuremath{\mathcal{C}_{#1}^{'\mathrm{(eff)}}}\xspace}       
\def\Ope#1    {\ensuremath{\mathcal{O}_{#1}}\xspace}                       
\def\Opep#1   {\ensuremath{\mathcal{O}_{#1}^{'}}\xspace}                    

\newcommand{\tev}{\ifthenelse{\boolean{inbibliography}}{\ensuremath{~T\kern -0.05em eV}\xspace}{\ensuremath{\mathrm{\,Te\kern -0.1em V}}}\xspace}
\newcommand{\gev}{\ensuremath{\mathrm{\,Ge\kern -0.1em V}}\xspace}
\newcommand{\mev}{\ensuremath{\mathrm{\,Me\kern -0.1em V}}\xspace}
\newcommand{\kev}{\ensuremath{\mathrm{\,ke\kern -0.1em V}}\xspace}
\newcommand{\ev}{\ensuremath{\mathrm{\,e\kern -0.1em V}}\xspace}
\newcommand{\gevc}{\ensuremath{{\mathrm{\,Ge\kern -0.1em V\!/}c}}\xspace}
\newcommand{\mevc}{\ensuremath{{\mathrm{\,Me\kern -0.1em V\!/}c}}\xspace}
\newcommand{\gevcc}{\ensuremath{{\mathrm{\,Ge\kern -0.1em V\!/}c^2}}\xspace}
\newcommand{\gevgevcccc}{\ensuremath{{\mathrm{\,Ge\kern -0.1em V^2\!/}c^4}}\xspace}
\newcommand{\mevcc}{\ensuremath{{\mathrm{\,Me\kern -0.1em V\!/}c^2}}\xspace}

\def\mum  {\ensuremath{{\,\upmu\mathrm{m}}}\xspace}

\def\invfb   {\ensuremath{\mbox{\,fb}^{-1}}\xspace}

\def\gsim{{~\raise.15em\hbox{$>$}\kern-.85em
          \lower.35em\hbox{$\sim$}~}\xspace}
\def\lsim{{~\raise.15em\hbox{$<$}\kern-.85em
          \lower.35em\hbox{$\sim$}~}\xspace}

\def\ptot       {\mbox{$p$}\xspace}
\def\pt         {\mbox{$p_{\mathrm{ T}}$}\xspace}

\def\evtgen     {\mbox{\textsc{EvtGen}}\xspace}

\def\geant      {\mbox{\textsc{Geant4}}\xspace}

\def\photos     {\mbox{\textsc{Photos}}\xspace}

\def\pythia     {\mbox{\textsc{Pythia}}\xspace}

\def\tell1  {TELL1\xspace}
\def\ukl1   {UKL1\xspace}

\newcommand{\ie}{\mbox{\itshape i.e.}\xspace}

\newcommand{\lblcpi}{\ensuremath{\Lb\to\Lc\pim}\xspace}

\newcommand{\lbdpk}{\ensuremath{\Lb\to \D \proton \Km}\xspace}

\newcommand{\lbdnpk}{\ensuremath{\Lb\to \Dz \proton K^-}\xspace}

\newcommand{\lbdnppi}{\ensuremath{\Lb\to \Dz \proton \pim}\xspace}

\newcommand{\lbdstppi}{\ensuremath{\Lb\to \Dstarz \proton \pim}\xspace}

\newcommand{\dnppi}{\ensuremath{\Dz\proton\pim}\xspace}
\newcommand{\dbppi}{\ensuremath{\Dzb\proton\pim}\xspace}

\newcommand{\Lcx}{\ensuremath{\Lz_{\cquark}(2860)^+}\xspace}
\newcommand{\Lcst}{\ensuremath{\Lz_{\cquark}(2880)^+}\xspace}
\newcommand{\Lcstst}{\ensuremath{\Lz_{\cquark}(2940)^+}\xspace}
\newcommand{\Lcstar}{\ensuremath{\Lz_{\cquark}^{*+}}\xspace}

\usepackage{cite} 
\usepackage{mciteplus}

\begin{document}

\renewcommand{\thefootnote}{\fnsymbol{footnote}}
\setcounter{footnote}{1}

\begin{titlepage}
\pagenumbering{roman}

\vspace*{-1.5cm}
\centerline{\large EUROPEAN ORGANIZATION FOR NUCLEAR RESEARCH (CERN)}
\vspace*{1.5cm}
\noindent
\begin{tabular*}{\linewidth}{lc@{\extracolsep{\fill}}r@{\extracolsep{0pt}}}
\ifthenelse{\boolean{pdflatex}}
{\vspace*{-2.7cm}\mbox{\!\!\!\includegraphics[width=.14\textwidth]{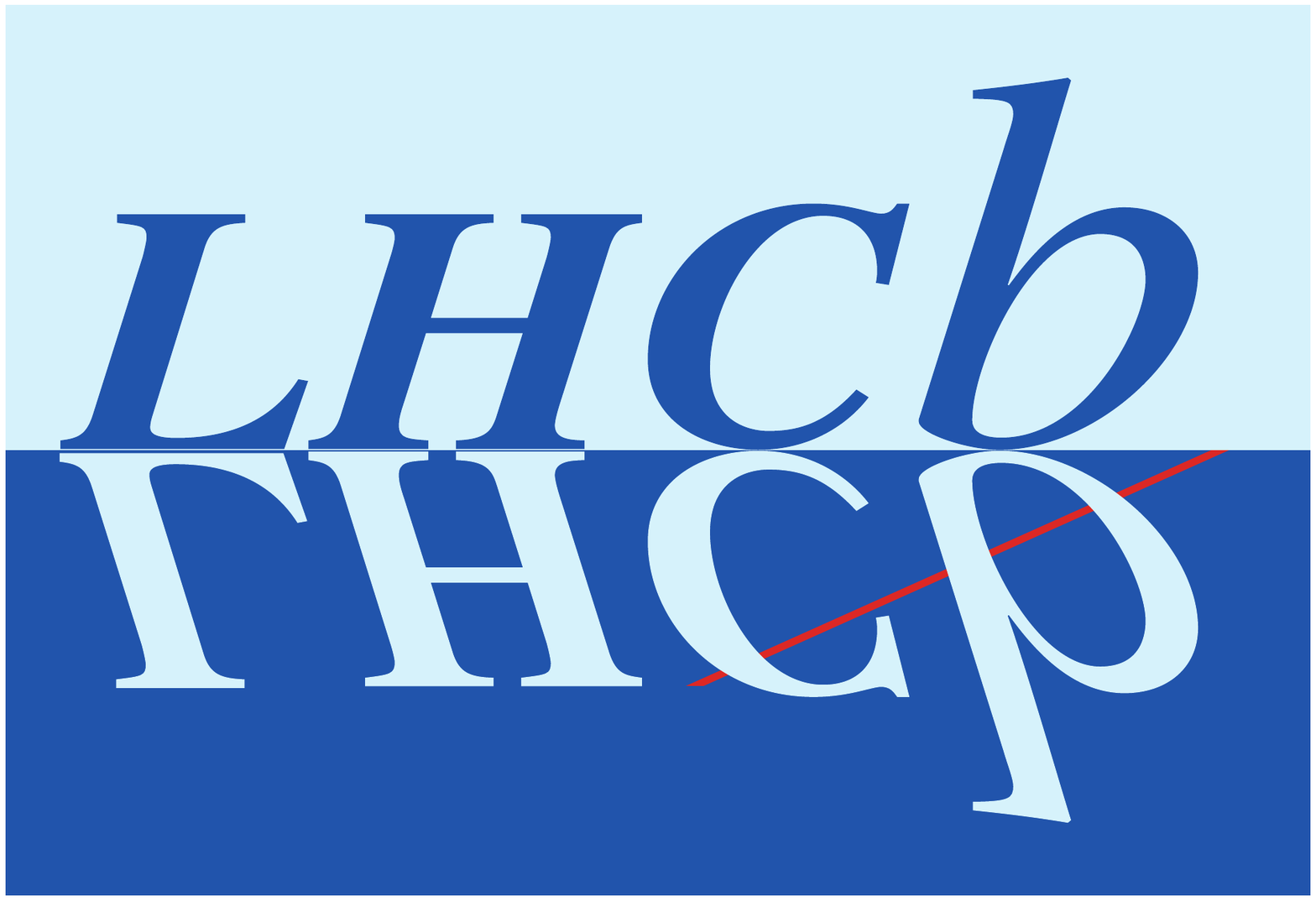}} & &}%
{\vspace*{-1.2cm}\mbox{\!\!\!\includegraphics[width=.12\textwidth]{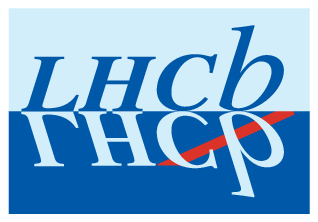}} & &}%
\\
 & & CERN-EP-2017-007 \\  
 & & LHCb-PAPER-2016-061 \\  
 & & 26 January 2017 \\ 
 & & \\
\end{tabular*}

\vspace*{4.0cm}

{\normalfont\bfseries\boldmath\huge
\begin{center}
  Study of the $\Dz\proton$ amplitude in $\lbdnppi$ decays
\end{center}
}

\vspace*{2.0cm}

\begin{center}
The LHCb collaboration\footnote{Authors are listed at the end of this paper.}
\end{center}

\vspace{\fill}

\begin{abstract}
  \noindent
  An amplitude analysis of the decay \lbdnppi is performed in the part of the phase space
  containing resonances in the $\Dz\proton$ channel. The study is based on a data sample 
  corresponding to an integrated luminosity of 3.0\invfb of $pp$ collisions recorded by the LHCb 
  experiment. The spectrum of excited \Lc states that decay into $\Dz\proton$ is studied. 
  The masses, widths and quantum numbers of the \Lcst and \Lcstst resonances are measured. 
  The constraints on the spin and parity for the \Lcstst state are obtained for the first time.
  A near-threshold enhancement in the $\Dz\proton$ amplitude 
  is investigated and found to be consistent with a new resonance, denoted the \Lcx, 
  of spin $3/2$ and positive parity.   
  
\end{abstract}

\vspace*{2.0cm}

\begin{center}
  Submitted to JHEP
\end{center}

\vspace{\fill}

{\footnotesize 
\centerline{\copyright~CERN on behalf of the \lhcb collaboration, licence \href{http://creativecommons.org/licenses/by/4.0/}{CC-BY-4.0}.}}
\vspace*{2mm}

\end{titlepage}

\newpage
\setcounter{page}{2}
\mbox{~}

\cleardoublepage


\newcommand{\lcstmass}{\ensuremath{2881.75\pm 0.29\mbox{(stat)}\pm 0.07\mbox{(syst)}{}^{+0.14}_{-0.20}\mbox{(model)}}\xspace}
\newcommand{\lcstmassstat}{\ensuremath{2881.75\pm 0.29}\xspace}
\newcommand{\lcstwidth}{\ensuremath{5.43{}^{+0.77}_{-0.71}\mbox{(stat)}\pm 0.29\mbox{(syst)}{}^{+0.75}_{-0.00}\mbox{(model)}}\xspace}
\newcommand{\lcstwidthstat}{\ensuremath{5.43{}^{+0.77}_{-0.71}}\xspace}
\newcommand{\lcstfracst}{\ensuremath{29.0{}^{+2.6}_{-3.6}\mbox{(stat)}\pm 1.1\mbox{(syst)}{}^{+3.0}_{-0.0}\mbox{(model)}}\xspace}
\newcommand{\lcstfracststat}{\ensuremath{29.0{}^{+2.6}_{-3.6}}\xspace}
\newcommand{\nronepfracst}{\ensuremath{11.3{}^{+2.2}_{-5.5}\mbox{(stat)}\pm 0.7\mbox{(syst)}{}^{+6.1}_{-1.1}\mbox{(model)}}\xspace}
\newcommand{\nronepfracststat}{\ensuremath{11.3{}^{+2.2}_{-5.5}}\xspace}
\newcommand{\nronemfracst}{\ensuremath{16.3{}^{+2.4}_{-2.6}\mbox{(stat)}\pm 0.6\mbox{(syst)}{}^{+0.0}_{-3.7}\mbox{(model)}}\xspace}
\newcommand{\nronemfracststat}{\ensuremath{16.3{}^{+2.4}_{-2.6}}\xspace}
\newcommand{\nrthreepfracst}{\ensuremath{38.2{}^{+5.0}_{-4.9}\mbox{(stat)}\pm 1.2\mbox{(syst)}{}^{+0.0}_{-3.3}\mbox{(model)}}\xspace}
\newcommand{\nrthreepfracststat}{\ensuremath{38.2{}^{+5.0}_{-4.9}}\xspace}
\newcommand{\nrthreemfracst}{\ensuremath{7.8{}^{+1.3}_{-3.1}\mbox{(stat)}\pm 0.8\mbox{(syst)}{}^{+0.0}_{-3.8}\mbox{(model)}}\xspace}
\newcommand{\nrthreemfracststat}{\ensuremath{7.8{}^{+1.3}_{-3.1}}\xspace}

\newcommand{\lcxmass}{\ensuremath{2856.1{}^{+2.0}_{-1.7}\mbox{(stat)}\pm 0.5\mbox{(syst)}{}^{+1.1}_{-5.6}\mbox{(model)}}\xspace}
\newcommand{\lcxmassstat}{\ensuremath{2856.1{}^{+2.0}_{-1.7}}\xspace}
\newcommand{\lcxwidth}{\ensuremath{67.6{}^{+10.1}_{-8.1}\mbox{(stat)}\pm 1.4\mbox{(syst)}{}^{+5.9}_{-20.0}\mbox{(model)}}\xspace}
\newcommand{\lcxwidthstat}{\ensuremath{67.6{}^{+10.1}_{-8.1}}\xspace}
\newcommand{\lcxfraclow}{\ensuremath{55.4{}^{+3.1}_{-2.7}\mbox{(stat)}\pm 1.1\mbox{(syst)}{}^{+0.6}_{-8.2}\mbox{(model)}}\xspace}
\newcommand{\lcxfraclowstat}{\ensuremath{55.4{}^{+3.1}_{-2.7}}\xspace}
\newcommand{\lcstfraclow}{\ensuremath{16.7{}^{+1.3}_{-1.1}\mbox{(stat)}\pm 0.4\mbox{(syst)}{}^{+0.1}_{-0.9}\mbox{(model)}}\xspace}
\newcommand{\lcstfraclowstat}{\ensuremath{16.7{}^{+1.3}_{-1.1}}\xspace}
\newcommand{\nronepfraclow}{\ensuremath{11.3{}^{+1.8}_{-2.8}\mbox{(stat)}\pm 0.5\mbox{(syst)}{}^{+2.7}_{-0.6}\mbox{(model)}}\xspace}
\newcommand{\nronepfraclowstat}{\ensuremath{11.3{}^{+1.8}_{-2.8}}\xspace}
\newcommand{\nronemfraclow}{\ensuremath{14.3{}^{+1.2}_{-1.6}\mbox{(stat)}\pm 0.8\mbox{(syst)}{}^{+4.0}_{-1.6}\mbox{(model)}}\xspace}
\newcommand{\nronemfraclowstat}{\ensuremath{14.3{}^{+1.2}_{-1.6}}\xspace}
\newcommand{\nrthreemfraclow}{\ensuremath{2.9{}^{+0.6}_{-1.2}\mbox{(stat)}\pm 0.4\mbox{(syst)}{}^{+2.2}_{-3.0}\mbox{(model)}}\xspace}
\newcommand{\nrthreemfraclowstat}{\ensuremath{2.9{}^{+0.6}_{-1.2}}\xspace}

\newcommand{\lcststmass}{\ensuremath{2944.8{}^{+3.5}_{-2.5}\mbox{(stat)}\pm 0.4\mbox{(syst)}{}^{+0.1}_{-4.6}\mbox{(model)}}\xspace}
\newcommand{\lcststmassstat}{\ensuremath{2944.8{}^{+3.5}_{-2.5}}\xspace}
\newcommand{\lcststwidth}{\ensuremath{27.7{}^{+8.2}_{-6.0}\mbox{(stat)}\pm 0.9\mbox{(syst)}{}^{+5.2}_{-10.4}\mbox{(model)}}\xspace}
\newcommand{\lcststwidthstat}{\ensuremath{27.7{}^{+8.2}_{-6.0}}\xspace}
\newcommand{\lcxfrac}{\ensuremath{47.2{}^{+2.9}_{-2.8}\mbox{(stat)}\pm 1.3\mbox{(syst)}{}^{+8.8}_{-6.0}\mbox{(model)}}\xspace}
\newcommand{\lcxfracstat}{\ensuremath{47.2{}^{+2.9}_{-2.8}}\xspace}
\newcommand{\lcstfrac}{\ensuremath{12.9{}^{+1.0}_{-0.9}\mbox{(stat)}\pm 0.3\mbox{(syst)}{}^{+0.9}_{-0.8}\mbox{(model)}}\xspace}
\newcommand{\lcstfracstat}{\ensuremath{12.9{}^{+1.0}_{-0.9}}\xspace}
\newcommand{\lcststfrac}{\ensuremath{8.2{}^{+2.3}_{-1.1}\mbox{(stat)}\pm 0.5\mbox{(syst)}{}^{+2.1}_{-4.1}\mbox{(model)}}\xspace}
\newcommand{\lcststfracstat}{\ensuremath{8.2{}^{+2.3}_{-1.1}}\xspace}

\newcommand{\lcxratio}{\ensuremath{4.54{}^{+0.51}_{-0.39}\mbox{(stat)}\pm 0.12\mbox{(syst)}{}^{+0.17}_{-0.58}\mbox{(model)}}\xspace}
\newcommand{\lcxratiostat}{\ensuremath{4.54{}^{+0.51}_{-0.39}}\xspace}
\newcommand{\lcststratio}{\ensuremath{0.83{}^{+0.31}_{-0.10}\mbox{(stat)}\pm 0.06\mbox{(syst)}{}^{+0.17}_{-0.43}\mbox{(model)}}\xspace}
\newcommand{\lcststratiostat}{\ensuremath{0.83{}^{+0.31}_{-0.10}}\xspace}

\renewcommand{\thefootnote}{\arabic{footnote}}
\setcounter{footnote}{0}



\pagestyle{plain} 
\setcounter{page}{1}
\pagenumbering{arabic}


%

\section{Introduction}
\label{sec:introduction}

Decays of beauty baryons to purely hadronic final states 
provide a wealth of information about the interactions between the fundamental 
constituents of matter. 
Studies of direct \CP violation in these decays can help constrain the parameters of the Standard Model 
and New Physics effects in a similar way as in decays of beauty 
mesons~\cite{Dunietz:1992ti, :1998upb, Giri:2001ju, Hsiao:2014mua, Bensalem:2000hq, Bensalem:2002ys, Bensalem:2002pz}. 
Studies of the decay dynamics of beauty baryons can provide important information on the spectroscopy of 
charmed baryons, since the known initial state provides strong constraints on the 
quantum numbers of intermediate resonances. 
The recent observation of pentaquark states at LHCb~\cite{LHCb-PAPER-2015-029} has renewed 
the interest in baryon spectroscopy.

The present analysis concerns the decay amplitude of the 
Cabibbo-favoured decay \mbox{\lbdnppi} 
(the inclusion of charge-conjugate processes is implied throughout this paper).
A measurement of the branching fraction of this decay 
with respect to the \mbox{$\Lb\to\Lc\pim$} mode
was reported by the LHCb collaboration using a data sample corresponding to 
$1.0\invfb$ of integrated luminosity~\cite{LHCb-PAPER-2013-056}. 
The \lbdnppi decay includes resonant contributions in the $\Dz\proton$ channel 
that are associated with intermediate excited $\Lc$ states, 
as well as contributions in the $\proton\pim$ channel due to excited nucleon ($N$) states. 
The study of the $\Dz\proton$ part of the amplitude will help to constrain the dynamics of the Cabibbo-suppressed
decay \lbdnpk, which is potentially sensitive to the angle $\gamma$ of the Cabibbo-Kobayashi-Maskawa 
quark mixing matrix~\cite{Cabibbo:1963yz, Kobayashi:1973fv}. 
The analysis of the $\Dz\proton$ amplitude is interesting in its own right. 
One of the states decaying to $\Dz\proton$, the $\Lcstst$, has a possible interpretation as a $D^*N$ molecule
\cite{Dong:2009tg, Dong:2010xv, He:2010zq, Dong:2011ys, Zhang:2012jk, Ortega:2012cx, Dong:2014ksa, Zhang:2014ska, Xie:2015zga}. 
There are currently no experimental constraints on the quantum numbers of the \Lcstst state. 

\begin{figure}
  \begin{center}
  \includegraphics[width=0.45\textwidth]{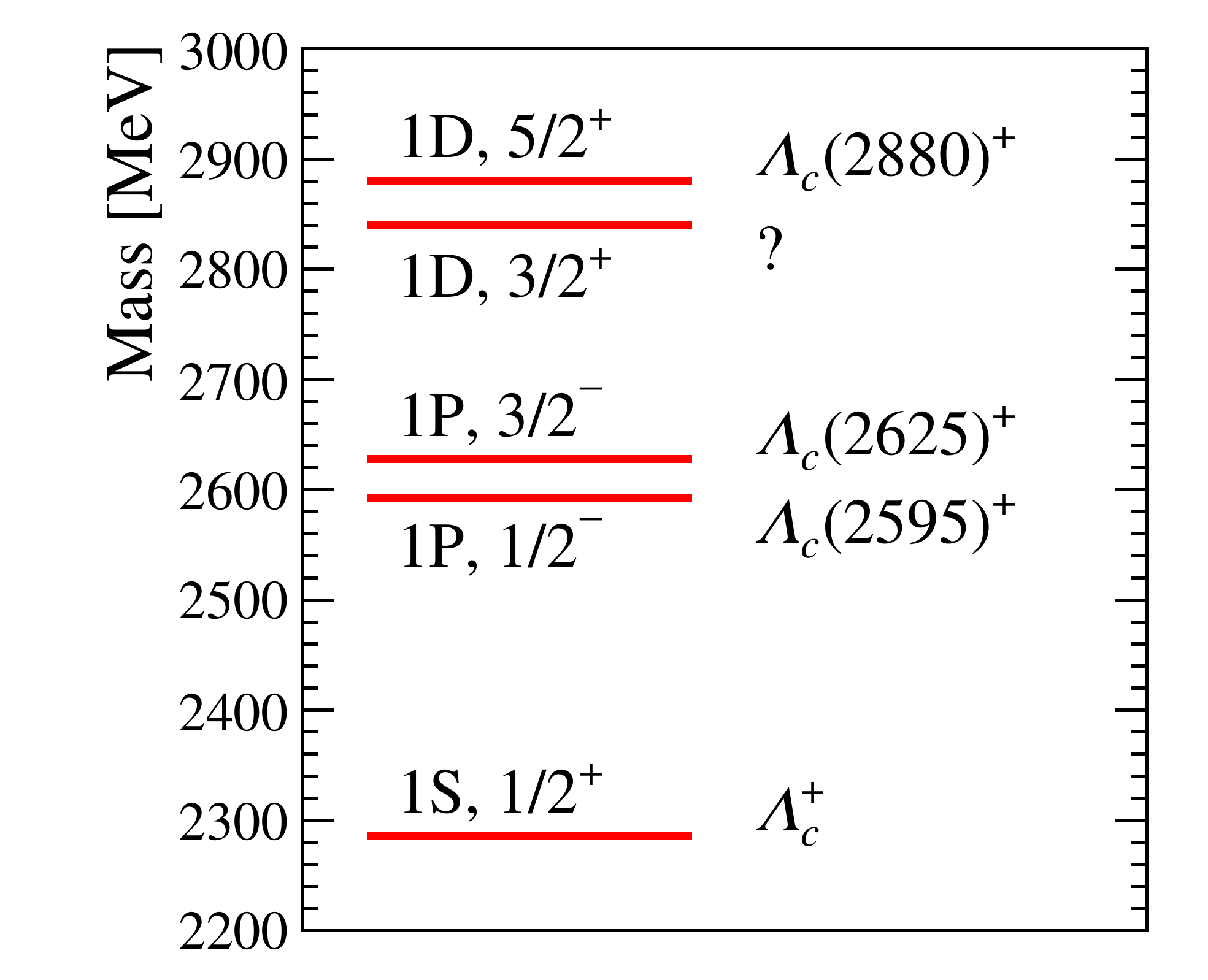}
  \end{center}
  \vspace{-\baselineskip}
  \caption{Expected spectrum of the \Lc ground state and its orbital excitations from 
           a study based on the nonrelativistic heavy quark - light diquark model~\cite{Chen:2014nyo}, 
           along with the observed resonances corresponding to those states~\cite{PDG2016}.}
  \label{fig:spectrum}
\end{figure}

The mass spectrum of the predicted and observed orbitally excited \Lc states~\cite{Chen:2014nyo} 
is shown in Fig.~\ref{fig:spectrum}. In addition to the ground state \Lc and to the $\Lambdares_{\cquark}(2595)^+$ and 
$\Lambdares_{\cquark}(2625)^+$ states, which are identified as the members of the $P$-wave doublet, 
a $D$-wave doublet with higher mass is predicted. One of the members of this doublet could be the state known as the $\Lcst$, 
which is measured to have spin and parity $J^P=5/2^+$~\cite{Abe:2006rz, PDG2016}, while 
no candidate for the other state has been observed yet. 
Several theoretical studies provide mass predictions for this state and other excited charm 
baryons~\cite{Chen:2016iyi, Lu:2016ctt, Chen:2016phw, Chen:2014nyo, Roberts:2007ni, Cheng:2006dk, Zhong:2007gp}.
The BaBar collaboration has previously reported indications of a structure in the $\Dz\proton$ mass spectrum 
close to threshold, at a mass around 
$2.84\gev$\footnote{Natural units with $\hbar=c=1$ are used throughout.}, 
which could be the missing member of the $D$-wave doublet~\cite{babardp}.

This analysis is based on a data sample corresponding to an integrated luminosity of 
3.0\invfb of $\proton\proton$ collisions recorded by the LHCb detector, with 1.0\invfb collected at 
centre-of-mass energy $\sqrt{s}=7\tev$ in 2011 and 2.0\invfb at $\sqrt{s}=8\tev$ in 2012. 

The paper is organised as follows. 
Section~\ref{sec:detector} gives a brief description of the LHCb experiment and its 
reconstruction and simulation software. 
The amplitude analysis formalism and fitting technique is introduced in Sec.~\ref{sec:formalism}. 
The selection of \lbdnppi candidates is described in Sec.~\ref{sec:selection}, followed by the measurement of 
signal and background yields (Sec.~\ref{sec:yields}), evaluation of the efficiency (Sec.~\ref{sec:efficiency}), 
determination of the shape of the background distribution (Sec.~\ref{sec:background}), and discussion of the
effects of momentum resolution (Sec.~\ref{sec:resolution}). 
Results of the amplitude fit are presented in Sec.~\ref{sec:amplitude} separately for four different 
regions of the \lbdnppi phase space, along with the systematic uncertainties for those fits. 
Section~\ref{sec:conclusion} gives a summary of the results.

\section{Detector and simulation}
\label{sec:detector}

The \lhcb detector~\cite{Alves:2008zz,LHCb-DP-2014-002} is a single-arm forward
spectrometer covering the \mbox{pseudorapidity} range $2<\eta <5$,
designed for the study of particles containing \bquark or \cquark
quarks. The detector includes a high-precision tracking system
consisting of a silicon-strip vertex detector surrounding the $pp$
interaction region, a large-area silicon-strip detector located
upstream of a dipole magnet with a bending power of about
$4{\mathrm{\,Tm}}$, and three stations of silicon-strip detectors and straw
drift tubes placed downstream of the magnet.
The tracking system provides a measurement of momentum, \ptot, of charged particles with
relative uncertainty that varies from 0.5\% at low momentum to 1.0\% at 200\gev.
The minimum distance of a track to a primary vertex (PV), the impact parameter (IP), 
is measured with a resolution of $(15+29/\pt)\mum$,
where \pt is the component of the momentum transverse to the beam, in\,\gev.
Different types of charged hadrons are distinguished using information
from two ring-imaging Cherenkov detectors. 
Photons, electrons and hadrons are identified by a calorimeter system consisting of
scintillating-pad and preshower detectors, an electromagnetic
calorimeter and a hadronic calorimeter. Muons are identified by a
system composed of alternating layers of iron and multiwire
proportional chambers.

The online event selection is performed by a trigger~\cite{LHCb-DP-2012-004}, 
which consists of a hardware stage, based on information from the calorimeter and muon
systems, followed by a software stage, which applies a full event
reconstruction.
At the hardware trigger stage, events are required to have a muon with high \pt or a
hadron, photon or electron with high transverse energy in the calorimeters. 
The software trigger requires a two-, three- or four-track
secondary vertex with significant displacement
from any PV in the event. At least one charged particle forming the vertex
must exceed a $\pt$ threshold in the range 1.6--1.7\gev
and be inconsistent with originating from a PV.
A multivariate algorithm~\cite{BBDT} is used for
the identification of secondary vertices consistent with the decay
of a \bquark hadron.

In the simulation, $pp$ collisions are generated using
\pythia 8~\cite{Sjostrand:2006za,*Sjostrand:2007gs} with a specific \lhcb
configuration~\cite{LHCb-PROC-2010-056}. Decays of hadronic particles
are described by \evtgen~\cite{Lange:2001uf}, in which final-state
radiation is generated using \photos~\cite{Golonka:2005pn}. The
interaction of the generated particles with the detector, and its response,
are implemented using the \geant
toolkit~\cite{Allison:2006ve, *Agostinelli:2002hh} as described in
Ref.~\cite{LHCb-PROC-2011-006}.

\section{Amplitude analysis formalism}

\label{sec:formalism}

The amplitude analysis is based on the helicity formalism used in previous LHCb analyses.
A detailed description of the formalism can be found in Refs.~\cite{LHCb-PAPER-2016-019, LHCb-PAPER-2016-015, LHCb-PAPER-2015-029}. This section gives details of the implementation specific to the decay \lbdnppi. 

\subsection{Phase space of the decay \boldmath{\lbdnppi}}

Three-body decays of scalar particles are described by the two-dimensional phase space 
of independent kinematic parameters, often represented as a Dalitz plot~\cite{Dalitz:1953cp}. 
For baryon decays, in general also the additional angular dependence of the decay products 
on the polarisation of the decaying particle has to be considered.

\begin{figure}
  \begin{center}
  \includegraphics[width=0.35\textwidth]{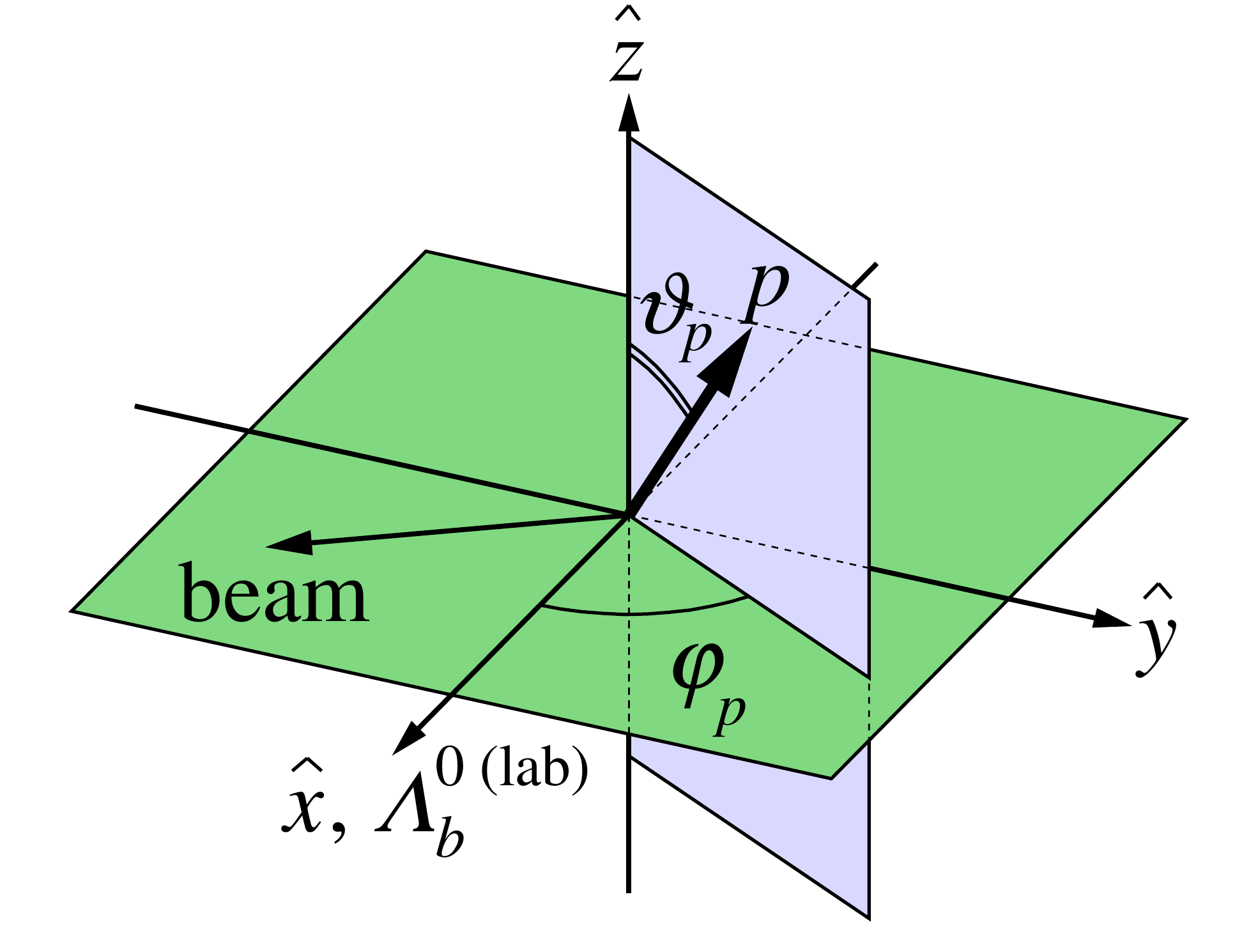}
  \put(-160,100){(a)}
  \hspace{0.02\textwidth}
  \includegraphics[width=0.35\textwidth]{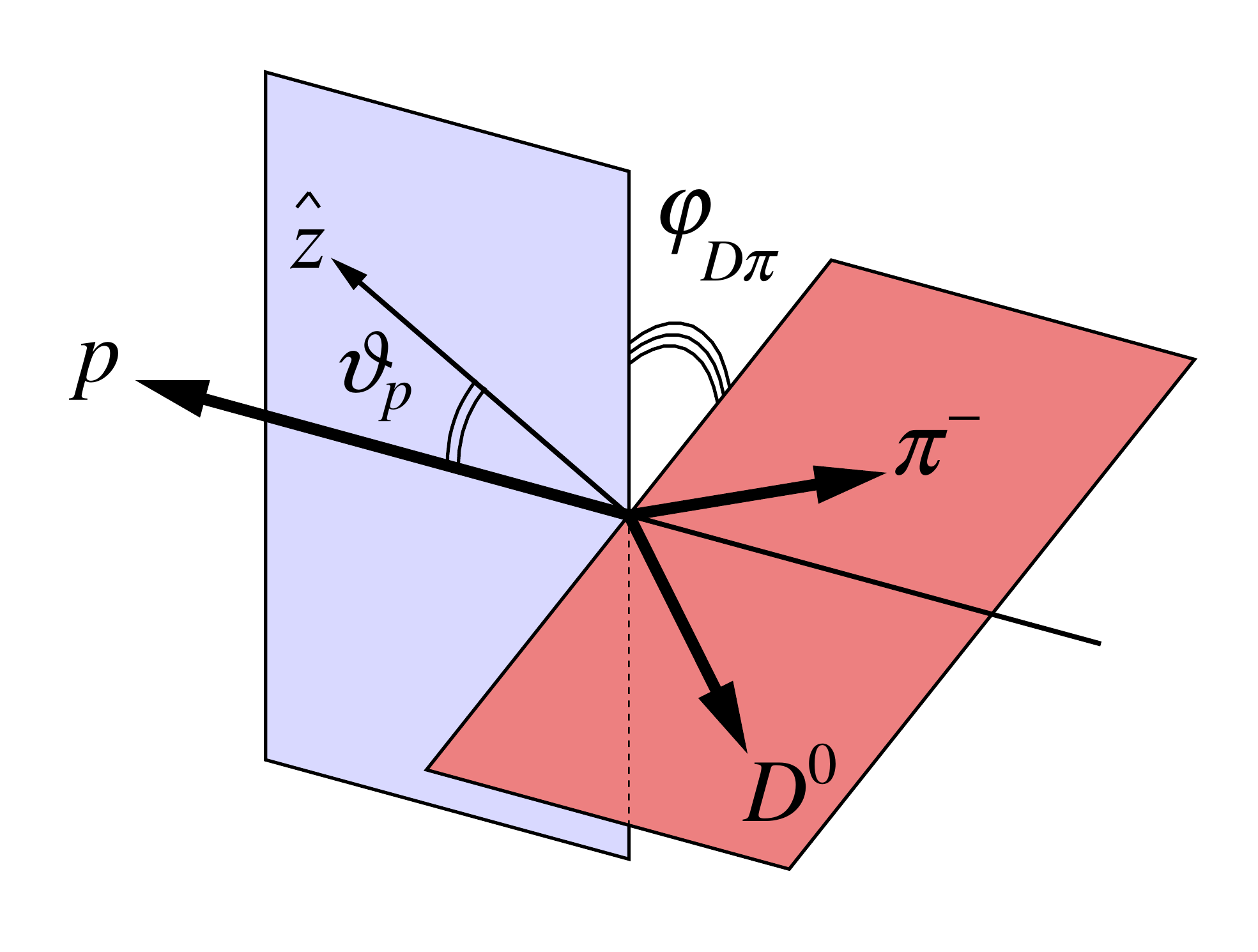}
  \put(-160,100){(b)}
  \end{center}
  \caption{Definition of the angles describing the orientation of the \lbdnppi decay in the reference frame where the \Lb baryon is at rest: 
           (a) $\vartheta_p$ and $\varphi_p$, and (b) $\varphi_{D\pi}$. }
  \label{fig:angles}
\end{figure}

A vector of five kinematic variables (denoted $\Omega$) describes the phase space of the decay \lbdnppi. 
The kinematic variables are the two Dalitz plot variables, namely the invariant masses squared 
of the $\Dz\proton$ and $\proton\pim$ combinations $M^2(\Dz\proton)$ and $M^2(\proton\pim)$, 
and three angles that determine the orientation of the three-body decay plane (Fig.~\ref{fig:angles}). 
These angles are defined in the rest frame of the decaying \Lb baryon with 
the $\hat{x}$ axis given by the direction of the \Lb baryon in the laboratory frame, 
the polarisation axis $\hat{z}$ given by the cross-product of beam direction and $\hat{x}$ axis, and the $\hat{y}$ axis 
given by the cross-product of the $\hat{z}$ and $\hat{x}$ axes. 
The angular variables are the cosine of the polar angle $\cos\vartheta_{\proton}$, 
and the azimuthal angle $\varphi_{\proton}$ of the proton momentum in the reference frame defined above (Fig.~\ref{fig:angles}(a)), 
and the angle $\varphi_{D\pi}$ between the $\Dz\pim$ plane and the plane formed by the proton direction 
and the polarisation axis $\hat{z}$ (Fig.~\ref{fig:angles}(b)). 

\subsection{Helicity formalism}

The baseline amplitude fit uses the helicity formalism where the interfering amplitude components 
are expressed as sequential quasi-two-body decays $\Lb\to R\pim$, $R\to\Dz\proton$ (where $R$ 
denotes the intermediate resonant or nonresonant state). 
The decay amplitude for a \Lb baryon with spin projection $\mu$ decaying via 
an intermediate state $R$ with helicity $\lambda_R$ into a 
final state with proton helicity $\lambda_p$ is
\begin{equation}
  \label{eq:hel_general}
  \begin{split}
  \mathcal{A}_{\mu, \lambda_R, \lambda_p} & [M^2(\Dz\proton), \theta_p, \phi_p, \theta_R, \phi_R] = \\
     & a_{\lambda_R}\,b_{\lambda_p}\,
       e^{i(\mu-\lambda_R)\phi_{R}}\,e^{i(\lambda_R-\lambda_p)\phi_p}\,
       d^{J_{\Lb}}_{\mu,\lambda_R}(\theta_R)\,d^{J_R}_{\lambda_R\lambda_p}(\theta_p)\,\mathcal{R}(M^2(\Dz\proton)), 
  \end{split}
\end{equation}
where $J_{\Lb}=1/2$ and $J_R$ are the spins of the \Lb baryon and the $R$ state, 
$d^{J}_{\lambda_1,\lambda_2}(\theta)$ are the reduced Wigner functions, and 
$a_{\lambda_R}$ and $b_{\lambda_p}$ are complex constants (couplings). 
The mass-dependent complex lineshape $\mathcal{R}(M^2)$ defines the dynamics of the $R$ decay. 
The angles defining the helicity amplitude are 
the polar ($\theta_R$) and azimuthal ($\phi_R$) angles of the intermediate state $R$ in the reference frame 
defined above, 
and the polar ($\theta_p$) and azimuthal ($\phi_p$) angles of 
the final-state proton in the frame where the intermediate state $R$ is at rest and
the polar axis points in the direction of $R$ in the \Lb rest frame. 
All of these angles 
are functions of the five phase space variables $\Omega$ defined previously and thus do not constitute 
additional degrees of freedom. 

The strong decay $R\to \Dz\proton$ conserves parity, which implies that 
\begin{equation}
  b_{\lambda_p} = (-1)^{J_p+J_D-J_{R}}\,\eta_{R}\,\eta_{D}\,\eta_{p}\,b_{-\lambda_p}, 
\end{equation}
where $J_p=1/2$, $J_D=0$ and $J_R$ are the spins of the proton, $\Dz$ meson and resonance $R$, respectively, and 
$\eta_p=+1$, $\eta_D=-1$ and $\eta_R$ are their parities. 
This relation reduces the number of free parameters in the helicity amplitudes: $|b_{\lambda_p}|$
is absorbed by $a_{\lambda_R}$, and each coefficient $a_{\lambda_R}$ 
enters the amplitude multiplied by a factor $\eta_{\lambda_p}=\pm 1$. The convention used is
\begin{equation}
  \eta_{\lambda_p} = \left\{ \begin{array}{ll}  
                             1 & \mbox{if $\lambda_p = +1/2$}, \\  
                             (-1)^{J_p+J_D-J_{R}}\,\eta_{R}\,\eta_{D}\,\eta_{p} & \mbox{if $\lambda_p = -1/2$}. \\  
                            \end{array} \right.
\end{equation}
As a result, only two 
couplings $a_{\lambda_R}$ remain for each intermediate state $R$, corresponding to its 
two allowed helicity configurations. The two couplings are denoted for brevity as $a^{\pm}$. 

The amplitude, for fixed $\mu$ and $\lambda_p$, after summation over the intermediate resonances $R_j$ 
and their two possible helicities $\lambda_{R_j}=\pm 1/2$ is  
\begin{equation}
 \begin{split}
  A_{\mu, \lambda_p}(\Omega) = e^{i(\mu\phi_R-\lambda_p\phi_p)}
                      \sum\limits_j \eta_{j, \lambda_p} & \left[ a^+_{j}\, d^{J_{\Lb}}_{\mu,+1/2}(\theta_R)\, 
                               d^{J_{R_i}}_{+1/2, \lambda_p}(\theta_p)\, \mathcal{R}_j(M^2(\Dz\proton)) +  \right. \\
                                             & \left. \;\, a^-_{j}\, d^{J_{\Lb}}_{\mu,-1/2}(\theta_R)\, 
                               d^{J_{R_i}}_{-1/2, \lambda_p}(\theta_p)\, \mathcal{R}_j(M^2(\Dz\proton))\, e^{i(\phi_R-\phi_p)} \right]. 
 \end{split}
 \label{eq:ampl_onechannel}
\end{equation}
To obtain the decay probability density, the amplitudes corresponding to different polarisations of the 
initial- and final-state particles have to be summed up incoherently. 
The \Lb baryons produced in $\proton\proton$ collisions can only have 
polarisation transverse to the production plane, \ie along the $\hat{z}$ axis. 
The longitudinal component is forbidden due to parity conservation in the strong processes that 
dominate \Lb production. In this case, the probability density function (PDF) 
of the kinematic variables that characterise the decay
of a \Lb with the transverse polarisation $P_{z}$, after summation over $\mu$ and $\lambda_p$, is
proportional to 
\begin{equation}
  p(\Omega, P_z) = \sum\limits_{\mu, \lambda_p=\pm 1/2} (1+2\mu P_z)|A_{\mu, \lambda_p}(\Omega)|^2. 
  \label{eq:dens_onechannel}
\end{equation}
Equations (\ref{eq:ampl_onechannel}) and (\ref{eq:dens_onechannel}) can be combined to yield the 
simplified expression: 
\begin{equation}
  p(\Omega, P_z) = 
       \sum\limits_{n=0}^{2J_{\rm max}}p_n(M^2(\Dz\proton))\cos(n\theta_p) + 
       P_z \cos\theta_R\sum\limits_{n=0}^{2J_{\rm max}}q_n(M^2(\Dz\proton))\cos(n\theta_p), 
\end{equation}
where $J_{\rm max}$ is the highest spin among the intermediate resonances and
$p_n$ and $q_n$ are functions of only $M^2(\Dz\proton)$. 
As a consequence, $p(\Omega, P_z)$ does not depend on the azimuthal angles $\phi_p$ and $\phi_R$. Dependence on 
the angle $\theta_R$ appears only if the \Lb is polarised. In the unpolarised case the density depends only on  
the internal degrees of freedom $M^2(\Dz\proton)$ and $\theta_p$ (which in turn can be expressed as
a function of the other Dalitz plot variable, 
$M^2(\proton\pim)$). Moreover, after integration over the angle $\theta_R$, the dependence on polarisation
cancels if the detection efficiency is symmetric over $\cos\theta_R$. Since \Lb polarisation in $\proton\proton$
collisions is measured to be small ($P_z=0.06\pm 0.07\pm 0.02$,~\cite{LHCb-PAPER-2012-057}) and the efficiency 
is highly symmetric in $\cos\theta_R$, the effects of polarisation can safely be neglected in the amplitude analysis, 
and only the Dalitz plot variables $\omega = (M^2(\Dz\proton), M^2(\proton\pim))$ need to be used to 
describe the probability density $p(\omega)$ of the decay. The density $p(\omega)$ is given by Eq.~(\ref{eq:dens_onechannel}) 
with $P_z=0$ such that no dependence on the angles $\vartheta_p$, $\varphi_{p}$ or $\varphi_{D\pi}$ remains.

Up to this point, the formalism has assumed that resonances are present only in the $\Dz\proton$ channel. 
While in the case of 
\lbdnppi decays the regions of phase space with contributions from $\Dz\proton$ and $\proton\pim$ resonances 
are generally well separated, there is a small region where they can overlap, and thus interference between resonances in 
the two channels has to be taken into account. In the helicity formalism, the proton spin-quantisation 
axes are different for the helicity amplitudes corresponding to $\Dz\proton$ and $\proton\pim$ 
resonances~\cite{LHCb-PAPER-2015-029}: they are parallel to the proton direction 
in the $\Dz\proton$ and $\proton\pim$ rest frames, and are thus antiparallel to the $\pim$ and $\Dz$ momenta, 
respectively. 
The rotation angle between the two spin-quantisation axes is given by
\begin{equation}
  \cos\theta_{\rm rot} = \frac{(\vec{p}^{\,(\proton)}_{\pim}\cdot \vec{p}^{\,(\proton)}_{\Dz})}{|\vec{p}^{\,(\proton)}_{\pim}| |\vec{p}^{\,(\proton)}_{\Dz})|}, 
\end{equation}
where $\vec{p}^{\,(\proton)}_{\pim}$ and $\vec{p}^{\,(\proton)}_{\Dz}$ are the momenta of the $\pim$ and $\Dz$ mesons, respectively, in the proton rest frame. 

If the proton spin-quantisation axis is chosen with respect to the $\Dz\proton$ resonances and the 
helicity basis is denoted as $|\lambda_p \rangle$, 
the helicity states $|\lambda'_p \rangle$ corresponding to $\proton\pim$ states are 
\begin{equation}
  |\lambda'_p \rangle = \sum\limits_{\lambda'_p = \pm 1/2} d^{1/2}_{\lambda_p, \lambda'_p}(\theta_{\rm rot})|\lambda_p \rangle
\end{equation}
and thus the additional terms in the amplitude (Eq.~(\ref{eq:ampl_onechannel})) related to the $\proton\pim$ channel 
are expressed as 
\begin{equation}
 \begin{split}
  A_{\mu, \lambda_p}^{(\proton\pim)}(\Omega) = & 
                      \sum\limits_{\lambda'_p = \pm 1/2} d^{1/2}_{\lambda_p, \lambda'_p}(\theta_{\rm rot})
                      \;e^{i(\mu\phi'_R-\lambda'_p\phi'_p)}
                      \sum\limits_j \eta_{j, \lambda'_p}\times \\ 
                                             & \left[ a^+_{j}\, d^{J_{\Lb}}_{\mu,+1/2}(\theta'_R)\, 
                               d^{J_{R_i}}_{+1/2, \lambda'_p}(\theta'_p)\, \mathcal{R}_j(M^2(\proton\pim)) +  \right. \\
                                             & \left. \;\, a^-_{j}\, d^{J_{\Lb}}_{\mu,-1/2}(\theta'_R)\, 
                               d^{J_{R_i}}_{-1/2, \lambda'_p}(\theta'_p)\, \mathcal{R}_j(M^2(\proton\pim))\, e^{i(\phi'_R-\phi'_p)} \right],  
 \end{split}
 \label{eq:ampl_ppi}
\end{equation}
where the angles $\theta'_p$, $\phi'_p$, $\theta'_R$ and $\phi'_R$ are defined in a 
similar way as $\theta_p$, $\phi_p$, $\theta_R$ and $\phi_R$, 
but with the intermediate state $R$ in the $\proton\pim$ channel. 

\subsection{Resonant and nonresonant lineshapes}

The part of the amplitude that describes the dynamics of the quasi-two-body decay, $\mathcal{R}(M^2)$, is given by one 
of the following functions. Resonances are parametrised with relativistic Breit--Wigner lineshapes multiplied 
by angular barrier terms and corrected by Blatt--Weisskopf form factors \cite{blatt-weisskopf}:
\begin{equation}
  \mathcal{R}_{\rm BW}(M^2) = \left[\frac{q(M)}{q_0}\right]^{L_{\Lb}}\left[\frac{p(M)}{p_0}\right]^{L_{R}}
                              \frac{F_{\Lb}(M,L_{\Lb}) F_R(M,L_R)}{m_R^2-M^2 - i m_R\Gamma(M)}
                    , 
\end{equation}
with mass-dependent width $\Gamma(M)$ given by
\begin{equation}
  \Gamma(M) = \Gamma_0\left[\frac{p(M)}{p_0}\right]^{2L_R+1}\frac{m_R}{M} F_R^2(M,L_R), 
\end{equation}
where $m_R$ and $\Gamma_0$ are the pole parameters of the resonance.
The Blatt--Weisskopf form factors for the resonance, $F_R(M,L_R)$, and for the $\Lb$, $F_{\Lb}(M,L_{\Lb})$, 
are parametrised as
\begin{equation}
  F_{R,\Lb}(M,L) = \left\{
   \begin{array}{ll}
     1                      & L=0 \\
     \sqrt{\frac{1+z_0^2}{1+z^2(M)}} & L=1 \\
     \sqrt{\frac{9+3z_0^2+z_0^4}{9+3z^2(M)+z^4(M)}} & L=2 \\
     \sqrt{\frac{225+45z_0^2+6z_0^4+z_0^6}{225+45z^2(M)+6z^4(M)+z^6(M)}} & L=3 \\
   \end{array}
  \right., 
\end{equation}
where the definitions of the terms $z(M)$ and $z_0$ depend on whether 
the form factor for the resonance $R$ or for the \Lb is being considered. For $R$ these terms are given by
$z(M)=p(M)d$ and $z_0=p_0d$, where $p(M)$ is the centre-of-mass momentum of the 
decay products in the two-body decay $R\to \Dz\proton$ with the mass 
of the resonance $R$ equal to $M$, $p_0\equiv p(m_R)$, and 
$d$ is a radial parameter taken to be $1.5\gev^{-1}$. 
For \Lb the respective functions are $z(M)=q(M)d$ and $z_0=q_0d$, where $q(M)$ is the centre-of-mass 
momentum of decay products in the two-body decay $\Lb\to R\pim$, $q_0=q(m_R)$, and $d=5.0\gev^{-1}$. 
The analysis is very weakly sensitive to the values of $d$, and these are 
varied in a wide range for assessing the associated systematic uncertainty (Sec.~\ref{sec:lc2880_amplitude}).

The mass-dependent width and form factors depend on the orbital angular momenta of the two-body decays. 
For the weak decay of the \Lb, the minimum possible angular momentum $L_{\Lb}=J-1/2$ 
(where $J$ is the spin of the resonance) is taken, while for the strong decay 
of the intermediate resonance, the angular momentum $L_R$ is fully determined 
by the parity of the resonance, $P=(-1)^{L_R+1}$, 
and conservation of angular momentum, which requires $L_R=J\pm 1/2$. 

Two parametrisations are used for nonresonant amplitudes: exponential and polynomial functions. 
The exponential nonresonant lineshape~\cite{Garmash:2004wa} used is
\begin{equation}
  \label{eq:nr_exp}
  \mathcal{R}_{\rm NRexp}(M^2) = \left[\frac{q(M)}{q_0}\right]^{L_{\Lb}}\left[\frac{p(M)}{p_0}\right]^{L_{R}}e^{-\alpha M^2}, 
\end{equation}
where $\alpha$ is a shape parameter.
The polynomial nonresonant lineshape~\cite{Lees:2012kxa} used is
\begin{equation}
  \label{eq:nr_poly}
  \mathcal{R}_{\rm NRpoly}(M^2) = \left[\frac{q(M)}{q_0}\right]^{L_{\Lb}}\left[\frac{p(M)}{p_0}\right]^{L_{R}}(a_2 \Delta M^2 + a_1 \Delta M + a_0), 
\end{equation}
where $\Delta M=M-M_0$, and $M_0$ is a constant that is chosen to minimise the correlations between the 
coefficients $a_i$ when they are treated as free parameters. In the case of the 
$\Dz\proton$ amplitude fit, $M_0$ is chosen to be near the middle of the fit range, $M_0\equiv 2.88\gev$. 
In both the exponential and the 
polynomial parametrisations, $M_0$ also serves as the resonance mass parameter in the definition of $p_0$ and $q_0$
in the angular barrier terms. Note that in Ref.~\cite{Lees:2012kxa} the polynomial 
form was introduced to describe the slow variations of a nonresonant amplitude across the large phase space of 
charmless $\B$ decays, and thus the parameters $a_i$ were defined as complex constants to allow slow phase motion over the 
wide range of invariant masses. In the present analysis, the phase space is much more constrained and 
no significant phase rotation is expected for the nonresonant amplitudes. 
The coefficients $a_i$ thus are taken to be real. 

To study the resonant nature of the $\Dz\proton$ states, model-independent parametrisations of the lineshape are used. 
One approach used here consists of interpolation with cubic splines, done independently for 
the real and imaginary parts of the amplitude (referred to as the ``complex spline'' lineshape)~\cite{LHCb-PAPER-2016-026}. 
The free parameters of such a fit are the real ${\rm Re}(\mathcal{R}_i)$ and 
imaginary ${\rm Im}(\mathcal{R}_i)$ parts of the amplitude at the spline knot positions. 
Alternatively, to assess the significance of the complex phase rotation in a model-independent 
way, a spline-interpolated shape is used in which the imaginary parts of the amplitude 
at all knots are fixed to zero (``real spline''). 

\subsection{Fitting procedure}

An unbinned maximum likelihood fit is performed in the two-dimensional phase space 
$\omega=(M^2(\Dz\proton), M^2(\proton\pim))$. 
Defining $\mathcal{L}$ as the likelihood function, the fit minimises
\begin{equation}
  \label{eq:likelihood}
  -2\ln\mathcal{L}=-2\sum\limits_{i=1}^{N}\ln p_{\rm tot}(\omega_i), 
\end{equation}
where the summation is performed over all candidates in the data sample and 
$p_{\rm tot}$ is the normalised PDF. It is given by
\begin{equation}
  \label{eq:total_density}
  p_{\rm tot}(\omega) = p(\omega)\epsilon(\omega)\frac{n_{\rm sig}}{\mathcal{N}} + p_{\rm bck}(\omega)\frac{n_{\rm bck}}{\mathcal{N}_{\rm bck}},  
\end{equation}
where $p(\omega)$ is the signal PDF, $p_{\rm bck}(\omega)$ is the background PDF, 
$\epsilon(\omega)$ is the efficiency, and $\mathcal{N}$ and $\mathcal{N}_{\rm bck}$ are the signal and background 
normalisations: 
\begin{equation}
  \mathcal{N}=\int\limits_{\mathcal{D}}p(\omega)\epsilon(\omega)\;d\omega, 
\end{equation}
and
\begin{equation}
  \mathcal{N}_{\rm bck}=\int\limits_{\mathcal{D}}p_{\rm bck}(\omega)\;d\omega, 
\end{equation}
where the integrals are taken over the part of the phase space $\mathcal{D}$ 
used in the fit (Section~\ref{sec:yields}), 
and $n_{\rm sig}$ and $n_{\rm bck}$ are the numbers of signal and background events in the signal region, respectively, 
evaluated from a fit to the $M(\Dz\proton\pim)$ invariant mass distribution. 
The normalisation integrals are calculated numerically using a fine grid with $400\times 400$ cells 
in the baseline fits; the numerical uncertainty is negligible compared with the other uncertainties in the analysis.

\subsection{Fit parameters and fit fractions}

The free parameters in the fit are the couplings $a^{\pm}$ for each of the amplitude components and certain parameters 
of the lineshapes (such as the masses and/or widths of the resonant states, or shape parameters of the nonresonant lineshapes). 
Since the overall normalisation of the density is arbitrary, one of the couplings can be set to unity. In this analysis, the 
convention $a^+\equiv 1$ for the \Lcst state is used. 
Additionally, the amplitudes corresponding to different helicity states of the initial- and final-state particles 
are added incoherently, so that the relative phase between $a^+$ and $a^-$ for one of the contributions is arbitrary. 
The convention ${\rm Im}(a^-)\equiv 0$ for the \Lcst is used. 

The definitions of the polynomial and spline-interpolated shapes already contain terms that characterise the 
relative magnitudes of the corresponding amplitudes. The couplings for them are defined in such a way as to remove the 
additional degree of freedom from the fit. For the polynomial and real spline lineshapes, the following couplings are used: 
\begin{equation}
  a^+ = r e^{i\phi_+},\;\;\; a^-=(1-r)e^{i\phi_-}, 
\end{equation}
where $r$, $\phi_+$ and $\phi_-$ are free parameters. For the complex spline lineshape, 
a similar parametrisation is used with $\phi_+$ fixed to zero, since the complex phase is 
already included in the spline definition. 

The observable decay density for an unpolarised particle in the initial state does not allow each polarisation 
amplitude to be obtained independently. As a result, the couplings $a^{\pm}$ in the fit can be strongly correlated. 
However, the size of each contribution can be characterised by its spin-averaged fit fraction
\begin{equation}
  \mathcal{F}_i=\frac{\sum\limits_{\mu, \lambda_p=\pm 1/2}\;\int\limits_{\mathcal{D}} |A^{(i)}_{\mu,\lambda_p}(\omega)|^2\, d\omega}
                     {\sum\limits_{\mu, \lambda_p=\pm 1/2}\;\int\limits_{\mathcal{D}} |\sum\limits_{i}A^{(i)}_{\mu,\lambda_p}(\omega)|^2\, d\omega}. 
\end{equation}
If all the components correspond to partial waves with different spin-parities, the sum of the spin-averaged fit 
fractions will be 100\%; 
otherwise it can differ from 100\% due to interference effects. The statistical uncertainties on the fit fractions are obtained 
from ensembles of pseudoexperiments. 

\subsection{Evaluation of fit quality}

To assess the goodness of each fit, a $\chi^2$ value is calculated by summing over the bins of the 
two-dimensional Dalitz plot. 
Since the amplitude is highly non-uniform and a meaningful $\chi^2$ test requires a certain minimum 
number of entries in each bin, 
an adaptive binning method is used to ensure that each bin 
contains at least 20 entries in the data. 

Since the fit itself is unbinned, some information is lost by the binning. 
The number of degrees of freedom for the $\chi^2$ test in such a case is not well defined. 
The effective number of degrees of freedom (${\rm ndf}_{\rm eff}$)
should be in the range $N_{\rm bins}-N_{\rm par}-1\leq{\rm ndf}_{\rm eff}\leq N_{\rm bins}-1$, 
where $N_{\rm bins}$ is the number of bins and $N_{\rm par}$ is the number of free parameters in the fit. 
For each fit, ${\rm ndf}_{\rm eff}$ is obtained from ensembles of 
pseudoexperiments by requiring that the probability value for the $\chi^2$ distribution with ${\rm ndf}_{\rm eff}$ degrees of freedom, 
$P(\chi^2, {\rm ndf}_{\rm eff})$, is distributed uniformly. 

Note that when two fits with different models have similar binned $\chi^2$ values, it does not necessarily follow that both models describe the data 
equally well. Since the bins in regions with low population density have large area, the binning can obscure features that could discriminate between 
the models. This information is preserved in the unbinned likelihood. 
Thus, discrimination between fit models is based on the difference $\Delta\ln\mathcal{L}$, the statistical significance of which is determined using 
ensembles of pseudoexperiments. The binned $\chi^2$ serves as a measure of the fit quality for individual models and is not used to discriminate
between them.

\section{Signal selection}

\label{sec:selection}

The analysis uses the decay $\lbdnppi$, where $\Dz$ mesons are reconstructed in the final state $\Km\pip$.
The selection of \Lb candidates is performed in three stages: a preliminary selection, a kinematic fit, 
and a final selection. 
The preliminary selection uses loose criteria on the kinematic and topological properties of the 
\Lb candidate. All tracks forming a candidate, as well as the \Lb and \Dz vertices, 
are required to be of good quality and be separated from every PV in the event. 
The separation from a PV is characterised by a quantity $\chi^2_{\rm IP}$, defined as 
the increase in the vertex-fit $\chi^2$ when the track (or combination of tracks corresponding 
to a short-lived particle) is included into the vertex fit. 
The tracks forming a $\Dz$ candidate are required to be positively identified as a pion 
and a kaon, and the $\Lb$ and $\Dz$ decay vertices are required to be downstream 
of their production vertices. All of the tracks are required to have no 
associated hits in the muon detector. 

For candidates passing this initial selection, a kinematic fit is performed~\cite{Hulsbergen:2005pu}. 
Constraints are imposed that the \Lb and \Dz decay products originate from the corresponding vertices, 
that the \Lb candidate originate from its associated PV (the one with the smallest value of $\chi^2_{\rm IP}$ for the \Lb), 
and that the mass of the \Dz candidate be equal to its known 
value~\cite{PDG2016}. The kinematic fit is required to converge with a good $\chi^2$, and 
the mass of the $\Lb$ candidate after the fit is required to be in the range $5400$--$5900\mev$. 
To suppress background from charmless $\Lb\to \proton\Km\pip\pim$ decays, 
the decay time significance of the $\Dz$ candidate obtained after the fit is required to be 
greater than one standard deviation. To improve the resolution of the squared invariant masses $M^2(\Dz\proton)$
and $M^2(\proton\pim)$ entering the amplitude fit, the additional constraint that the
invariant mass of the $\dnppi$ combination be equal to the known \Lb mass~\cite{PDG2016} 
is applied when calculating these variables. 

After the initial selection, the background in the region of the \lbdnppi signal is dominated 
by random combinations of tracks. The final selection is based on a boosted decision tree 
(BDT) algorithm~\cite{Breiman,AdaBoost} designed to separate signal from this background. 
The selection is trained using simulated \lbdnppi events
generated uniformly across the phase space as the signal sample, and the sample of opposite-flavour 
\dbppi, $\Dzb\to\Kp\pim$ combinations from data as background. In total, 12 discriminating variables 
are used in the BDT selection: the $\chi^2$ of the kinematic fit, the angle between the 
momentum and the direction of flight of the \Lb candidate, the $\chi^2$ of the \Lb and \Dz vertex fits, 
the lifetime significance of the \Dz candidate with respect to the \Lb vertex, 
the $\chi^2_{\rm IP}$ of the final-state tracks and the \Dz candidate, 
and the particle identification (PID) information 
of the proton and pion tracks from the \Lb vertex. 
Due to differences between simulation and 
data, corrections are applied to all the variables from the simulated sample used in the BDT training, except for the PID variables.
These corrections are typically about $10\%$ and are obtained from a large and clean sample of \lblcpi decays. 
The simulated proton and pion PID variables are replaced with values generated using 
distributions obtained from calibration samples of $\Dstarp\to\Dz\pip$ and $\Lc\to\proton\Km\pip$ decays in data. 
For these calibration samples, the four-dimensional distributions of PID variable, 
$\pt$, $\eta$ and the track multiplicity of the event are described using a nonparametric kernel-based procedure~\cite{meerkat}. 
The resulting distributions are used to generate PID variables for each pion or proton track given its $\pt$, 
$\eta$ and the track multiplicity in the simulated event.

The BDT requirement is chosen such that the fraction of background in the signal region used for the 
subsequent amplitude fit, $|M(\dnppi)-m(\Lb)|<30\mev$, does not exceed 15\%. This corresponds to a signal
efficiency of 66\% and a background rejection of 96\% with respect to the preliminary selection. 
After all selection requirements are applied, fewer than 1\% of
selected events contain a second candidate. All multiple candidates are retained; 
the associated systematic uncertainty is negligible.

\section{Fit regions and event yields}

\label{sec:yields}

\begin{figure}
  \includegraphics[width=0.47\textwidth]{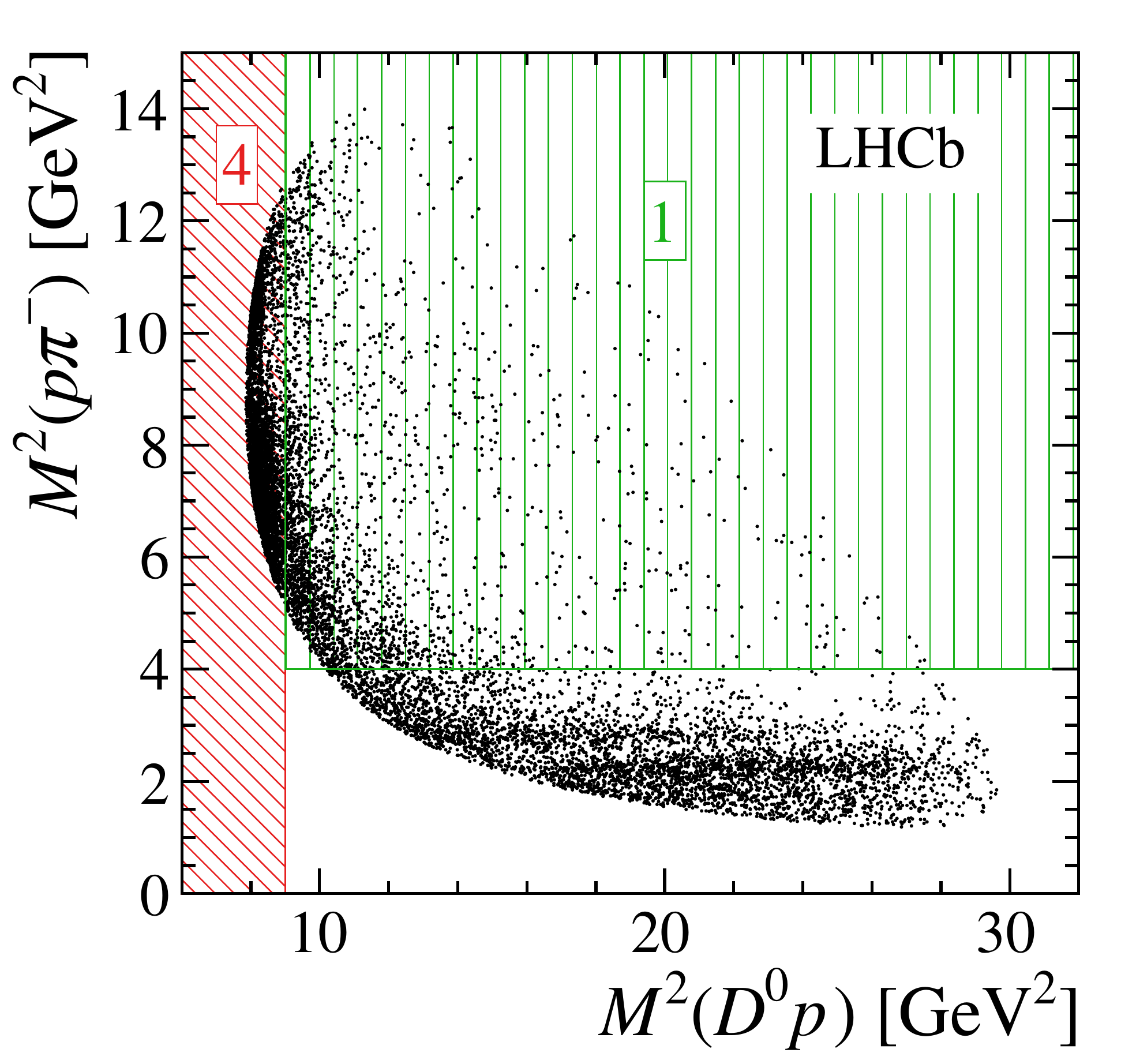}
  \put(-156, 50){(a)}
  \hfill
  \includegraphics[width=0.47\textwidth]{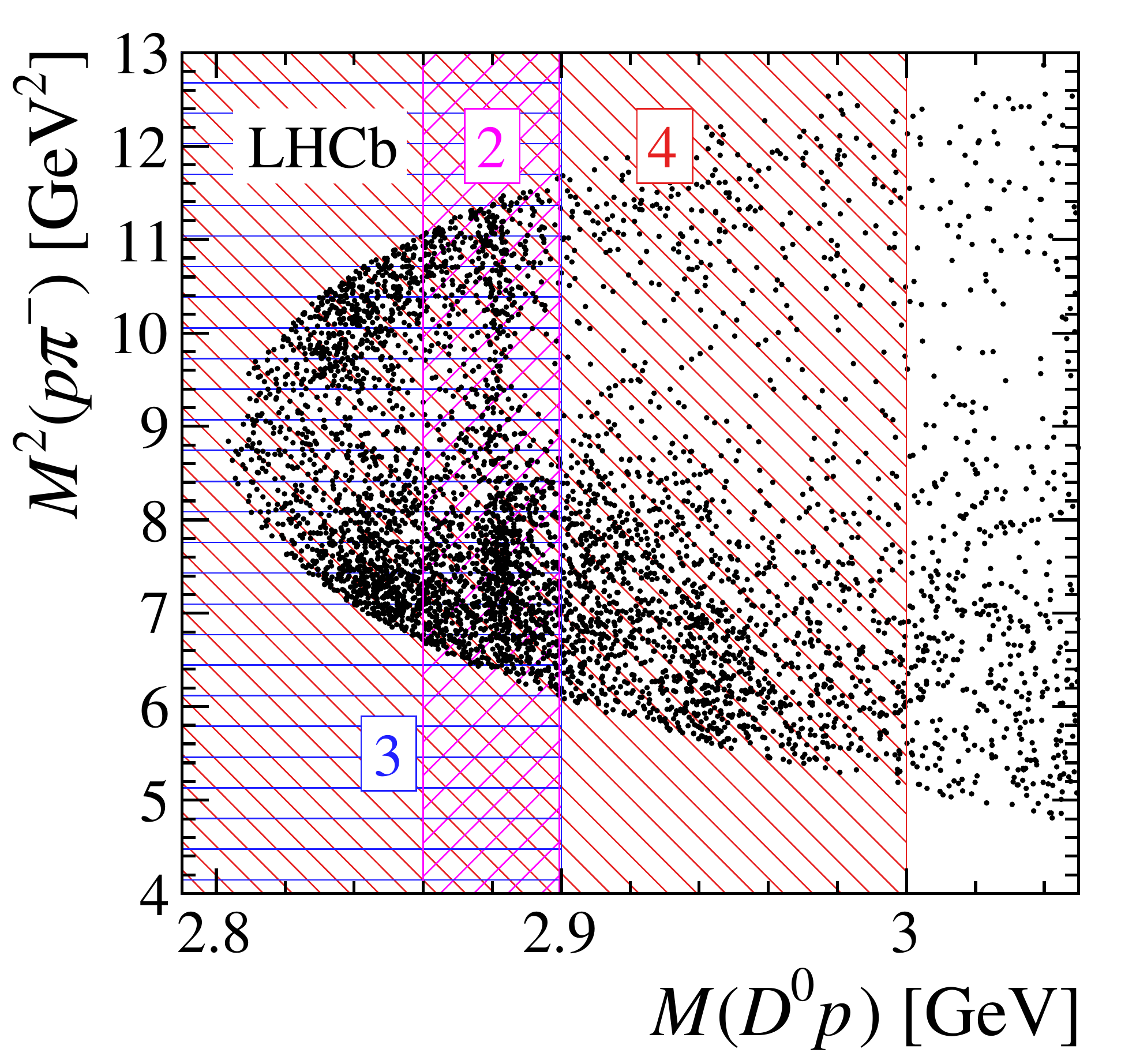}
  \put(-171, 50){\colorbox{white}{(b)}}
  \caption{Distributions of \lbdnppi candidates in data: (a) the full Dalitz plot as a function of $M^2(D^0p)$ and $M^2(p\pi^-)$, 
           and (b) the part of the phase space including the resonances in the $D^0p$ channel 
           (note the change in variable on the horizontal axis). 
           The distributions are neither background-subtracted nor efficiency-corrected.
           The hatched areas 1--4 are described in the text. }
  \label{fig:phase_space}
\end{figure}

The Dalitz plot of selected events, without background subtraction or efficiency correction, 
in the signal $\dnppi$ invariant mass range defined in Sec.~\ref{sec:selection} is shown in Fig.~\ref{fig:phase_space}(a).
The part of the phase space near the $\Dz\proton$ threshold that contains contributions from \Lcstar 
resonances is shown in Fig.~\ref{fig:phase_space}(b). The latter uses 
$M(\Dz\proton)$ as the horizontal axis instead of $M^2(\Dz\proton)$. 

In Fig.~\ref{fig:phase_space}, the four amplitude fit regions of the \lbdnppi phase space are indicated.
These are denoted regions 1--4. Region~1, $M(\Dz\proton)>3\gev$ and $M(\proton\pim)>2\gev$, 
is the part of the phase space that does not include resonant contributions and is used only to constrain 
the nonresonant $\proton\pim$ amplitude in the $\Dz\proton$ regions. Region~2, 
$2.86<M(\Dz\proton)<2.90\gev$, contains the well-known \Lcst state and is used to measure its 
parameters and to constrain the slowly varying amplitude underneath it in a model-independent way. 
The fit in region~3 near the $\Dz\proton$ threshold, $M(\Dz\proton)<2.90\gev$, provides additional 
information about the slowly-varying $\Dz\proton$ amplitude. 
Finally, the fit in region~4, $M(\Dz\proton)<3.00\gev$, 
which includes the \Lcstst state, gives information about the properties of this resonance and the relative 
magnitudes of the resonant and nonresonant contributions. 
Note that region 2 is fully contained in region 3, while region 3 is fully contained in region 4.

The signal and background yields in each region are obtained from extended unbinned maximum likelihood fits 
of the $\dnppi$ invariant mass distribution in the range $5400$--$5900\mev$. The fit model 
includes the signal component, a contribution from random combinations of tracks (combinatorial 
background) and the background from partially reconstructed \lbdstppi decays 
(where \Dstarz decays into $\Dz\piz$ or $\Dz\g$ and the $\piz$ or $\g$ are not included in the 
reconstruction). 

The signal component is modelled as the sum of two Crystal Ball functions~\cite{CB} with the same most
probable value and power-law tails on both sides. 
All parameters of the model are fixed from simulation except for the peak position  
and a common scale factor for the core widths, which are floated in the fit to data. The combinatorial background 
is parametrised by an exponential function, and the partially reconstructed 
background is described by a bifurcated Gaussian distribution. The shape parameters of the background 
distributions are free parameters of the fit.  

\begin{figure}
  \begin{center}
  \includegraphics[width=0.45\textwidth]{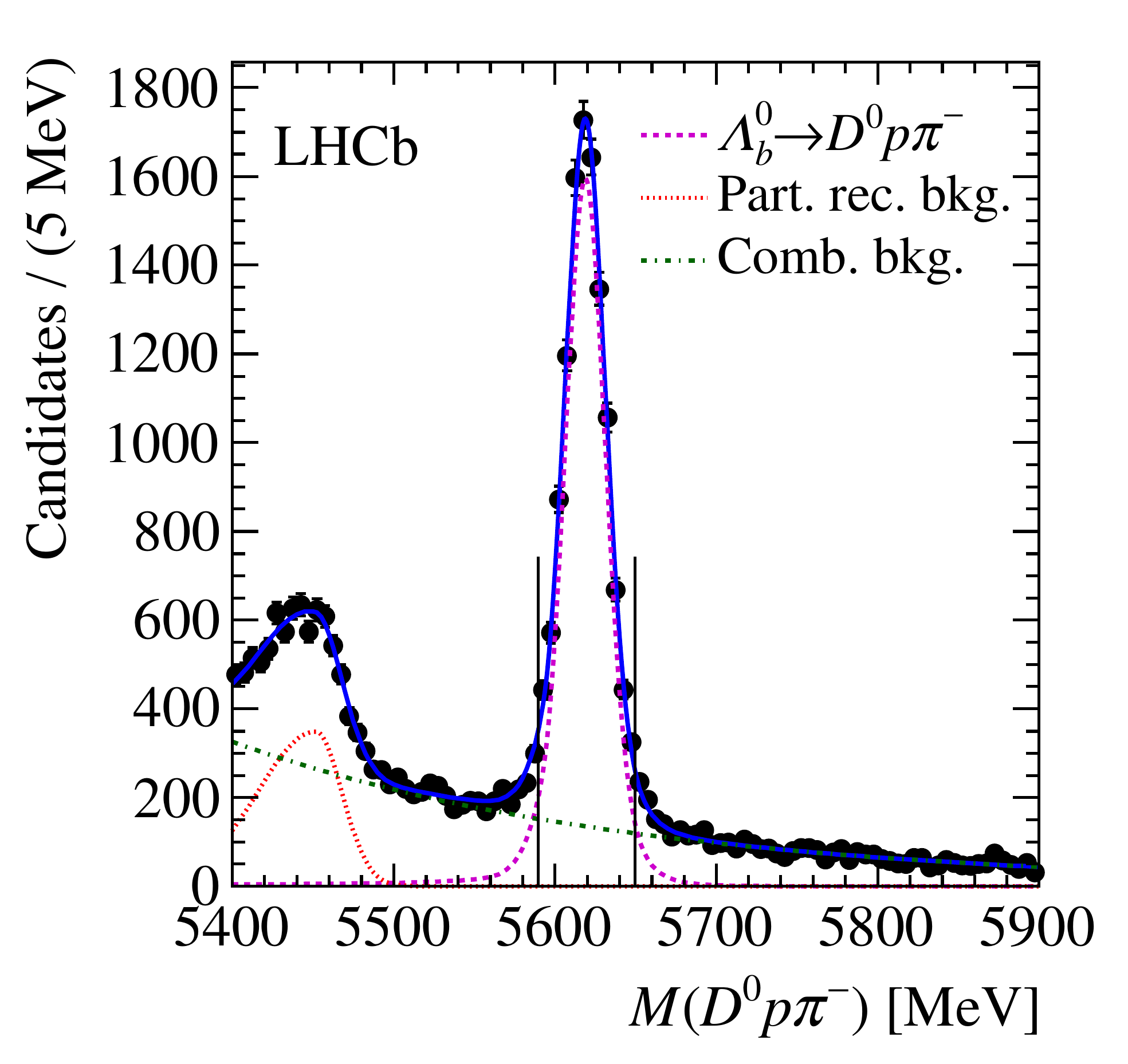}
  \end{center}
  \caption{Invariant mass distribution for the \dnppi candidates in the entire \dnppi phase space. 
           The blue solid line is the fit result. 
           Signal, partially reconstructed and combinatorial background components are shown with different line styles. 
           Vertical lines indicate the boundaries of the signal region used in the amplitude fit. }
  \label{fig:mass_fit}
\end{figure}

The results of the fit for candidates in the entire \dnppi phase space are shown in Fig.~\ref{fig:mass_fit}. 
The background and signal yields in the entire 
$\dnppi$ phase space, as well as in the regions used in the amplitude fit, are given in 
Table~\ref{tab:mass_fit}. 

\begin{table}
  \caption{Results of the fits to the \lbdnppi mass distribution in 
           the entire \lbdnppi phase space and in the four phase space regions 
           used in the amplitude fits. The signal and background yields
           for the full $M(\dnppi)$ range, as well as for the amplitude 
           fit region $|M(\dnppi)-m(\Lb)|<30\mev$ (``box''), are reported.
  }
  \label{tab:mass_fit}
  \begin{tabular}{l|rrrrr}
            & \multicolumn{5}{c}{Phase space region} \\
  \cline{2-6}
  Yield & \multicolumn{1}{c}{Full} & \multicolumn{1}{c}{1} & \multicolumn{1}{c}{2} & \multicolumn{1}{c}{3} & \multicolumn{1}{c}{4} \\
  \hline
  \lbdnppi         & $11\,212 \pm  126 $  & $2\,250 \pm   61 $\phantom{0}  
                 & $1\,674 \pm   46 $   & $3\,141 \pm   63 $  
                 & $4\,750 \pm   79 $\phantom{0} \\
Combinatorial    & $14\,024 \pm  224 $  & $4\,924 \pm  132 $  
                 & $ 968 \pm   78 $   & $2\,095 \pm   96 $  
                 & $4\,188 \pm  127 $ \\
Partially rec.   & $4\,106 \pm  167 $   & $1\,344 \pm   96 $\phantom{0}  
                 & $ 321 \pm   64 $   & $ 691 \pm   75 $  
                 & $1\,204 \pm   96 $\phantom{0} \\
\hline 
Signal in box    & 10\,233\phantom{$0 \pm 00$}   & 2\,061\phantom{$0 \pm 00$}
                 &  1\,500\phantom{$0 \pm 0$}   & 2\,803\phantom{$0 \pm 0$}
                 &  4\,261\phantom{$0 \pm 00$} \\
Background in box& 1\,616\phantom{$0 \pm 00$}    & 598\phantom{$0 \pm 00$}
                 &     89\phantom{$0 \pm 0$}    & 192\phantom{$0 \pm 0$}
                 &    427\phantom{$0 \pm 00$}  \\

  \end{tabular}
\end{table}

\section{Efficiency variation over the Dalitz plot}

\label{sec:efficiency}

The same sample of simulated events as in the selection training (Sec.~\ref{sec:selection}) 
is used to determine the variation of the efficiency across the Dalitz plot. 
The sample is generated uniformly in the decay phase space and consists of approximately 
$8\times 10^4$ \lbdnppi events satisfying the selection requirements.
Each simulated event is assigned a weight, 
derived from control samples of data, to correct 
for known differences in track reconstruction and hardware trigger efficiency between data and simulation. 
Since the PID variables in the sample are replaced by those generated from calibration data, 
the efficiency of PID requirements is included in the efficiency calculation and 
does not need to be treated separately.

\begin{figure}
  \includegraphics[width=0.49\textwidth]{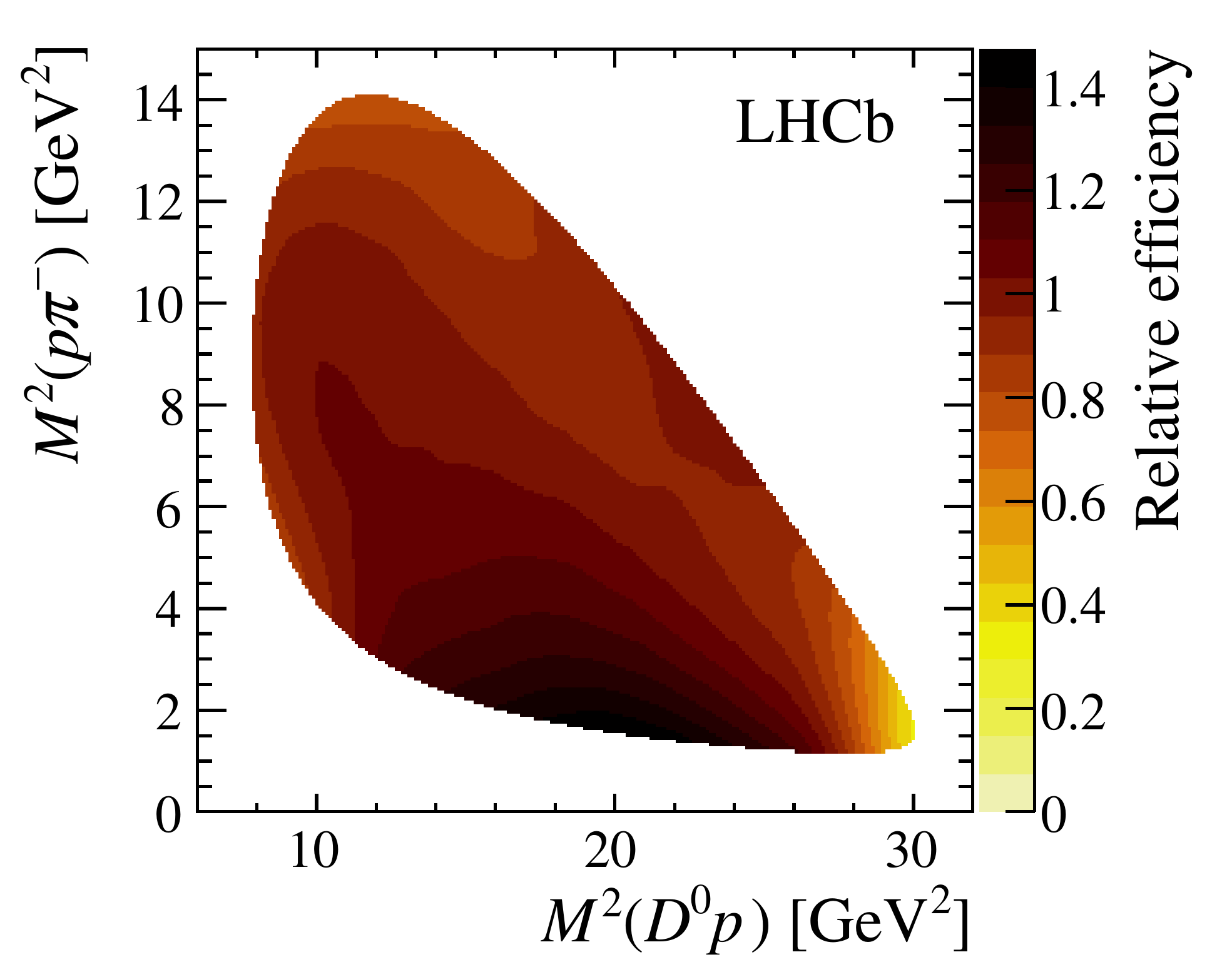}
  \put(-75,130){(a)}
  \hfill
  \includegraphics[width=0.49\textwidth]{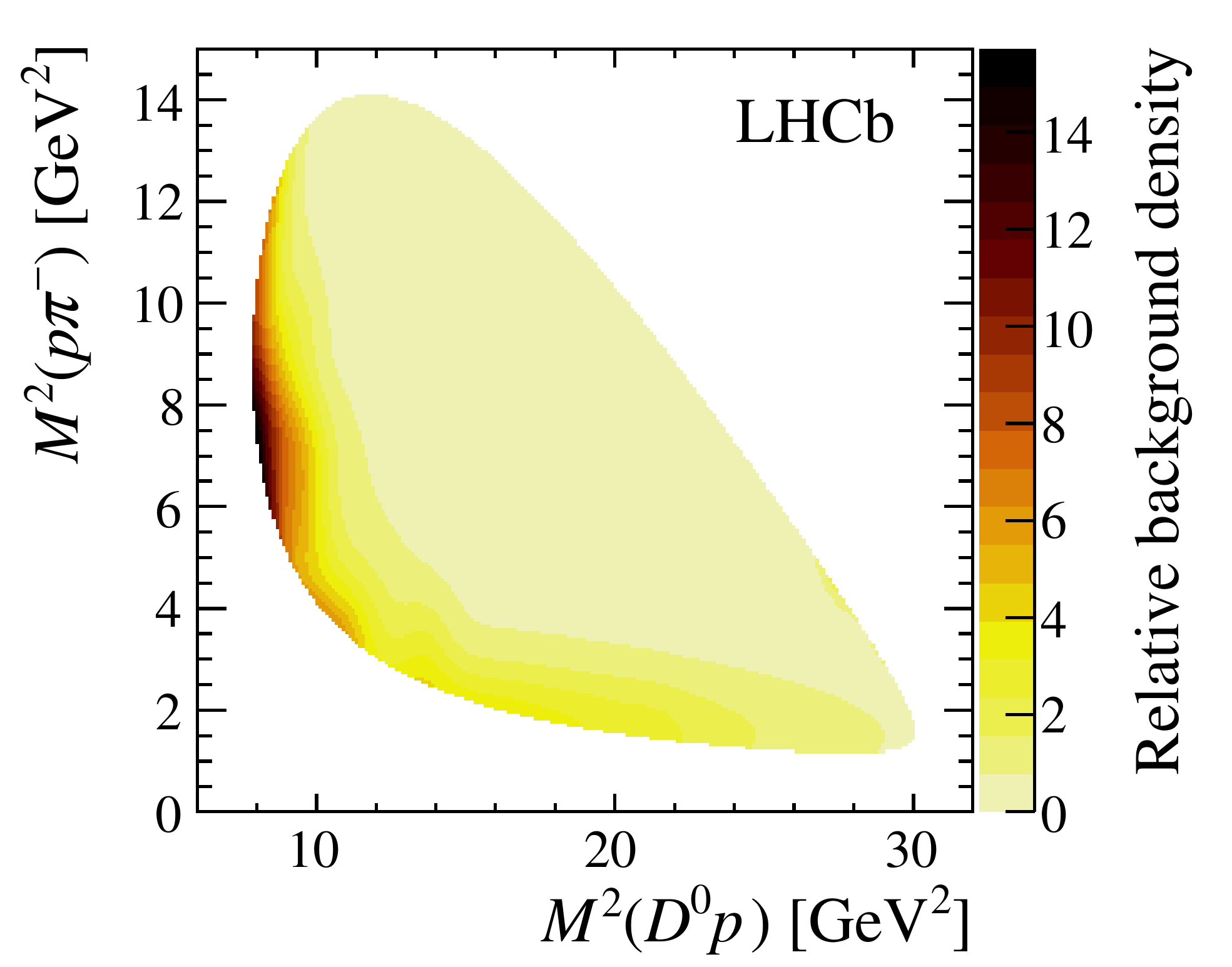}
  \put(-75,130){(b)}
  \caption{(a) Relative selection efficiency and (b) background density 
           over the \lbdnppi phase space. 
           The normalisations are such that the average over the phase space is unity. }
  \label{fig:eff_bck}
\end{figure}

The Dalitz plot efficiency profile is calculated separately for two disjoint sets of candidates, defined 
according to whether the hardware trigger was activated by one of the \Lb decay products or by other particles in the event. 
For each of those samples, a kernel-based density estimation procedure with a correction for boundary effects~\cite{meerkat} 
is used to obtain a description of the relative 
efficiency as a function of the Dalitz plot variables. The overall efficiency is then given by the average of the two 
profiles, weighted according to the ratio of yields of the two classes of events in data. 
The resulting profile is shown in Fig.~\ref{fig:eff_bck}(a). 
The normalisation of the efficiency profile used in the amplitude fit likelihood 
(Eqs.~(\ref{eq:likelihood}) and (\ref{eq:total_density})) does not affect the result. 
The efficiency profile shown in Fig.~\ref{fig:eff_bck}(a) 
is normalised such that the average efficiency over the phase space is equal to unity. 

\section{Background distribution}

\label{sec:background}

Background in the vicinity of the \lbdnppi invariant mass peak is dominated by random 
combinations of \Dz mesons, proton, and pion tracks. 
To determine the background shape as a function of Dalitz plot variables $M^2(\Dz\proton)$ and $M^2(\proton\pim)$, 
the $\Lb$ mass sidebands are used: $5500<M(\dnppi)<5560\mev$ and $5680<M(\dnppi)<5900\mev$.
The same procedure is applied to the opposite-flavour $\Dzb\proton\pim$ sample 
to verify that the background shape in the mass sidebands is representative of that in the signal window. 
Good agreement is found. 

The background distribution as a function of the Dalitz plot variables is estimated using  
a Gaussian mixture model, describing the background as a sum of several two-dimensional 
Gaussian distributions, whose parameters are allowed to vary in the fit. 
For the limited-size sample of background events this approach appears more suitable
than a kernel-based technique. The parametrisation is
obtained using an iterative procedure where Gaussian components are added to the model 
one by one; at each iteration the parameters of all components are adjusted using an unbinned 
maximum likelihood fit. 
The result of the procedure is shown in Fig.~\ref{fig:eff_bck}(b). 
The baseline parametrisation is a sum of 25 two-dimensional Gaussian components. 
The normalisation of the background density used in the fit is arbitrary; 
for the purposes of illustration in Fig.~\ref{fig:eff_bck}(b) it is set such 
that the average density across the phase space is unity.

\section{Effect of momentum resolution}

\label{sec:resolution}

Finite momentum resolution smears the structures in the Dalitz plot. The use of the  
kinematic fit with \Lb and \Dz mass constraints significantly improves the resolution near the edges of the phase space, but less so in the central region. 
The only structure in the \lbdnppi amplitude that is expected to be affected by the finite resolution is the 
resonance $\Lcst$, which has a natural width of approximately 6\mev. Therefore, only the $M(\Dz\proton)$ resolution 
is considered, and is obtained from a sample of simulated events
by comparing the generated and reconstructed values of $M(\Dz\proton)$. 
The width of the resolution function at $M(\Dz\proton)=2.88\gev$ is $1.1\mev$, \ie significantly 
smaller than the natural width of the $\Lcst$. 
However, simulation shows that neglecting the resolution would lead to a bias on the $\Lcst$ width of about 10\%. 
Therefore, the $M(\Dz\proton)$ resolution is taken into account in the fit by convolving the signal PDF  
with a Gaussian resolution function, where the width of the Gaussian is a function of $M(\Dz\proton)$.

\section{Amplitude analysis}

\label{sec:amplitude}

The amplitude fit is performed in the four phase space regions defined in Fig.~\ref{fig:phase_space}. 
This approach has been chosen instead of performing the fit to the entire Dalitz plot since 
the amplitude contains many unexplored contributions. The full fit would include too many 
degrees of freedom and a very large range of systematic variations would need to be considered. 
Instead, the fit is first performed around the well-known 
resonance \Lcst and then the fitting region is gradually extended 
to include a larger portion of the phase space. 

\subsection{Fit in the nonresonant region}

\label{sec:lc2880}

The fit in region~1, where no significant resonant contributions are expected, 
provides constraints on the high-mass 
behaviour of the $\proton\pim$ amplitude, and thus on the
$\proton\pim$ partial waves in the $\Dz\proton$ fit regions. 
The fit model includes four exponential nonresonant components (Eq.~(\ref{eq:nr_exp})) in each of the $\Dz\proton$
and $\proton\pim$ spectra, corresponding to the four combinations of spin ($1/2$ and $3/2$) and parity
(negative and positive). 
Since there is no reference amplitude with known parity in this region, 
there is an ambiguity: all parities can be reversed simultaneously without changing the amplitude. 
The shape parameters $\alpha$ of all eight nonresonant components are varied in the fit. 

\begin{figure}
\centering
\includegraphics[width=0.35\textwidth]{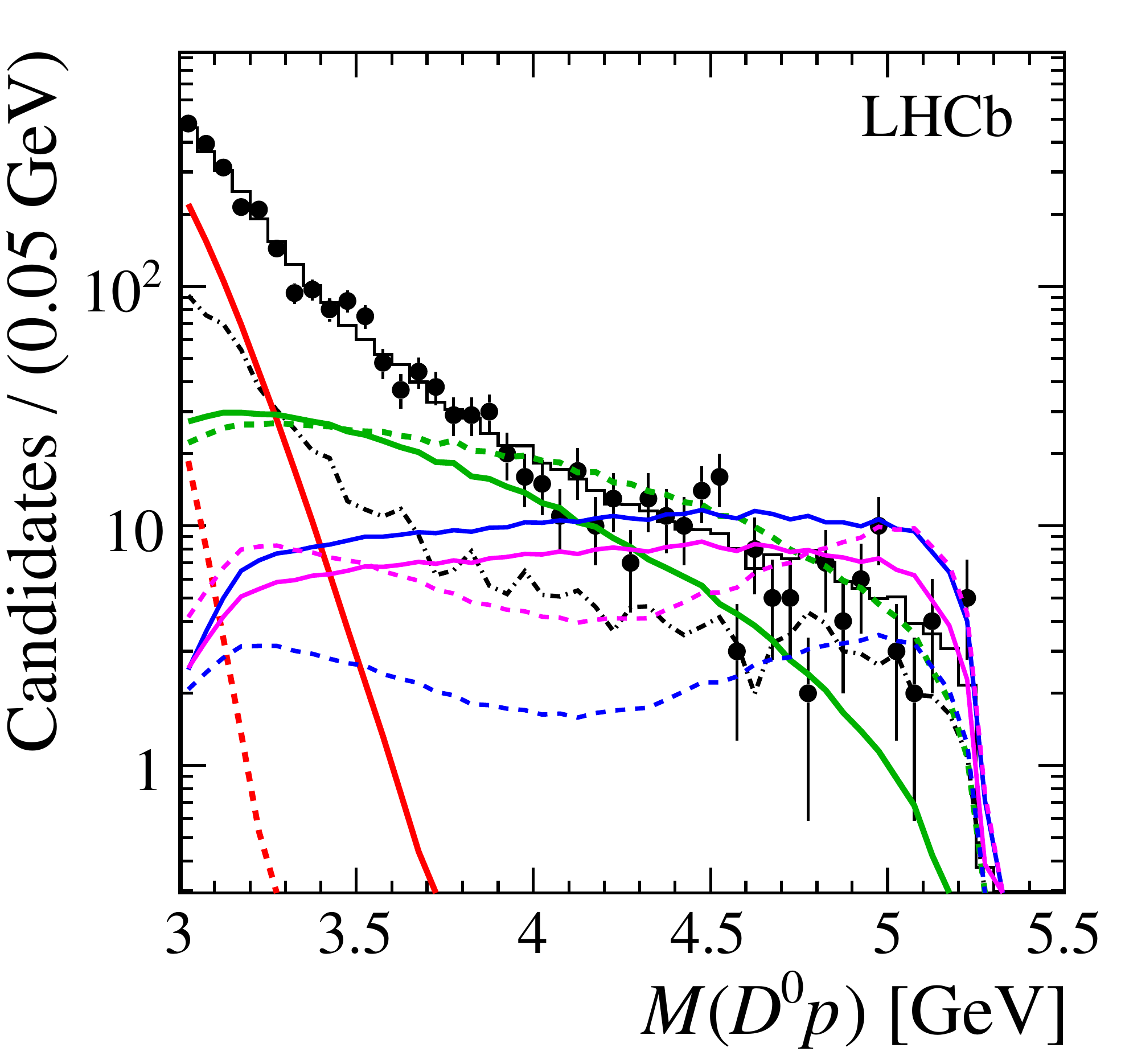} 
  \put(-32,117){(a)}
\hspace{0.01\textwidth}
\includegraphics[width=0.35\textwidth]{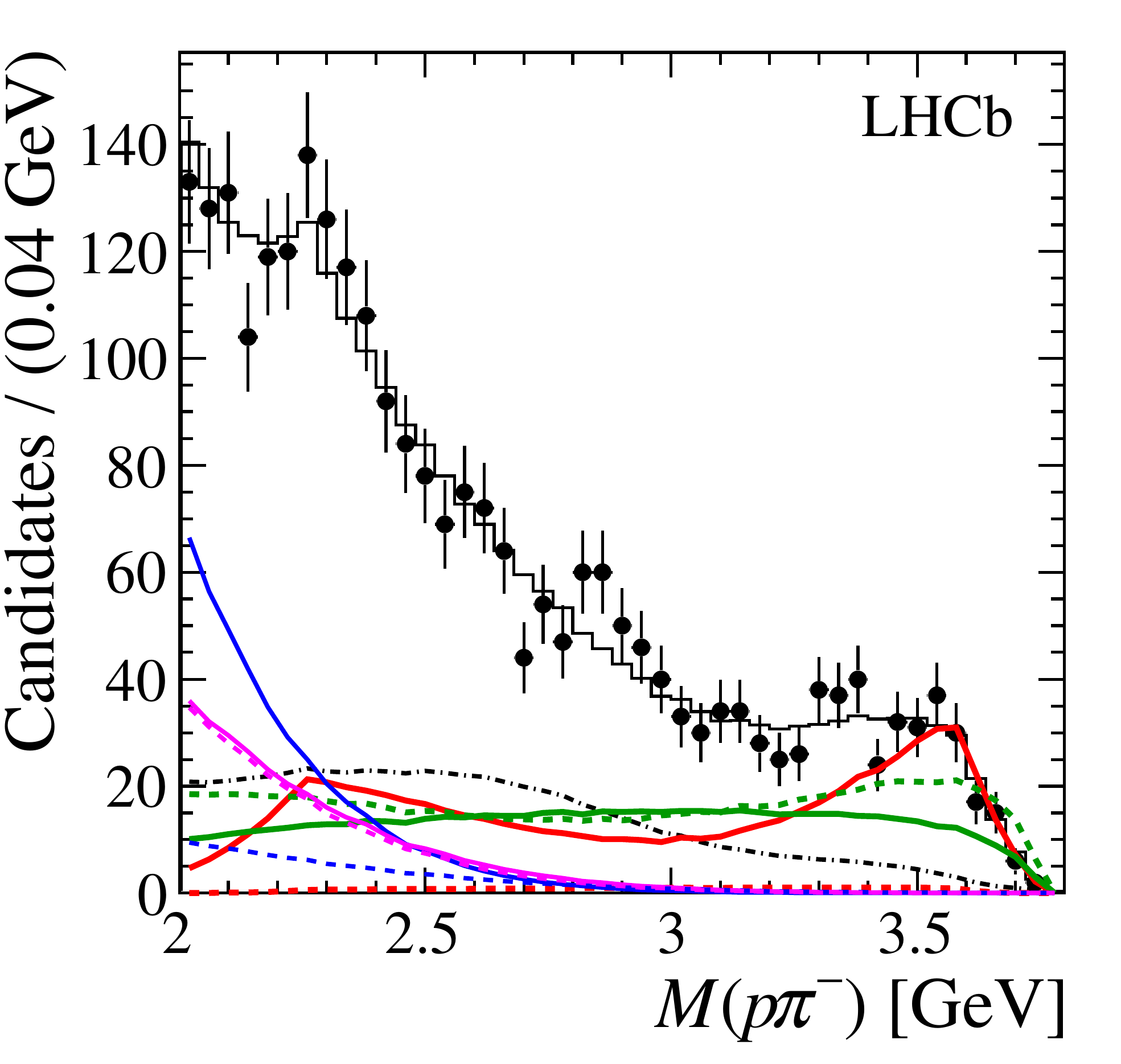} 
  \put(-32,117){(b)}
\hspace{0.01\textwidth}
\raisebox{0.03\textwidth}{\includegraphics[width=0.20\textwidth]{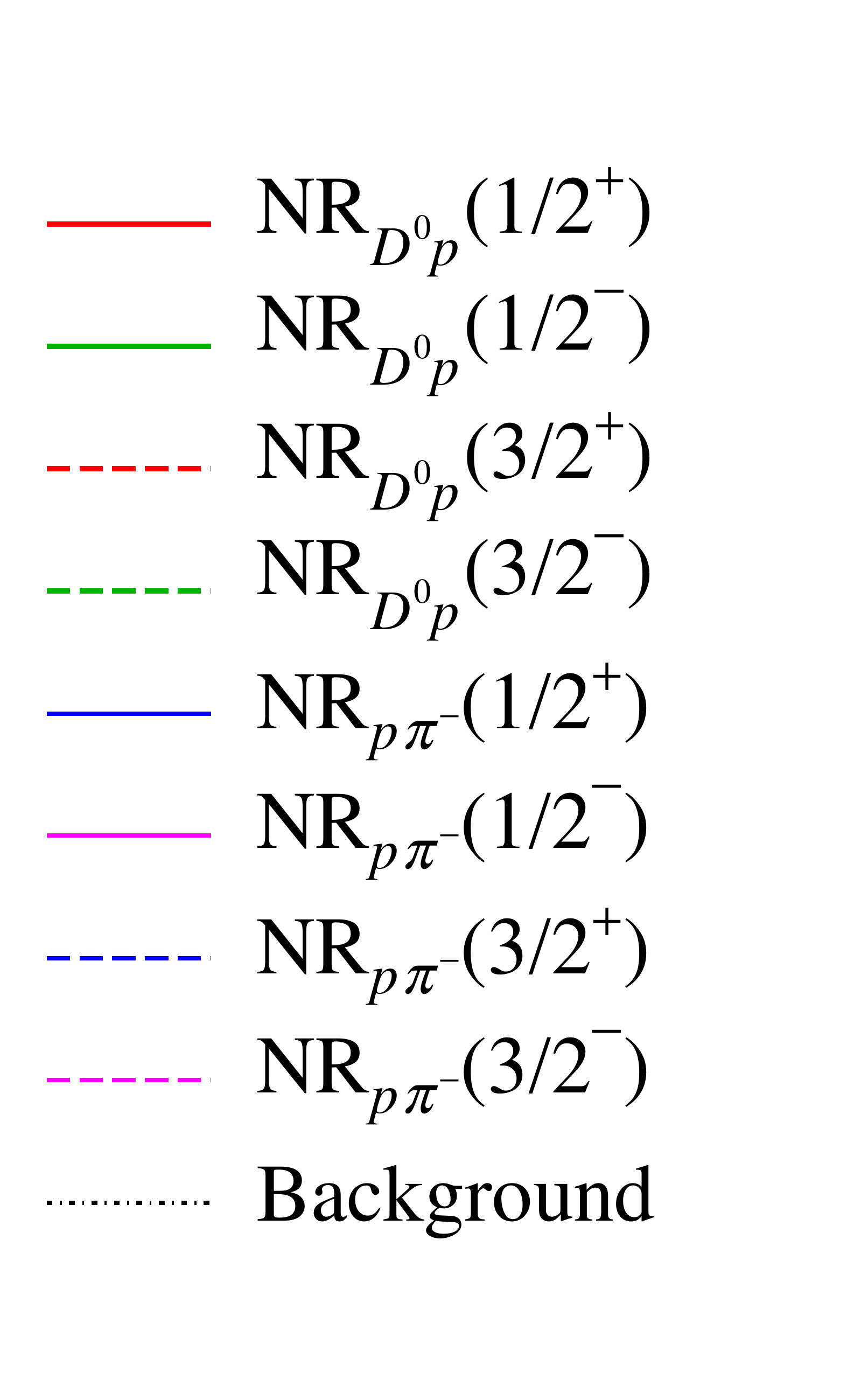}}
\caption{Fit results for the \lbdnppi amplitude in the nonresonant region (region 1)
(a) $M(\Dz\proton)$ projection and (b) $M(\proton\pim)$ projection.  
The points with error bars are data, the black histogram is the fit result, and coloured curves show 
the components of the fit model taking into account the efficiency.  
The dash-dotted line represents the background. 
Due to interference effects the total is not necessarily equal to the sum of the components. 
}
\label{fig:nonres_ampl_fit}
\end{figure}

The projections of the fitted data are shown in Fig.~\ref{fig:nonres_ampl_fit}. The fitted 
$\proton\pim$ amplitude is extrapolated into the regions 2--4 of the $\lbdnppi$ phase space
using the fitted helicity distributions. 
The estimated contributions of the $\proton\pim$ nonresonant components in the $\Dz\proton$
mass regions are given in Table~\ref{tab:crossfeeds} and compared with the total 
numbers of signal events in those regions. They amount to less than $1\%$ of the signal yield for 
the regions 2 and 3, and around 1.5\% for region 4. Therefore, the baseline fit models for regions 
2 and 3 do not include $\proton\pim$ crossfeed (although it is taken into account as a part of 
the uncertainty due to modelling of nonresonant amplitudes), 
while for region 4 the $\proton\pim$ nonresonant component is included in the model. Since only a 
small part of the $\proton\pim$ helicity distribution enters the $\Dz\proton$ fit region, the spin and parity assignment 
of the $\proton\pim$ amplitude should have a very small effect. Thus only one partial 
wave ($J^P=1/2^-$) of the nonresonant $\proton\pim$ component is included for the $\Dz\proton$ amplitude fit. 

\begin{table}
  \caption{Estimated contributions from the $\proton\pim$ nonresonant components in different phase space regions. 
           The signal yields from Table~\ref{tab:mass_fit} are also included for comparison. }
  \label{tab:crossfeeds}
  \begin{center}
  \begin{tabular}{c|cc}
  Region & Signal yield & ~$\proton\pim$ yield \\
  \hline
2 & $1\;500$ &   \phantom{0}9 \\
3 & $2\;803$ &   16 \\
4 & $4\;261$ &   61 \\
  \end{tabular}
  \end{center}
\end{table}

\subsection{\boldmath Fit in the region of $\Lcst$}

\label{sec:lc2880_amplitude}

Next, an amplitude fit is performed in region 2, in the vicinity of the well-established 
$\Lcst$ resonance. 
The quantum numbers of this state have been measured by the Belle collaboration to be $J^P=5/2^+$~\cite{Abe:2006rz, PDG2016}. 
The fit probes the structure of the wide $\Dz\proton$ amplitude component underneath 
the $\Lcst$ peak using the shape of the latter as a reference. 
Other $\Lcst$ spin assignments from $1/2$ to $7/2$ are also tried (spin $7/2$ was not tested in the Belle 
analysis~\cite{Abe:2006rz}).
Since the amplitude is not sensitive 
to the absolute parities of the components, the parity of the \Lcst is always fixed to be 
positive; the parities of the other amplitude components are determined relative to its parity. 

As for region 1, the nonresonant amplitude 
model consists of four contributions with spins $1/2$ and $3/2$ and both parities. The nonresonant 
components are parametrised either with the exponential model of Eq.~\ref{eq:nr_exp} (``Exponential''), 
or the amplitude with both real and imaginary parts varying linearly in $M^2(\Dz\proton)$ (``Linear'', 
which is a special case of the spline-interpolated shape with only two knots). The mass and width 
of the $\Lcst$ state are free parameters. 

\begin{figure}
\centering
\includegraphics[width=0.33\textwidth]{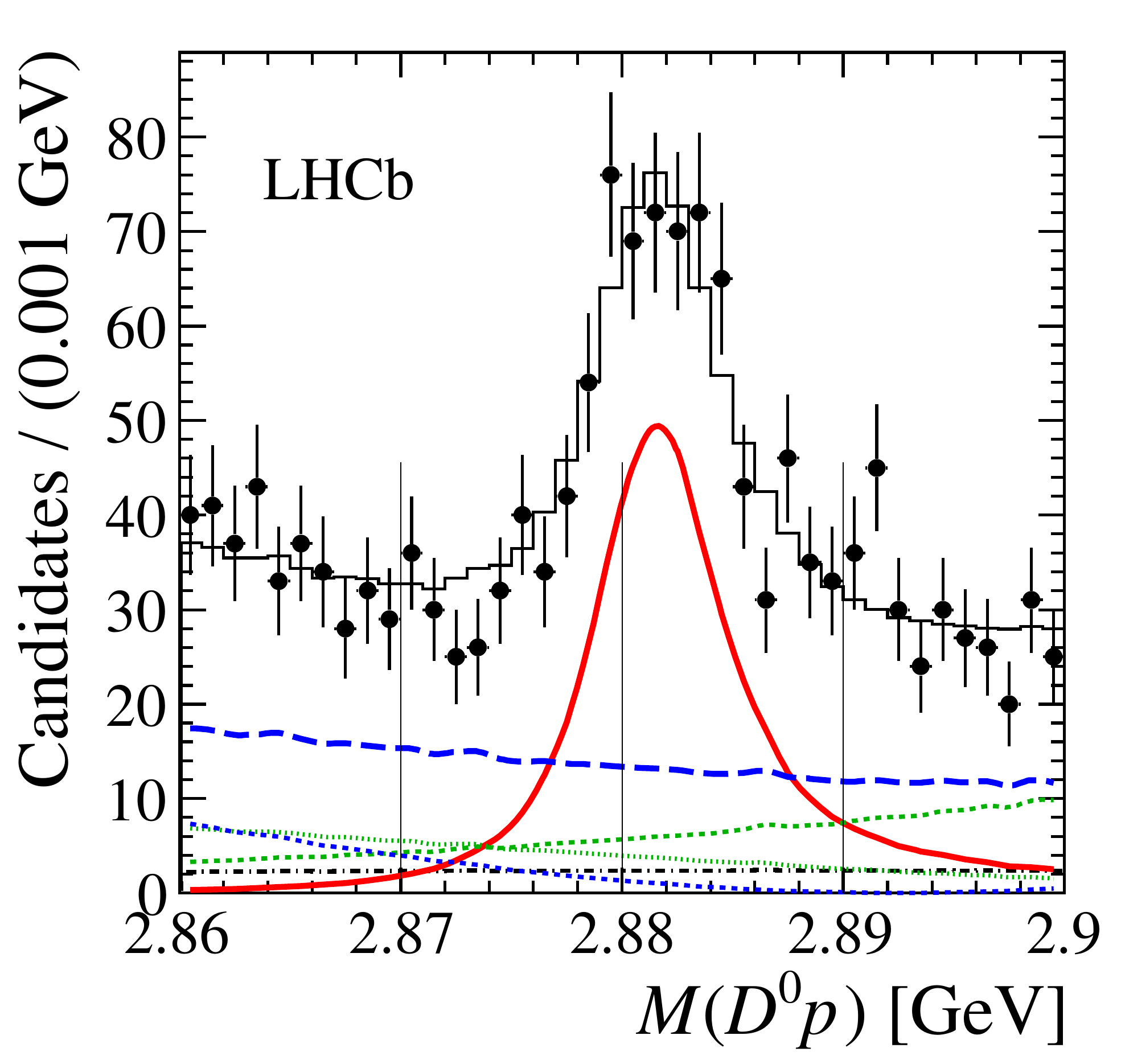} 
  \put(-112,103){(a)}
\includegraphics[width=0.33\textwidth]{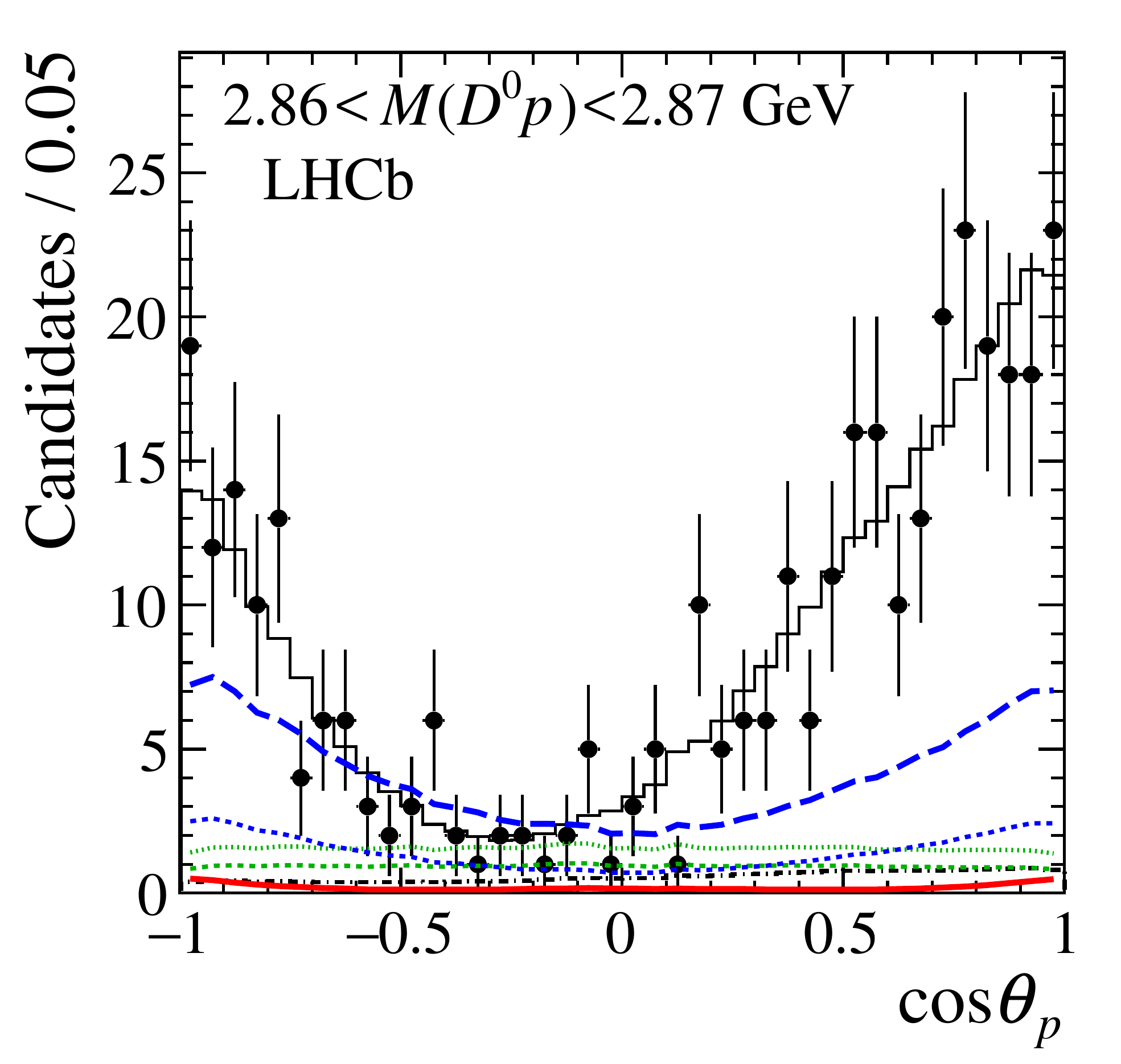} 
  \put(-112,103){(b)}
\includegraphics[width=0.33\textwidth]{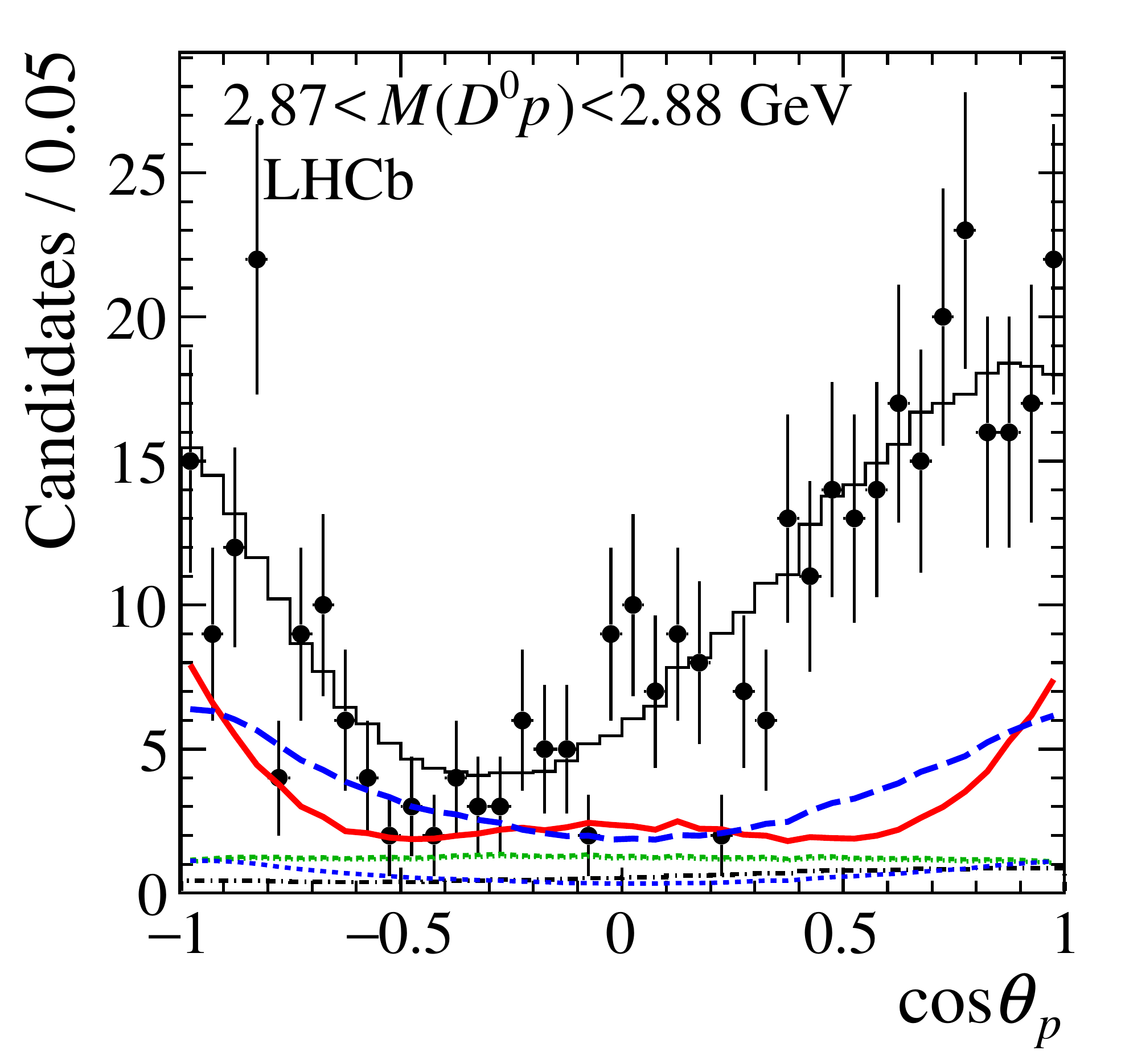} 
  \put(-112,103){(c)}

\includegraphics[width=0.33\textwidth]{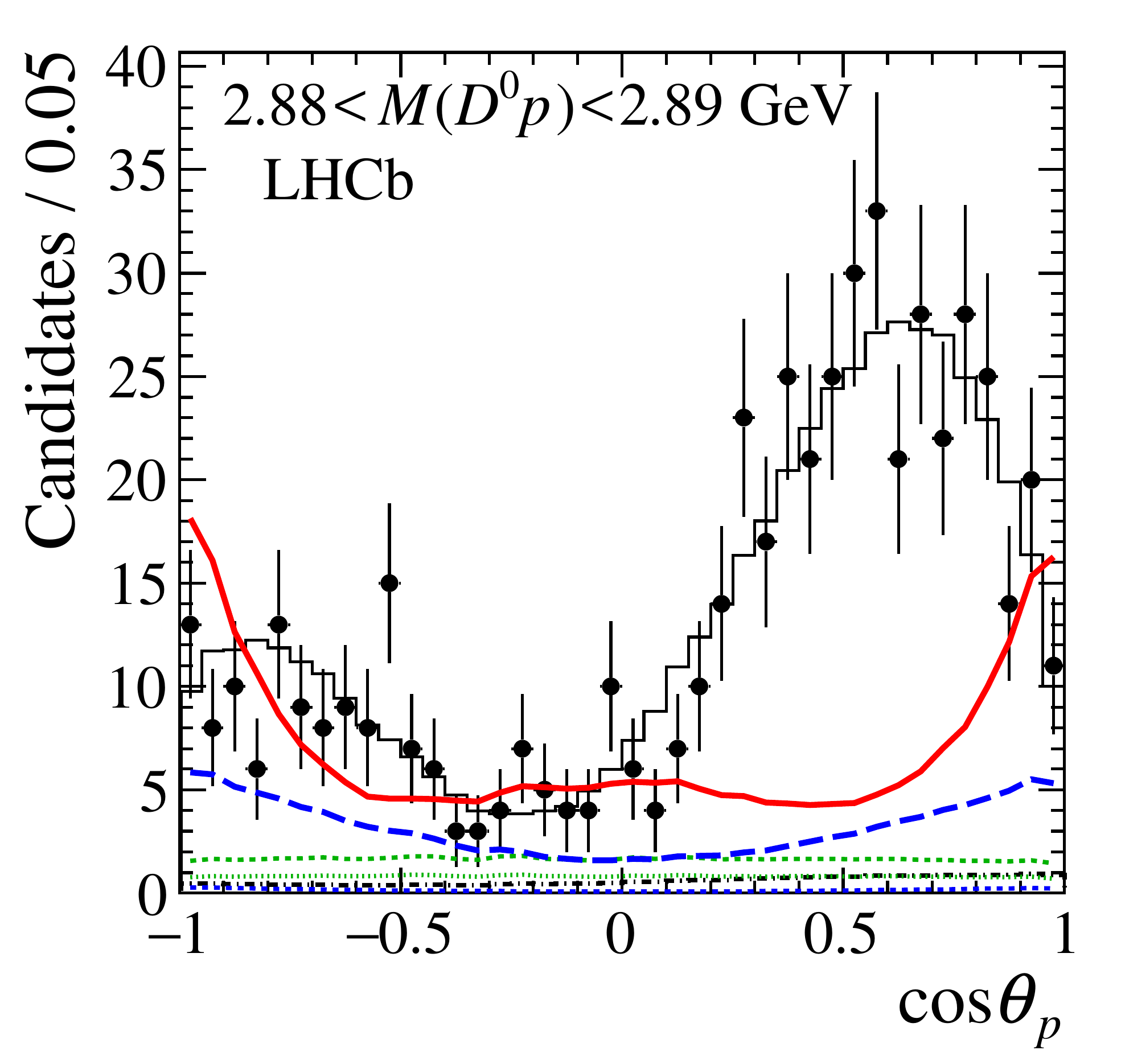} 
  \put(-112,103){(d)}
\includegraphics[width=0.33\textwidth]{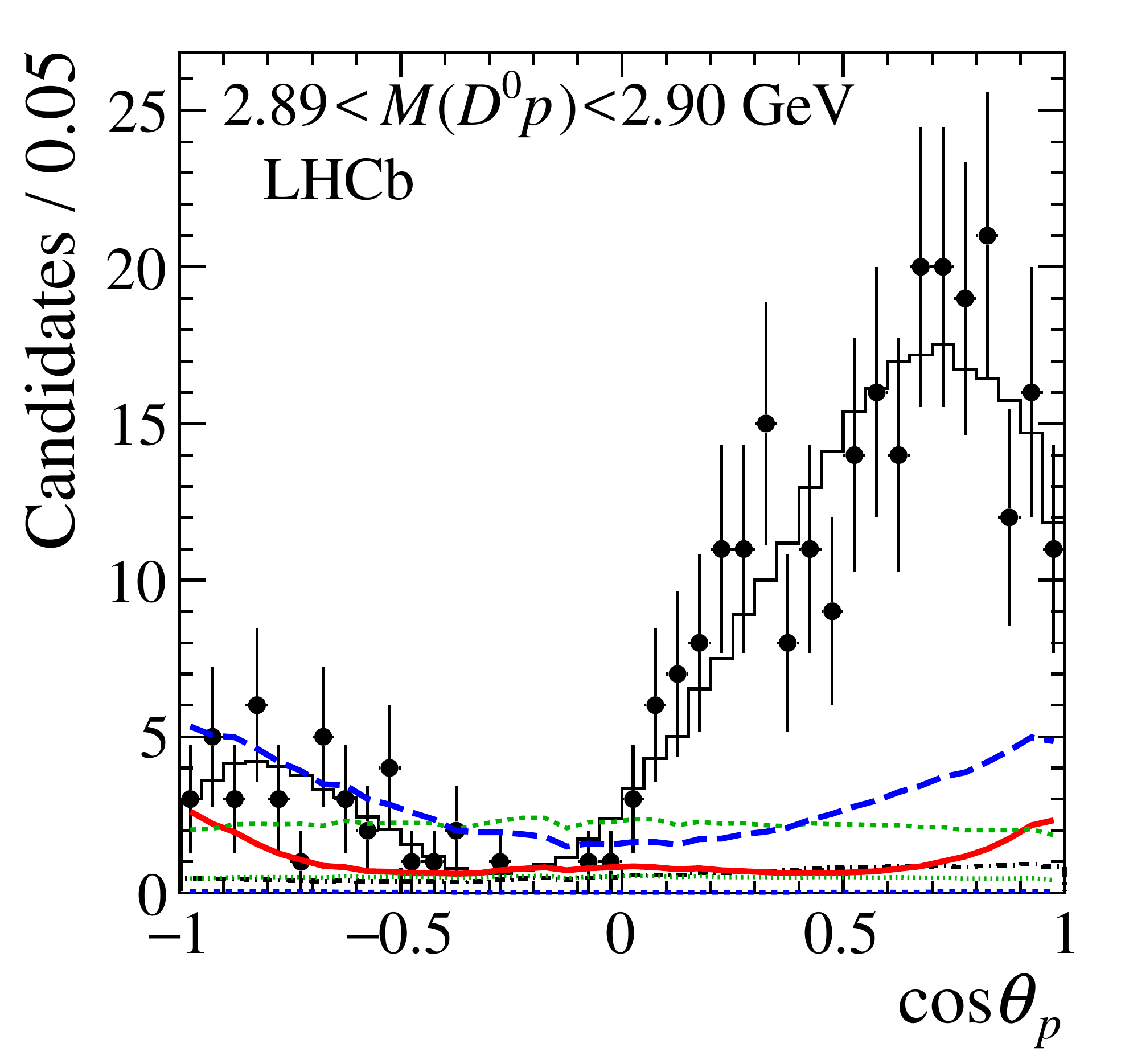} 
  \put(-112,103){(e)}
\includegraphics[width=0.33\textwidth]{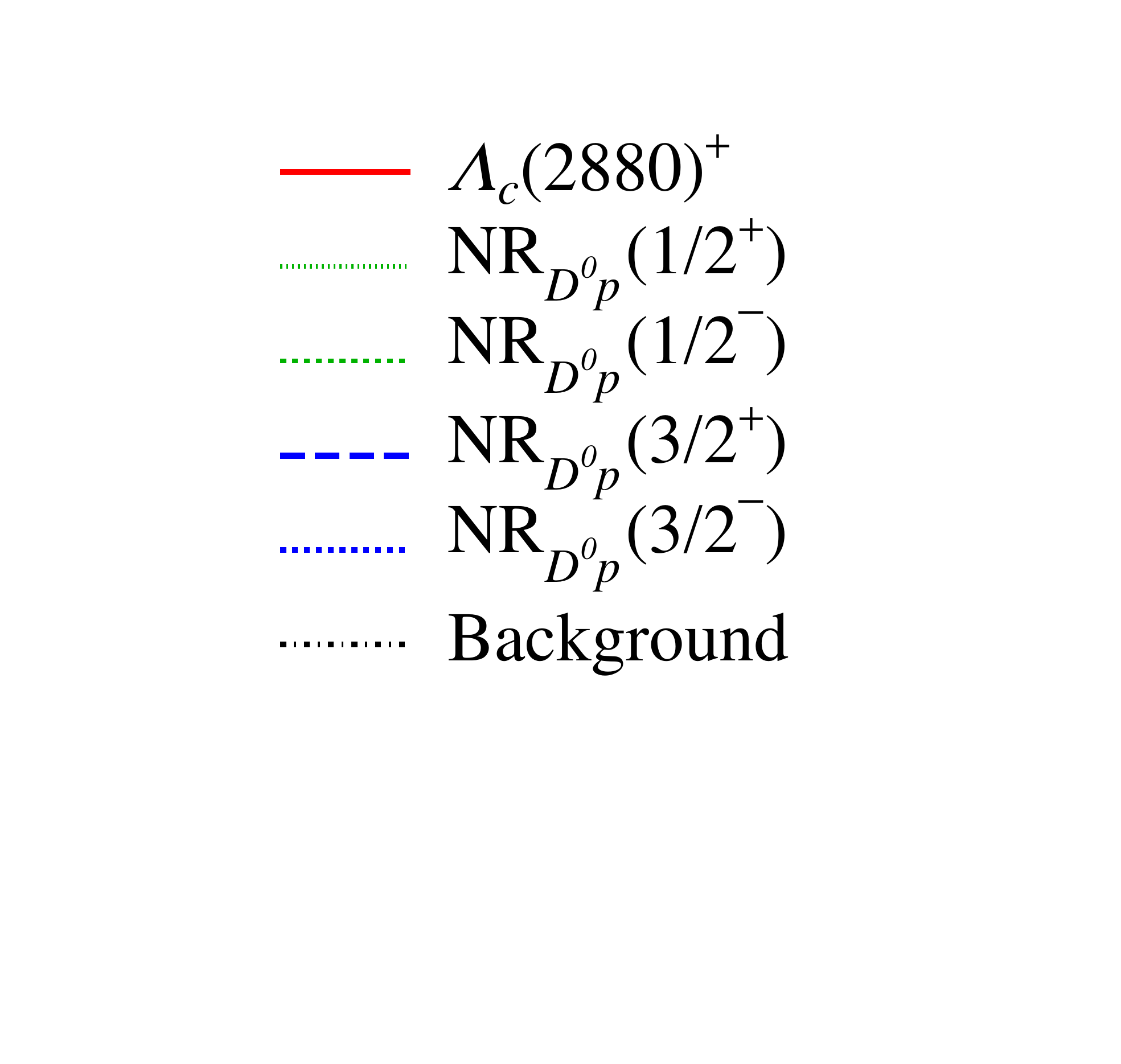} 

\caption{Results of the \lbdnppi amplitude fit in the $\Lcst$ mass region with spin-parity assignment 
$J^P=5/2^+$ for the \Lcst resonance: 
(a) $M(\Dz\proton)$ projection and (b--e) $\cos\theta_{\proton}$ projections in 
slices of the $\Dz\proton$ invariant mass. The linear nonresonant model is used. 
Points with error bars are data, the black histogram is the fit result, coloured curves show 
the components of the fit model. The dash-dotted line represents the background. 
Vertical lines in (a) indicate the boundaries of the $\Dz\proton$ invariant mass slices. 
Due to interference effects the total is not necessarily equal to the sum of the components. 
}
\label{fig:lc2880_ampl_fit}
\end{figure}

The model in which the \Lcst has spin $5/2$ is preferred for both 
nonresonant models, while the difference between exponential and linear models is negligible. 
The model with spin $5/2$ and linear nonresonant amplitude parametrisation is taken as the baseline. 
Table~\ref{tab:lc2880_lh} gives the differences in $\ln\mathcal{L}$ compared to the baseline, along with 
the $\chi^2$ values and the associated probabilities. 
The quality of the fit is obtained using the adaptive binning approach with at least 20 data entries in each bin
and with the effective number of degrees of freedom ${\rm ndf}_{\rm eff}$ obtained from pseudoexperiments. 
The results of the fit with the baseline model are shown in Fig.~\ref{fig:lc2880_ampl_fit}. 

\begin{table}[b!]
  \caption{Values of the $\Delta\ln\mathcal{L}$ and fit quality for various \Lcst spin assignments 
           and nonresonant amplitude models. The baseline model is shown in bold face. }
  \label{tab:lc2880_lh}
  \centering
  \begin{tabular}{l|c|rcc}
  Nonresonant model & $\Lcst$ $J^P$ & $\Delta\ln\mathcal{L}$ & $\chi^2/$ndf & $P(\chi^2, \mbox{ndf})$, \% \\
  \hline
  Exponential  & $1/2^+$ & $41.5$ & 108.9/70 &  \phantom{0}0.2 \\ 
     & $3/2^+$ & $35.5$ &  \phantom{0}99.4/70 & \phantom{0}1.2 \\ 
     & $5/2^+$ & $-0.2$ &  \phantom{0}65.6/70 & 62.7 \\ 
     & $7/2^+$ &  $8.4$ &  \phantom{0}76.8/70 & 27.0 \\ 
\hline 
Linear & $1/2^+$ & $40.3$ & 107.4/71 &  \phantom{0}0.3 \\ 
       & $3/2^+$ & $35.7$ &  \phantom{0}98.8/71 &  \phantom{0}1.6 \\ 
       & \boldmath{$5/2^+$} &  {\bf 0.0} &  {\bf \phantom{0}69.2/71} & {\bf 53.8} \\ 
       & $7/2^+$ &  $8.6$ &  \phantom{0}76.2/71 & 31.5 \\ 

  \end{tabular}
\end{table}

\begin{figure}
\begin{center}
  \includegraphics[width=0.35\textwidth]{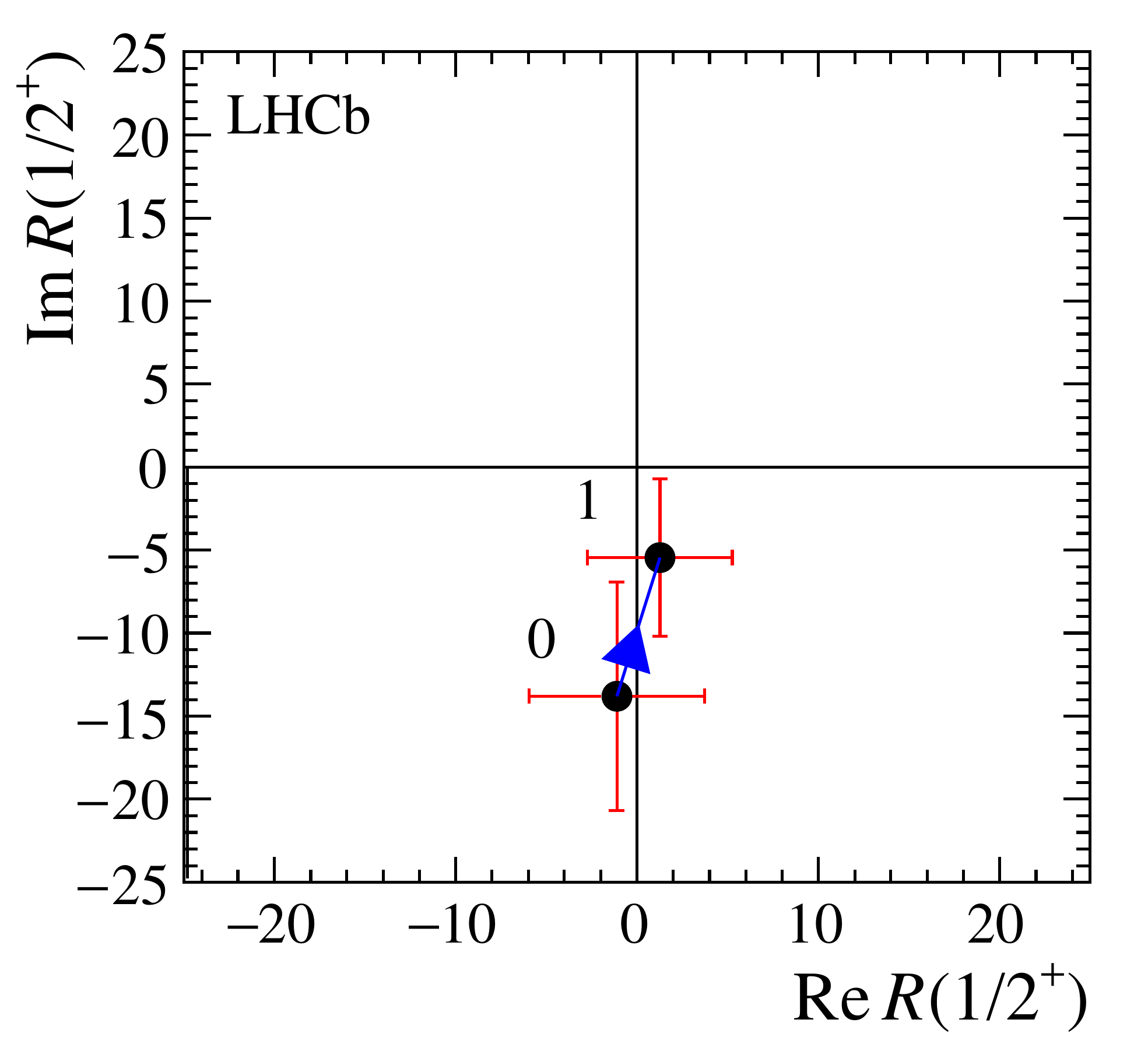}
  \includegraphics[width=0.35\textwidth]{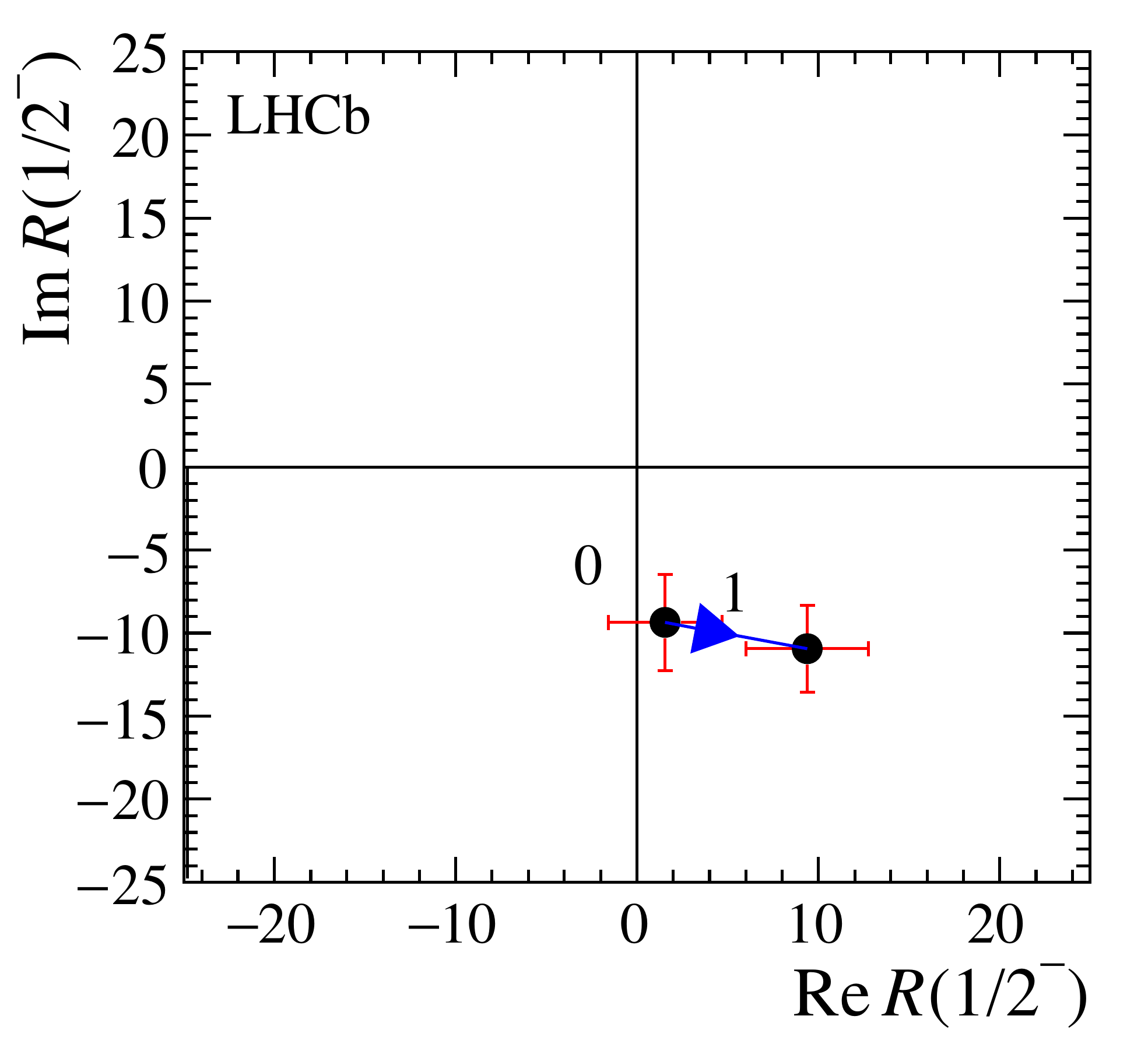}

  \includegraphics[width=0.35\textwidth]{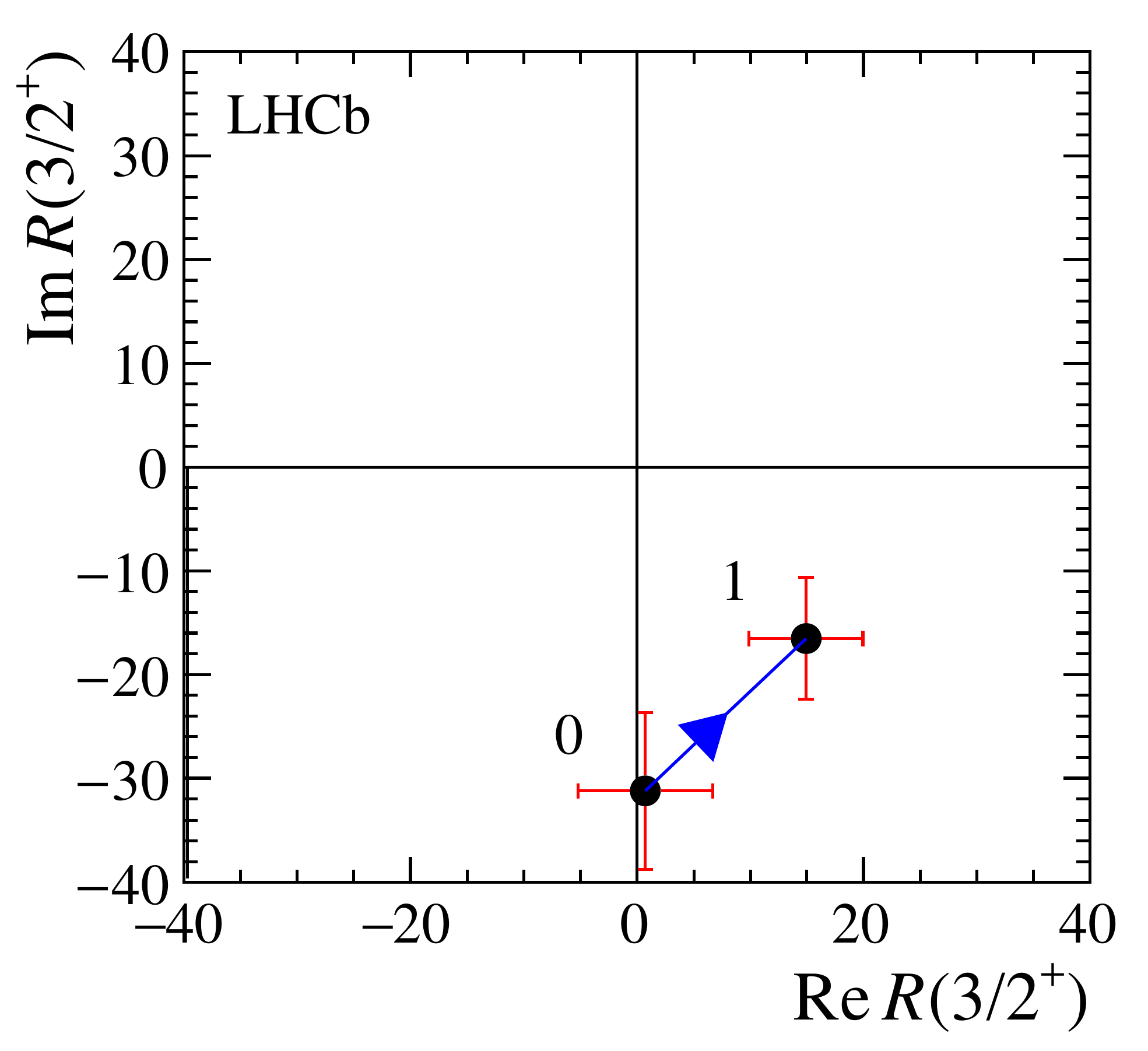}
  \includegraphics[width=0.35\textwidth]{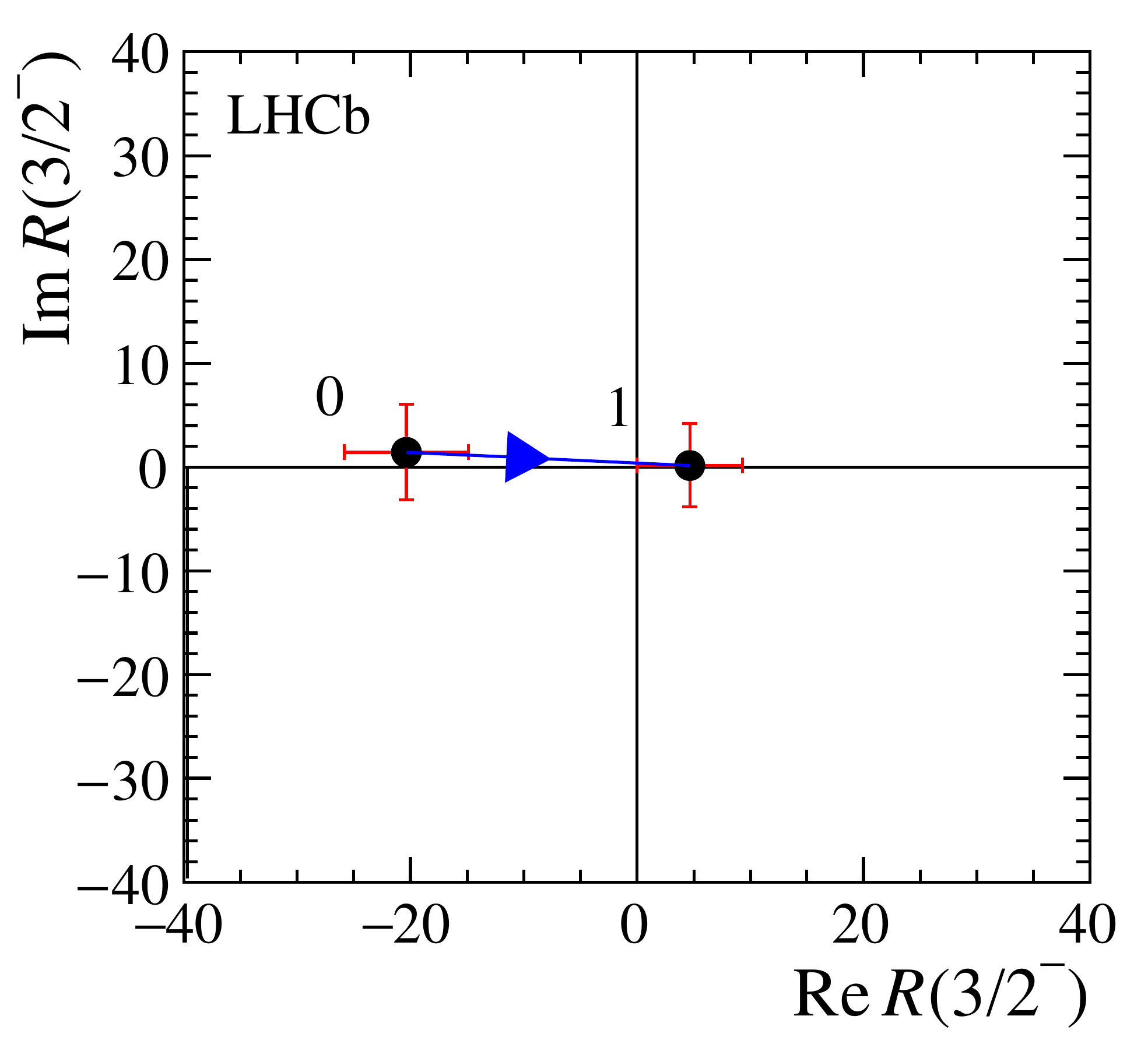}
\end{center}
\caption{Argand diagrams for the four amplitude components underneath the $\Lcst$ peak in the linear 
nonresonant model. In each diagram, point 0 corresponds to $M(\Dz\proton)=2.86$\gev, and 
point 1 to $M(\Dz\proton)=2.90$\gev. }
\label{fig:lc2880_argand}
\end{figure}

Argand diagrams illustrating the amplitude and phase motion of the fit components are shown in Fig.~\ref{fig:lc2880_argand}. 
The plots contain a hint of phase rotation for the $J^P=3/2^+$ partial wave in a counter-clockwise direction, 
consistent with the resonance-like phase motion observed in the near-threshold fit (Sec.~\ref{sec:lowerdp_amplitude}). 
The statistical significance of this effect is studied with 
a series of pseudoexperiments where the samples are generated according to the fit where the complex 
phase in all the nonresonant components is constant. Each is fitted with two models, with the complex 
phase constrained to be the same for both endpoints, and floated freely. The distribution of the 
logarithmic likelihood difference $\Delta\ln\mathcal{L}$ between the two fits is studied and compared to the value obtained in data. 
The study shows that around $55\%$ of the samples have $\Delta\ln\mathcal{L}$ greater than the 
value observed in data (1.4), \ie this effect is not statistically significant with the data in region 2 alone. 

Ensembles of pseudoexperiments, where the baseline model is used both to generate and to fit
samples of the same size as in the data, are used to validate the statistical uncertainties 
obtained from the fit, check for systematic biases due to the fitting procedure, evaluate the 
statistical uncertainties on the fit fractions, and obtain the effective number of degrees of freedom 
for the fit quality evaluation based on a binned $\chi^2$ measure. 

The unbinned maximum likelihood fit is unbiased only in the limit of a large data sample; in general a fit to a  
finite sample can exhibit a bias that is usually significantly smaller than the statistical uncertainty. 
Pseudoexperiments are used to evaluate and correct for such biases on the mass and the width of the \Lcst state, 
as well as on the fit fractions of the amplitude components obtained from the fit. The corrected values are
\begin{equation*}
\begin{split}
m(\Lcst)&=\lcstmassstat\mev, \\ 
\Gamma(\Lcst)&=\lcstwidthstat\mev, \\
\mathcal{F}(\Lcst)&=(\lcstfracststat)\%, \\
\mathcal{F}(1/2^+)&=(\nronepfracststat)\%, \\
\mathcal{F}(1/2^-)&=(\nronemfracststat)\%, \\
\mathcal{F}(3/2^+)&=(\nrthreepfracststat)\%, \\
\mathcal{F}(3/2^-)&=(\nrthreemfracststat)\%. \\
\end{split}
\end{equation*}
The uncertainties are statistical only. 
Correlations between the fit parameters do not exceed 20\%.
Since all the amplitude components have different quantum numbers, the 
interference terms cancel out after integrating over the phase space, and the
sum of uncorrected fit fractions is exactly 100\%.  
After the bias correction is applied individually for each fit fraction, 
statistical fluctuations in the corrections lead to a small, statistically not significant, difference from 100\%
(in this case, the sum of fit fractions increases to 102.6\%).

A number of experimental systematic uncertainties on the \Lcst mass and width and on the 
difference $\Delta\ln\mathcal{L}$ between the baseline ($5/2$) and the next-best ($7/2$) spin assignments 
are considered and are given in Table~\ref{tab:lc2880_syst}. These arise from: 
\begin{enumerate}

  \item Uncertainty on the background fraction in the signal region (Sec.~\ref{sec:yields}). 
        The statistical uncertainty is obtained from the fit to the $M(\dnppi)$ distribution, 
        and a systematic uncertainty arising from the modelling of the signal and 
        background $M(\dnppi)$ distributions is estimated 
        by performing fits with modified $M(\dnppi)$ models. 
        The sum in quadrature of these contributions is taken as the systematic uncertainty.

  \item Uncertainty on the efficiency profile (Sec.~\ref{sec:efficiency}). 
        The statistical uncertainty is evaluated via a 
        bootstrapping procedure~\cite{efron1979}. 
        The uncertainty related to the kernel density estimation procedure is obtained by varying the kernel size. 
        The uncertainty due to differences between data and simulation in 
        the input variables of the BDT is estimated by varying the scaling 
        factors for these variables. 
        In addition, the replacement of simulated proton and pion PID variables with 
        values drawn from control samples in the data with matching kinematics, 
        described in Section 4, introduces further systematic uncertainties. 
        The uncertainty associated with the limited size of these control samples 
        is evaluated again with a bootstrapping procedure, and the uncertainty 
        associated with the kinematic matching process is assessed by changing the kernel 
        size in the nonparametric algorithm used to estimate the PID response as a 
        function of the kinematic properties of the track.

  \item Uncertainty on the background shape (Sec.~\ref{sec:background}). This is assessed by 
        varying the density estimation procedure (changing the number of Gaussian cores 
        in the mixture model, or 
        using kernel density estimation instead of a Gaussian mixture model), and by 
        using only a narrower upper sideband of the $M(\dnppi)$ distribution, $5680<M(\dnppi)<5780\mev$. 
        The statistical uncertainty due to the finite size of the background sample 
        is estimated by bootstrapping. 

  \item Uncertainty on the momentum resolution (Sec.~\ref{sec:resolution}). This is 
        estimated by varying the $M^2(\Dz\proton)$ resolution by $15\%$. 
        It mainly affects the width of the \Lcst resonance. 

  \item Uncertainties on the mass scale. Due to the constraints on the hadron masses, the momentum scale 
        uncertainty of the detector has a negligible effect on the fit. However, the uncertainties 
        on the assigned mass 
        values themselves do contribute. For $M(\Dz\proton)$ amplitudes the dominant 
        contribution comes from the $\Dz$ mass uncertainty. 

  \item Uncertainty on the fit procedure itself. This is assessed by fitting ensembles of pseudoexperiments, 
        where the baseline amplitude model is used for both generation and fitting, and the number of 
        events generated for each pseudoexperiment is equal to the number of events in the data sample.
        The mean value for each fitted parameter is used as a correction for fitting bias, while the 
        statistical uncertainty on the mean is taken as the uncertainty due to the fit procedure.

\end{enumerate}
The uncertainties on the $\Dz$ mass and the fit procedure do not affect the significance of the quantum 
number assignment and are thus not included in $\Delta\ln\mathcal{L}$ uncertainty.

Also reported in Table~\ref{tab:lc2880_syst} is the uncertainty related to the amplitude model. 
It consists of two contributions, corresponding to the 
uncertainties in the modelling of the resonant \Lcst shape and the nonresonant amplitudes. 
The model uncertainties are asymmetric, and the positive and negative uncertainties for the two components 
are combined in quadrature separately to obtain the total model uncertainty. 

The uncertainty due to the Breit--Wigner parametrisation of the \Lcst amplitude is estimated by 
        varying the radial parameters $r_{\Lb}$ and $r_{\Lcst}$ between $0$ and $10\gev^{-1}$ and $0$ and $3\gev^{-1}$, 
        respectively, and by removing the angular barrier factor from the Breit--Wigner amplitude. 
        The maximum deviation is taken as the uncertainty.

The uncertainty due to the modelling of the nonresonant amplitudes is estimated
        by taking the difference between the fit results obtained with the default linear nonresonant model
        and the alternative exponential model. The possible crossfeed from the $\proton\pim$ channel 
        is estimated by adding a 
        $J^P=1/2^-$ component in the $\proton\pim$ channel to the amplitude. This component has a fixed 
        exponential lineshape with shape parameter $\alpha=0.5\gev^{-2}$ (obtained in the fit to region 1 data)
        and its complex couplings are free parameters in the fit. 

The helicity formalism used to describe the amplitudes is inherently non-relativistic. 
        To assess the model uncertainty due to this limitation, an alternative description is 
        obtained with covariant tensors using the {\tt qft++} framework~\cite{Williams:2008wu}, but it is much 
        more expensive from a computational point of view and is therefore not used for the baseline 
        fits. Differences between the helicity and the covariant formalism are mainly associated with the 
        broad amplitude components and are therefore treated as a part of the uncertainty due to the nonresonant model. 
        Although this contribution is included in the nonresonant model uncertainty in Table~\ref{tab:lc2880_syst}, 
        it is also reported separately. 
        
\begin{table}
  \caption{Systematic and model uncertainties on the \Lcst parameters and on the value of $\Delta\ln\mathcal{L}$ between the 
           $5/2$ and $7/2$ spin assignments. The uncertainty due to the nonresonant model 
           includes a component associated with the helicity formalism, 
           which for comparison is given explicitly in the table, too.}
  \label{tab:lc2880_syst}
  \centering
  \begin{tabular}{l|ccc}
         & \multicolumn{3}{c}{Uncertainty} \\
  \cline{2-4}
  Source & $m(\Lcst)$ & $\Gamma(\Lcst)$ & $\Delta\ln\mathcal{L}$ \\
         & $[\mev\,]$ & $[\mev\,]$        & \\
  \hline
       Background fraction & $0.01$ & $0.02$ & $0.11$  \\
      Efficiency profile & $0.01$ & $0.10$ & $0.35$  \\
        Background shape & $0.02$ & $0.11$ & $0.28$  \\
     Momentum resolution & $0.02$ & $0.24$ & $0.29$  \\
              Mass scale & $0.05$ & $-$ & $-$  \\
           Fit procedure & $0.03$ & $0.08$ & $-$  \\
\hline
        Total systematic & $0.07$ & $0.29$ & $0.54$  \\
\hline
      Breit--Wigner model & $${\small $+0.01$/$-0.00$}$$ & $${\small $+0.01$/$-0.00$}$$ & $0.01$  \\
       Nonresonant model & $${\small $+0.14$/$-0.20$}$$ & $${\small $+0.75$/$-0.00$}$$ & $0.62$  \\
--- of which helicity formalism & $${\small $+0.14$/$-0.00$}$$ & $${\small $+0.36$/$-0.00$}$$ & $0.62$  \\
\hline
             Total model & $${\small $+0.14$/$-0.20$}$$ & $${\small $+0.75$/$-0.00$}$$ & $0.88$  \\

  \end{tabular}
\end{table}

The significance of the spin assignment $J=5/2$ with respect to the next most likely hypothesis 
$J=7/2$ for the \Lcst state is evaluated with a series of pseudoexperiments, where the samples are 
generated from the model with $J=7/2$ and then fitted with both $J=5/2$ and $7/2$ hypotheses. 
The difference of the logarithmic likelihoods $\Delta\ln\mathcal{L}$ is used as the test statistic. 
The distribution in $\Delta\ln\mathcal{L}$ is fitted with a Gaussian function and compared 
to the value of $\Delta\ln\mathcal{L}$ observed in data. The statistical significance is 
expressed in terms of a number of standard deviations ($\sigma$). 
The uncertainty in $\Delta\ln\mathcal{L}$ due to systematic effects 
is small compared to the statistical uncertainty; 
combining them in quadrature results in an overall significance of $4.0\sigma$. 
The fits with spins $1/2$ and $3/2$ for the \Lcst state yield large $\Delta\ln\mathcal{L}$
and poor fit quality, as seen from Table~\ref{tab:lc2880_lh}. These spin assignments 
are thus excluded. 

In conclusion, the mass and width of the \Lcst resonance are found to be
  \begin{equation*}
   \begin{split}
    m(\Lcst) & = \lcstmass\mev, \\
    \Gamma(\Lcst) & = \lcstwidth\mev. \\
   \end{split}
  \end{equation*}
These are consistent with the current world averages, and have comparable precision. 
The preferred value for the spin of this state is confirmed to be $5/2$, with a significance of 
$4\sigma$ over the next most likely hypothesis, $7/2$. The spin assignments $1/2$ and $3/2$ are excluded. 
The largest nonresonant contribution underneath the \Lcst state comes from a partial wave with spin $3/2$ 
and positive parity. 
With a larger dataset, it would be possible to constrain the phase motion of the nonresonant amplitude in a model-independent way
using the \Lcst amplitude as a reference. 

\subsection{Fit in the near-threshold region}

\label{sec:lowerdp_amplitude}

\begin{figure}
\centering
\includegraphics[width=0.35\textwidth]{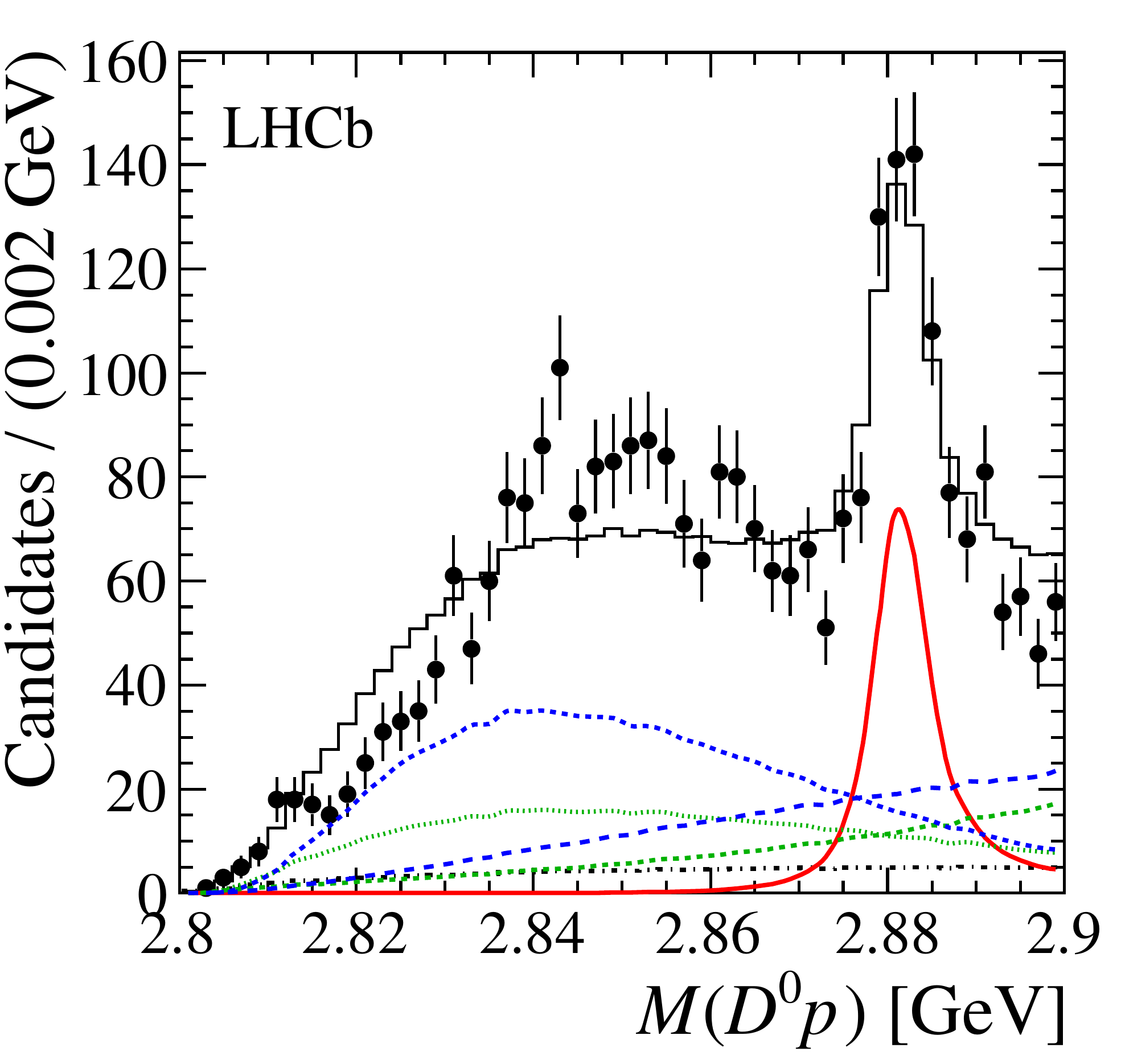} 
\hspace{0.01\textwidth}
\includegraphics[width=0.22\textwidth]{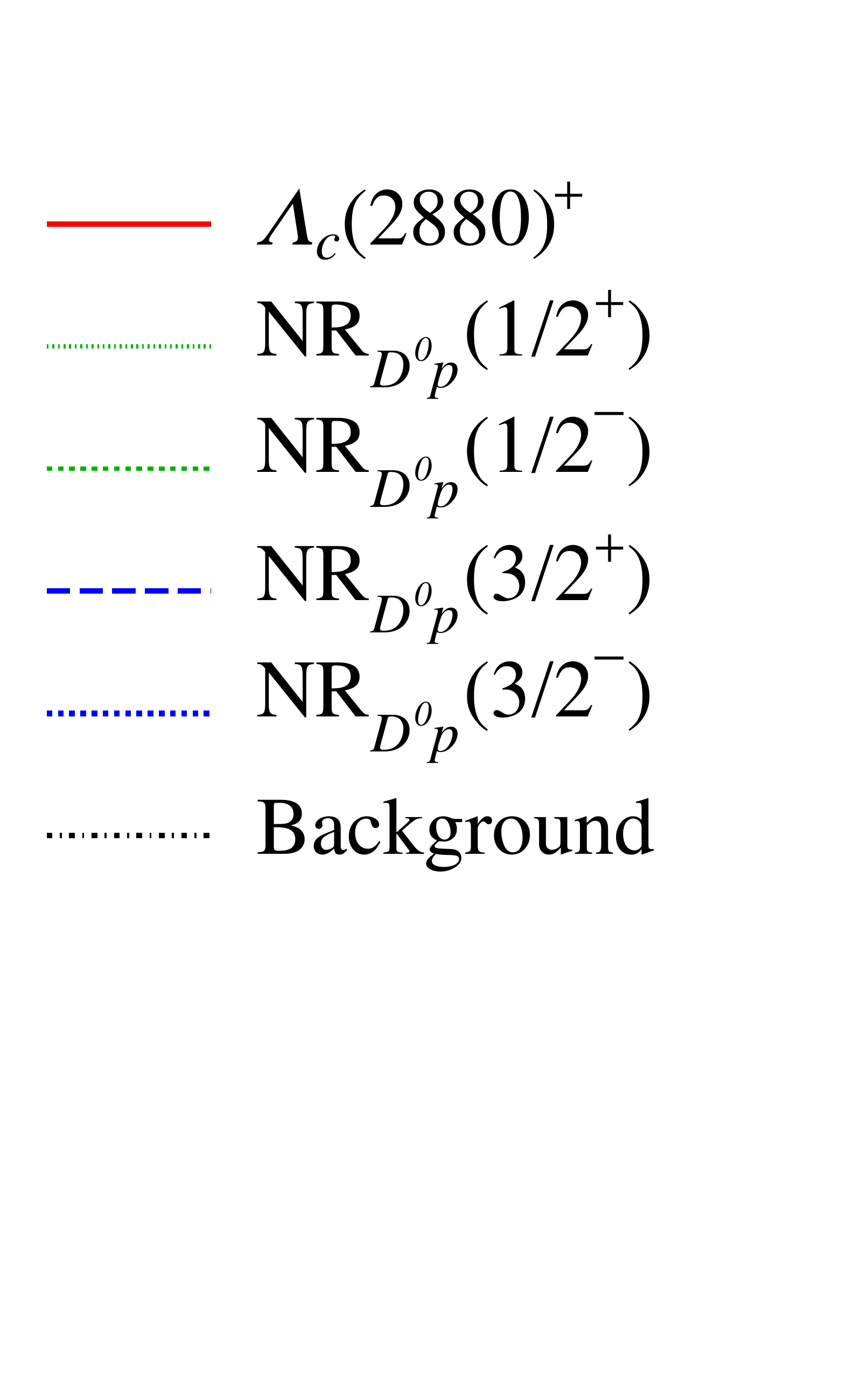} 
\caption{$M(\Dz\proton)$ projections for the fit including the \Lcst state and four exponential 
         nonresonant amplitudes. 
}
\label{fig:lowerdp_bad_fit}
\end{figure}

Extending the $M(\Dz\proton)$ range down to the $\Dz\proton$ threshold (region~3), it becomes evident that a simple model 
for the broad amplitude components, such as an exponential lineshape, cannot describe the data (Fig.~\ref{fig:lowerdp_bad_fit}). 
The hypothesis that an additional resonance is present in the amplitude is tested in a model-dependent way
by introducing a Breit--Wigner resonance in each of the $\Dz\proton$ partial waves. 
Model-independent tests are also performed via fits in which one or more partial waves 
are parametrised with a spline-interpolated shape. The results of these tests 
are summarised in Table~\ref{tab:lowerdp_lh}. The mass and width of the \Lcst state are fixed 
to their known values~\cite{PDG2016} in these fits. 

There are no states with mass around the $\Dz\proton$ threshold ($2800\mev$) 
that are currently known to decay to the $\Dz\proton$ final state. A broad structure has been seen 
previously in the $\Lc\pip\pim$ final state that is referred to as 
the $\Lambdares_{\cquark}(2765)^{+}$~\cite{Artuso:2000xy}. 
It could contribute to the $\Dz\proton$ amplitude if its width is large. 
Since neither the quantum numbers nor the width of this structure have been measured, 
fits are carried out in which this structure is included, modelled as a Breit--Wigner amplitude 
with spin-parity $1/2^{\pm}$ or $3/2^{\pm}$, 
and with a width that is free to vary; its mass is fixed to $2765\mev$.
In addition, four exponential 
nonresonant components with $J^P=1/2^+$, $1/2^-$, $3/2^+$, and $3/2^-$ are included. 
None of these fits are of acceptable quality, as shown in Table~\ref{tab:lowerdp_lh}. 
A Flatt\'{e} parametrisation of the line shape~\cite{Flatte:1976xu} with couplings to 
$\Lc\pip\pim$ and $\Dz\proton$ channels is also considered, but does not produce a fit 
of acceptable quality either.
Therefore, a resonance with a fixed mass of $2765\mev$ is not sufficient to explain the data. 

\begin{figure}
\centering
\includegraphics[width=0.33\textwidth]{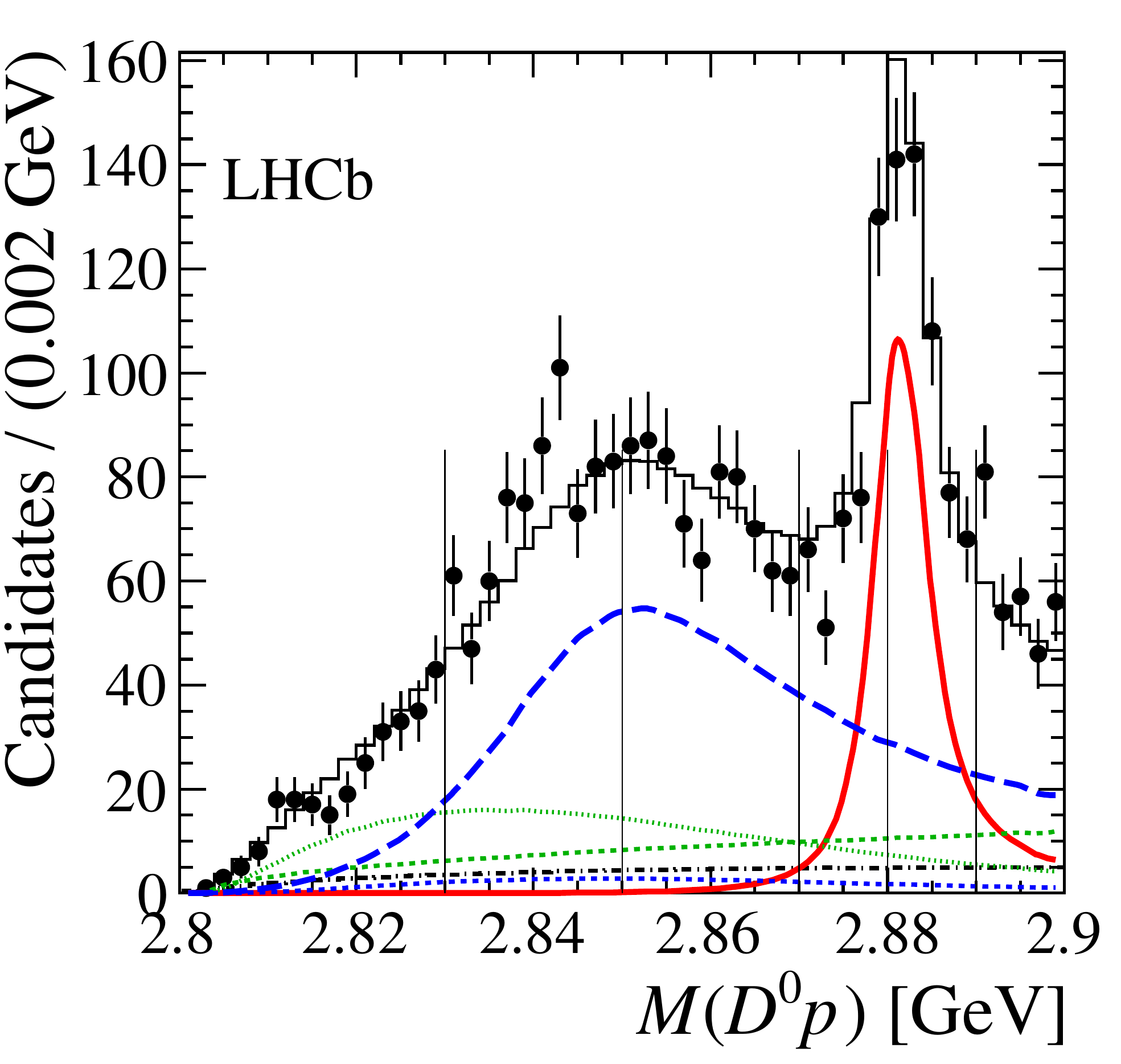} 
  \put(-113,104){(a)}
\includegraphics[width=0.33\textwidth]{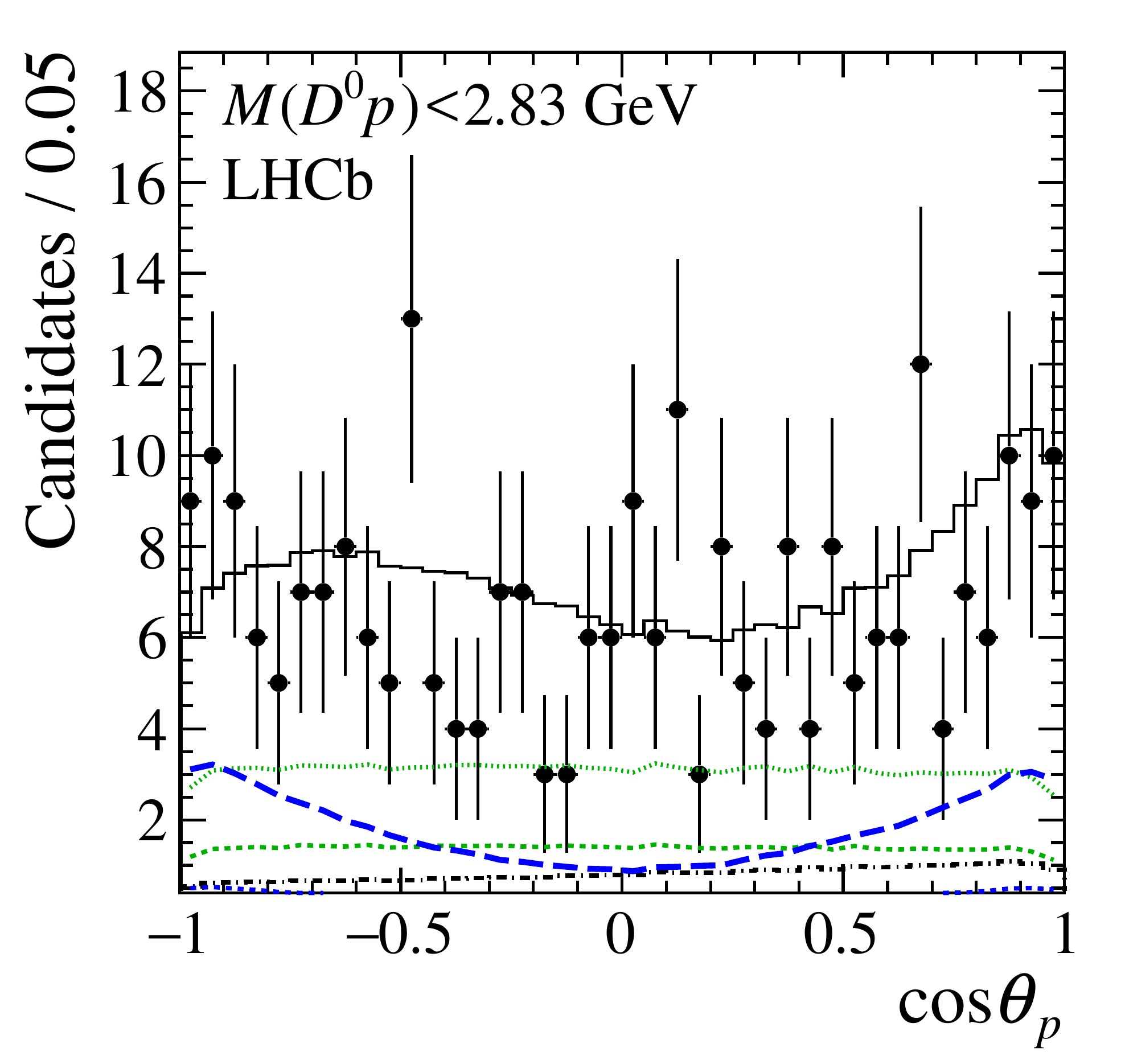} 
  \put(-113,104){(b)}
\includegraphics[width=0.33\textwidth]{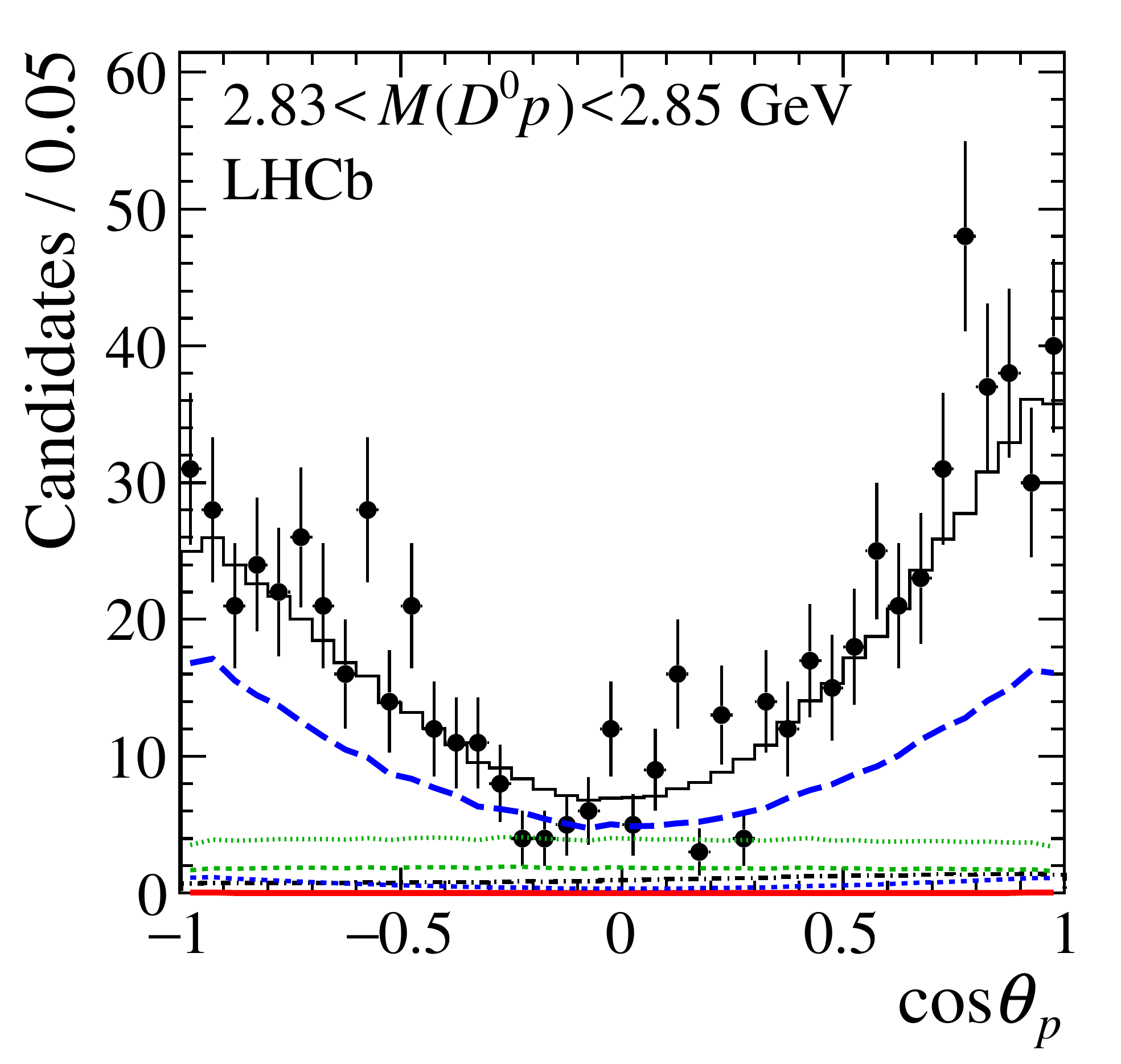} 
  \put(-113,104){(c)}

\includegraphics[width=0.33\textwidth]{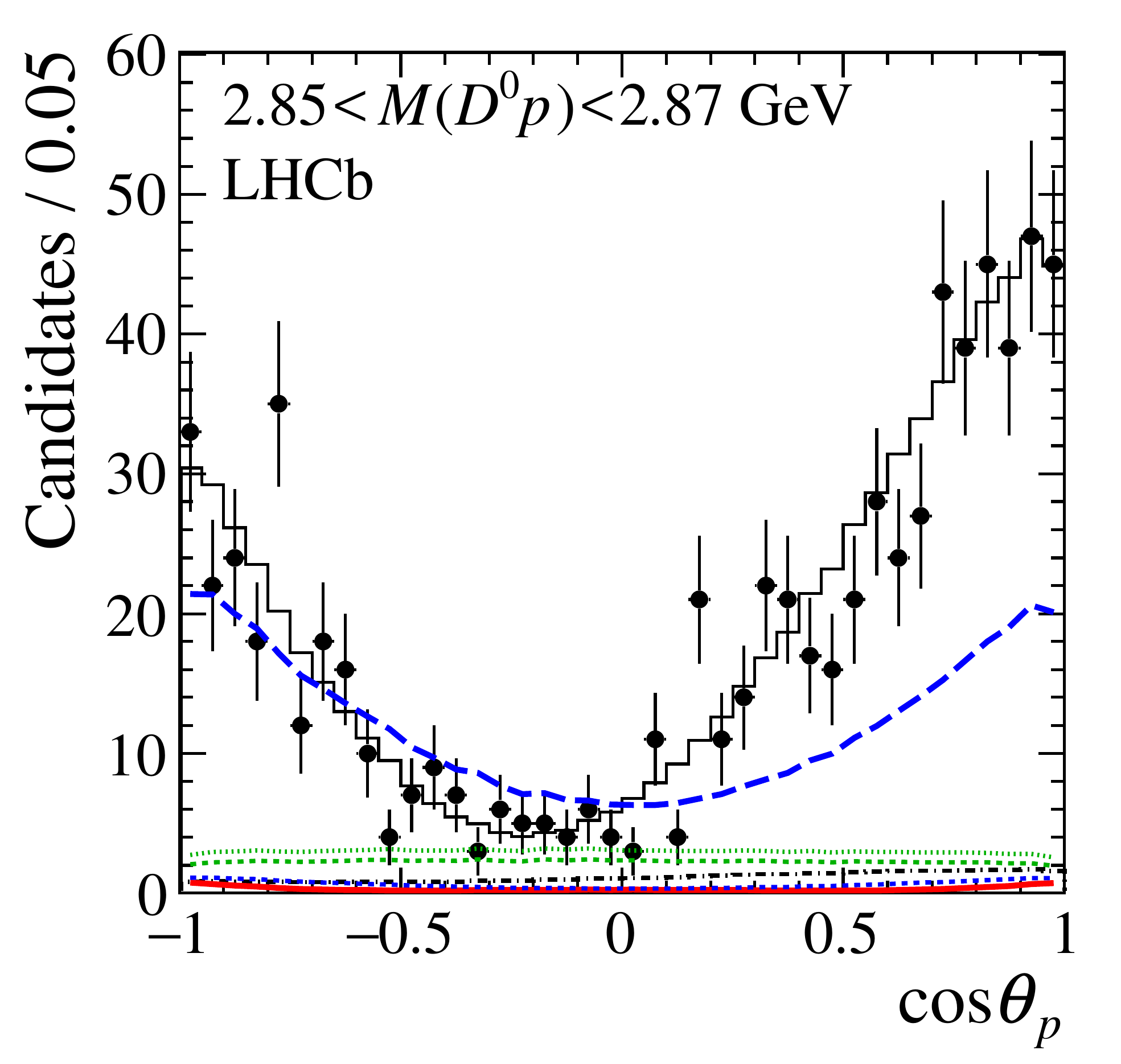} 
  \put(-113,104){(d)}
\includegraphics[width=0.33\textwidth]{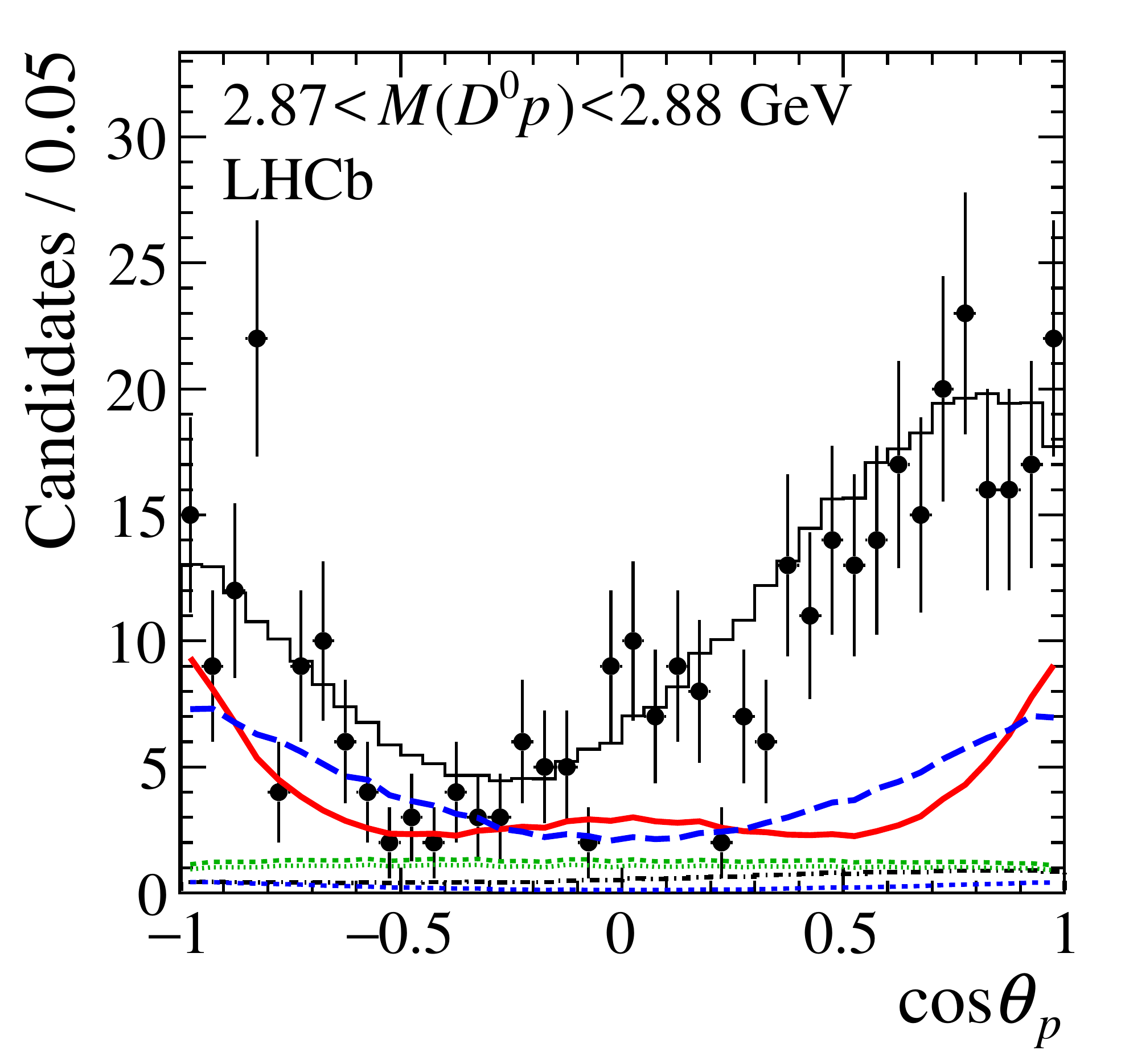} 
  \put(-113,104){(e)}
\includegraphics[width=0.33\textwidth]{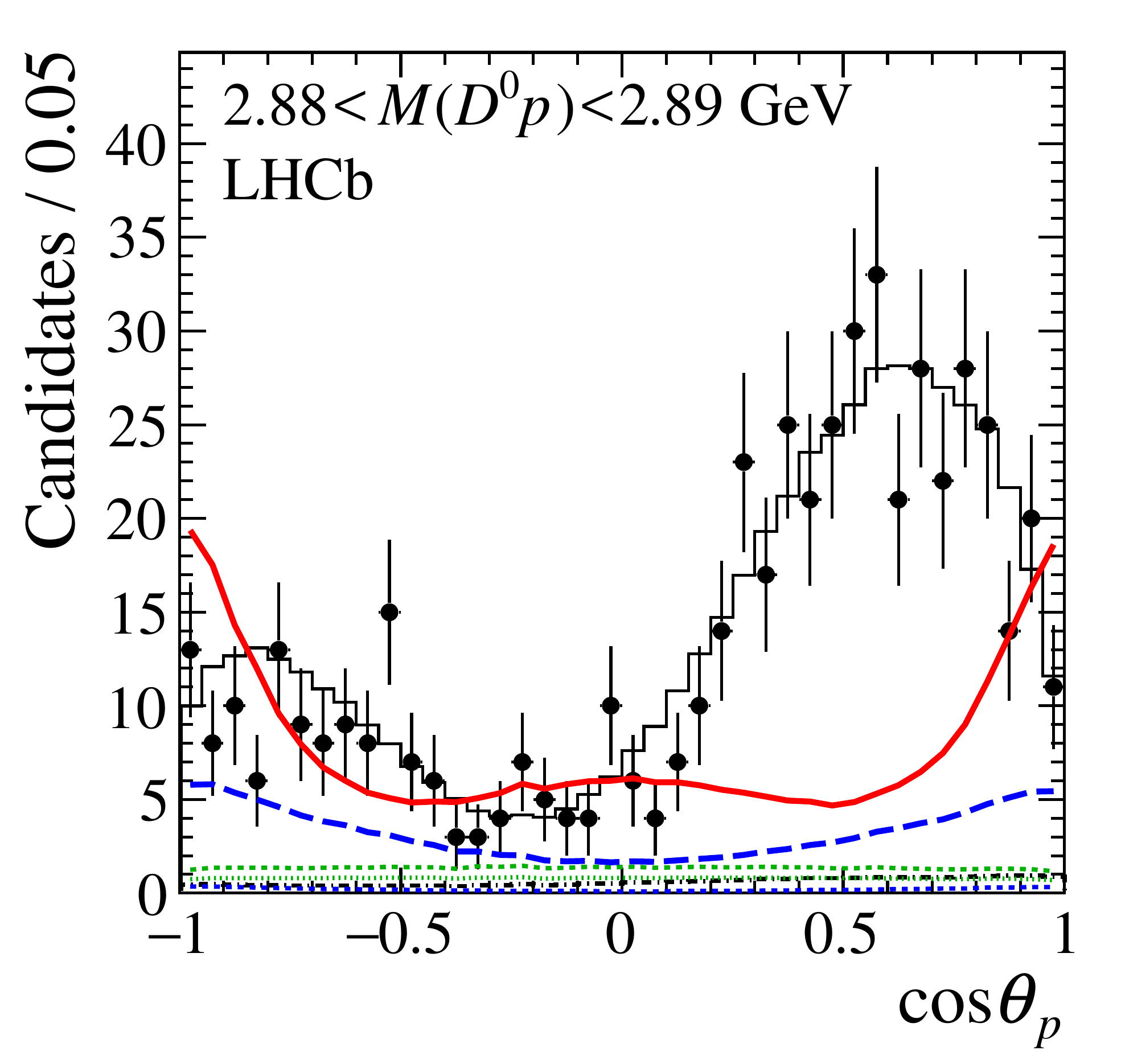} 
  \put(-113,104){(f)}

\includegraphics[width=0.33\textwidth]{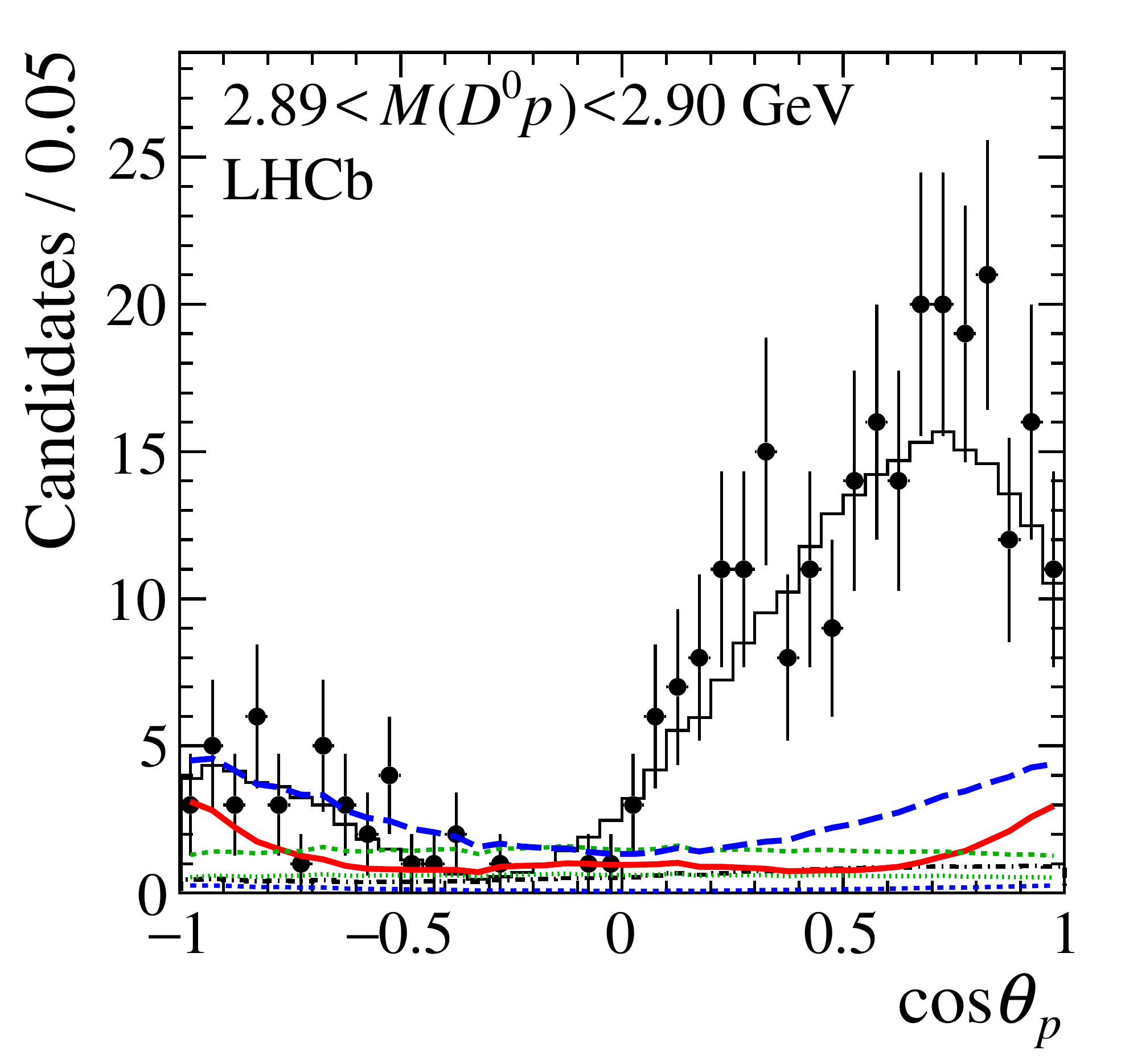} 
  \put(-113,104){(g)}
\includegraphics[width=0.33\textwidth]{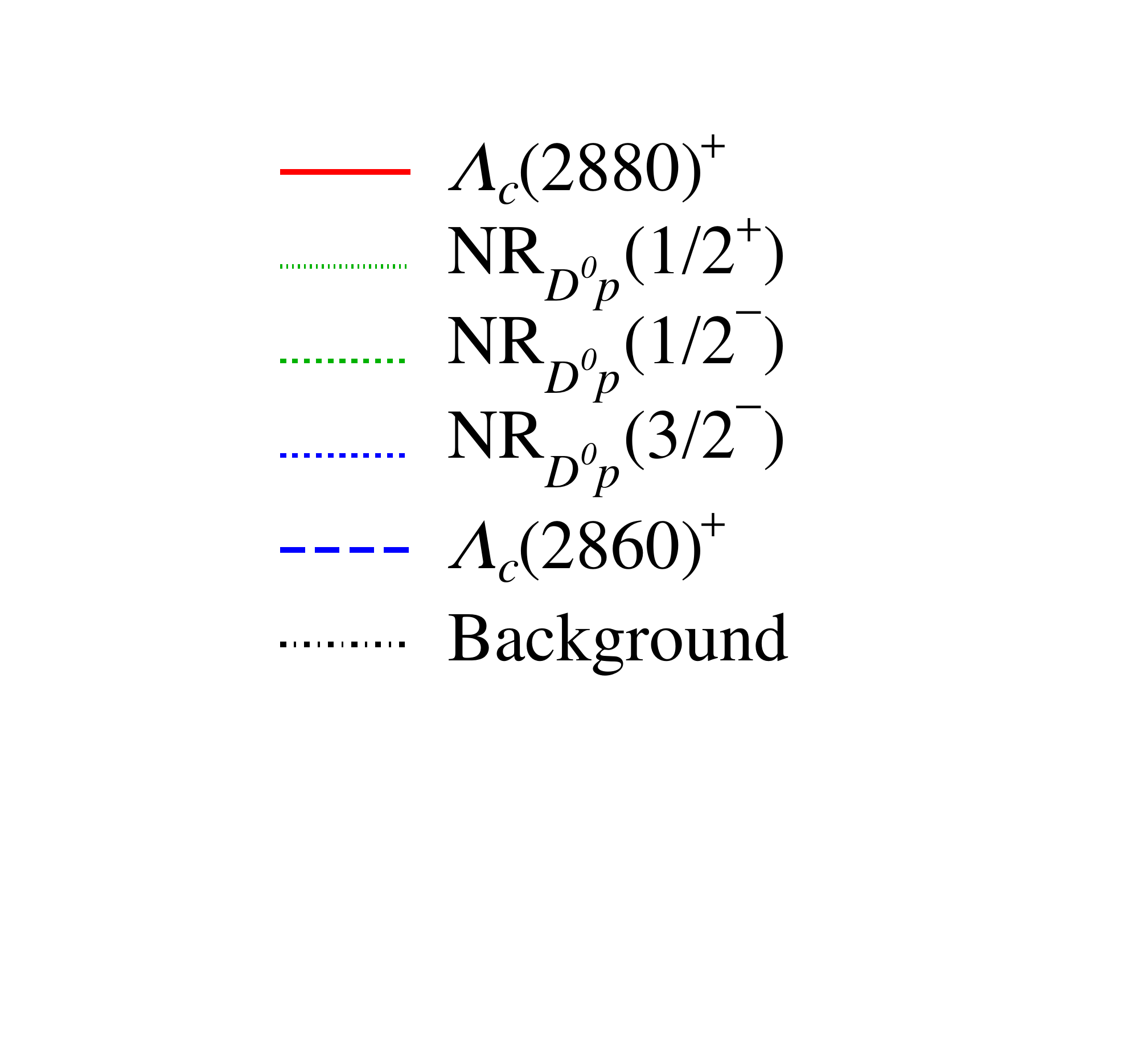} 

\caption{Results for the fit of the \lbdnppi Dalitz plot distribution in the near-threshold $\Dz\proton$ 
mass region (region 3): (a) $M(\Dz\proton)$ projection, and (b--g) $\cos\theta_{\proton}$ projections for slices 
in $\Dz\proton$ invariant mass. An exponential model is used for the nonresonant partial waves. A broad 
$\Lcx$ resonance and the \Lcst state are also present. 
Vertical lines in (a) indicate the boundaries of the $\Dz\proton$ invariant mass slices. 
Due to interference effects the total is not necessarily equal to the sum of the components. 
}
\label{fig:lowerdp_ampl_fit}
\end{figure}

If the mass of the Breit--Wigner resonance is allowed to vary in the fit, 
good agreement with data can be obtained for the spin-parity assignment $J^P=3/2^+$. 
Moreover, if the resonance is assumed to have $J^P=3/2^+$, the exponential 
nonresonant component with $J^P=3/2^+$ can be removed from the 
amplitude model without loss of fit quality. This model is taken as the baseline for this fit region.  
The mass and the width of the resonance obtained from the fit are around $2856\mev$
and $65\mev$, respectively, and therefore this structure will be referred to as $\Lcx$
hereafter. The results of this fit are shown in Fig.~\ref{fig:lowerdp_ampl_fit}. 

\begin{table}
  \caption{Quality of various fits to the 
           near-threshold $\Dz\proton$ data. 
           The models include nonresonant components for partial waves with $J\leq 3/2$
           with or without a resonant component, whose mass is fixed to 2765\mev or allowed to vary (``Float''). 
           ``Exp'' denotes an exponential nonresonant lineshape, ``CSpl'' a 
           complex spline parametrisation, and ``RSpl'' a real spline parametrisation multiplied by a constant phase. 
           The baseline model is shown in bold face. }
  \label{tab:lowerdp_lh}
  \centering
  \scalebox{0.95}{
  \begin{tabular}{cccc|cc|rcc}
  \multicolumn{4}{c|}{Nonresonant model}& \multicolumn{2}{c|}{Resonance} &                  &              &                       \\
  $1/2^-$ & $1/2^+$ & $3/2^-$ & $3/2^+$ & Mass $[\mev\,]$ & $J^P$ & $\Delta\ln\mathcal{L}$ & $\chi^2/$ndf & $P(\chi^2, \mbox{ndf})$ [\%] \\
  \hline
  Exp & Exp & Exp & Exp &     $-$ &       $-$  & $72.2$ & 287.4/150 & \phantom{0}0.0 \\ 
Exp & Exp & Exp & Exp &  2765 & $1/2^-$  & $53.6$ & 247.2/146 & \phantom{0}0.0 \\ 
Exp & Exp & Exp & Exp &  2765 & $1/2^+$  & $52.8$ & 254.8/146 & \phantom{0}0.0 \\ 
Exp & Exp & Exp & Exp &  2765 & $3/2^-$  & $45.8$ & 240.5/146 & \phantom{0}0.0 \\ 
Exp & Exp & Exp & Exp &  2765 & $3/2^+$  & $38.5$ & 226.0/146 & \phantom{0}0.0 \\ 
\hline 
Exp & Exp & Exp & Exp & Float & $1/2^-$  & $8.2$  & 162.7/145 & 14.9 \\ 
Exp & Exp & Exp & Exp & Float & $1/2^+$  & $15.2$ & 170.2/145 & \phantom{0}7.5 \\ 
Exp & Exp & Exp & Exp & Float & $3/2^-$  & $9.3$  & 162.1/145 & 15.7 \\ 
Exp & Exp & Exp & Exp & Float & $3/2^+$  & $-3.3$ & 139.5/145 & 61.3 \\ 
Exp & Exp & $-$   & $-$   & Float & $3/2^+$  & $12.8$ & 169.7/153 & 16.9 \\ 
{\bf Exp} & {\bf Exp} & {\bf Exp} & \boldmath{$-$}   & {\bf Float} & \boldmath{$3/2^+$}  & {\bf 0.0}  & \boldmath{$143.1/149$} & {\bf 62.1} \\ 
\hline 
CSpl & Exp & Exp & Exp &      $-$ &     $-$  & $16.1$ & 181.3/140 & \phantom{0}1.1 \\ 
Exp & CSpl & Exp & Exp &      $-$ &     $-$  & $2.0$  & 154.8/140 & 18.5 \\ 
Exp & Exp & CSpl & Exp &      $-$ &     $-$  & $16.6$ & 172.9/140 & \phantom{0}3.1 \\ 
Exp & Exp & Exp & CSpl &      $-$ &     $-$  & $-0.4$ & 146.6/140 & 33.4 \\ 
\hline 
Exp & Exp & CSpl & $-$ &        $-$ &     $-$  & $63.1$ & 234.8/143 & \phantom{0}0.0 \\ 
Exp & Exp & $-$ & CSpl &        $-$ &     $-$  & $10.8$ & 165.7/143 & \phantom{0}9.4 \\ 
Exp & Exp & CSpl & CSpl &     $-$ &     $-$  & $-4.7$ & 146.1/130 & 15.8 \\ 
\hline 
Exp & Exp & RSpl &  Exp &     $-$ &     $-$  & $17.4$ & 177.0/143 & \phantom{0}2.8 \\ 
Exp & Exp &  Exp & RSpl &     $-$ &     $-$  & $15.4$ & 174.5/143 & \phantom{0}3.8 \\ 
Exp & Exp & RSpl & RSpl &     $-$ &     $-$  & $-0.4$ & 145.1/138 & 32.3 \\ 

  \end{tabular}
  }
\end{table}

\begin{figure}
\begin{center}
 \includegraphics[width=0.40\textwidth]{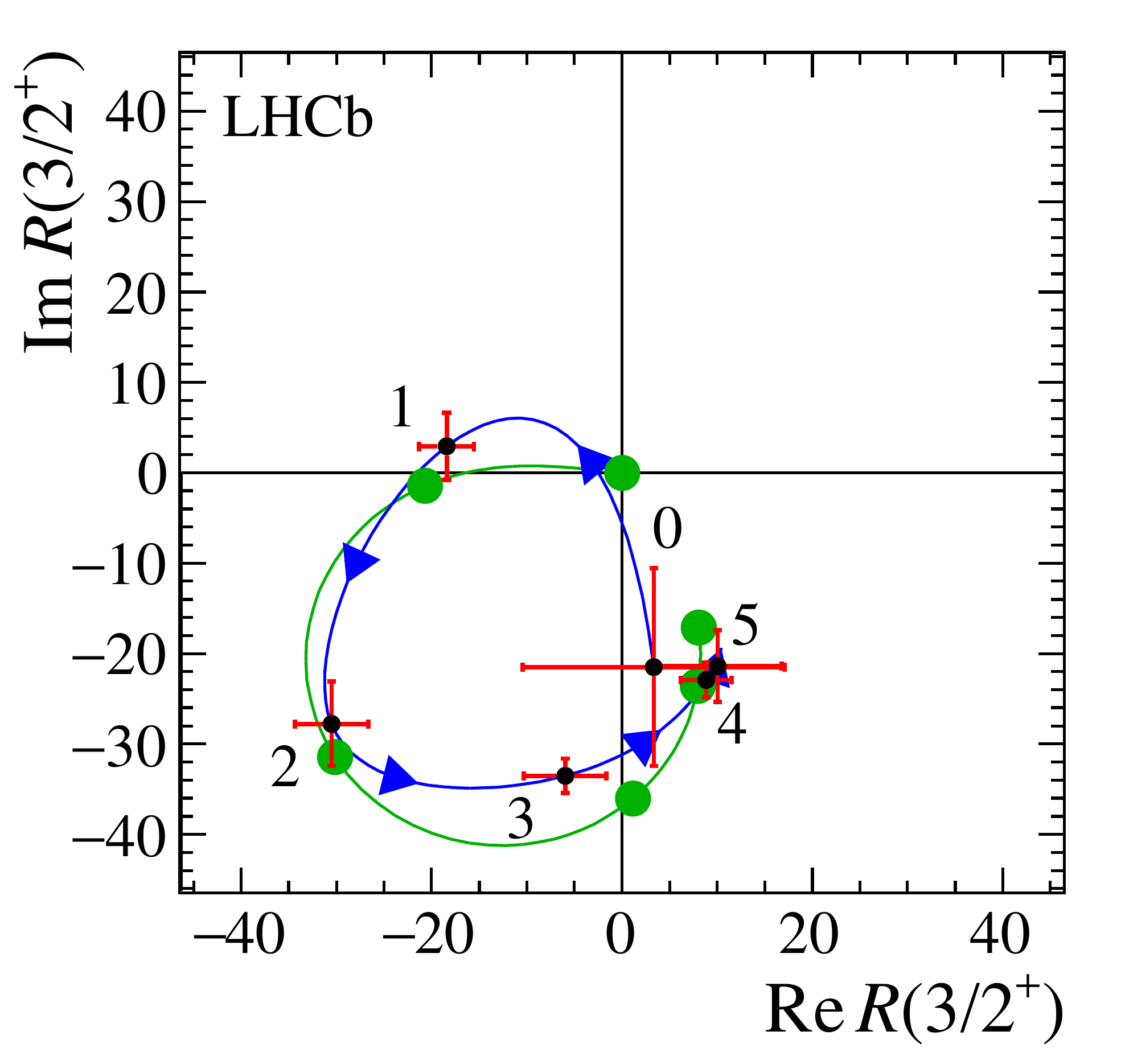}
  \put(-30,145){(a)}
 \hspace{0.04\textwidth}
 \includegraphics[width=0.40\textwidth]{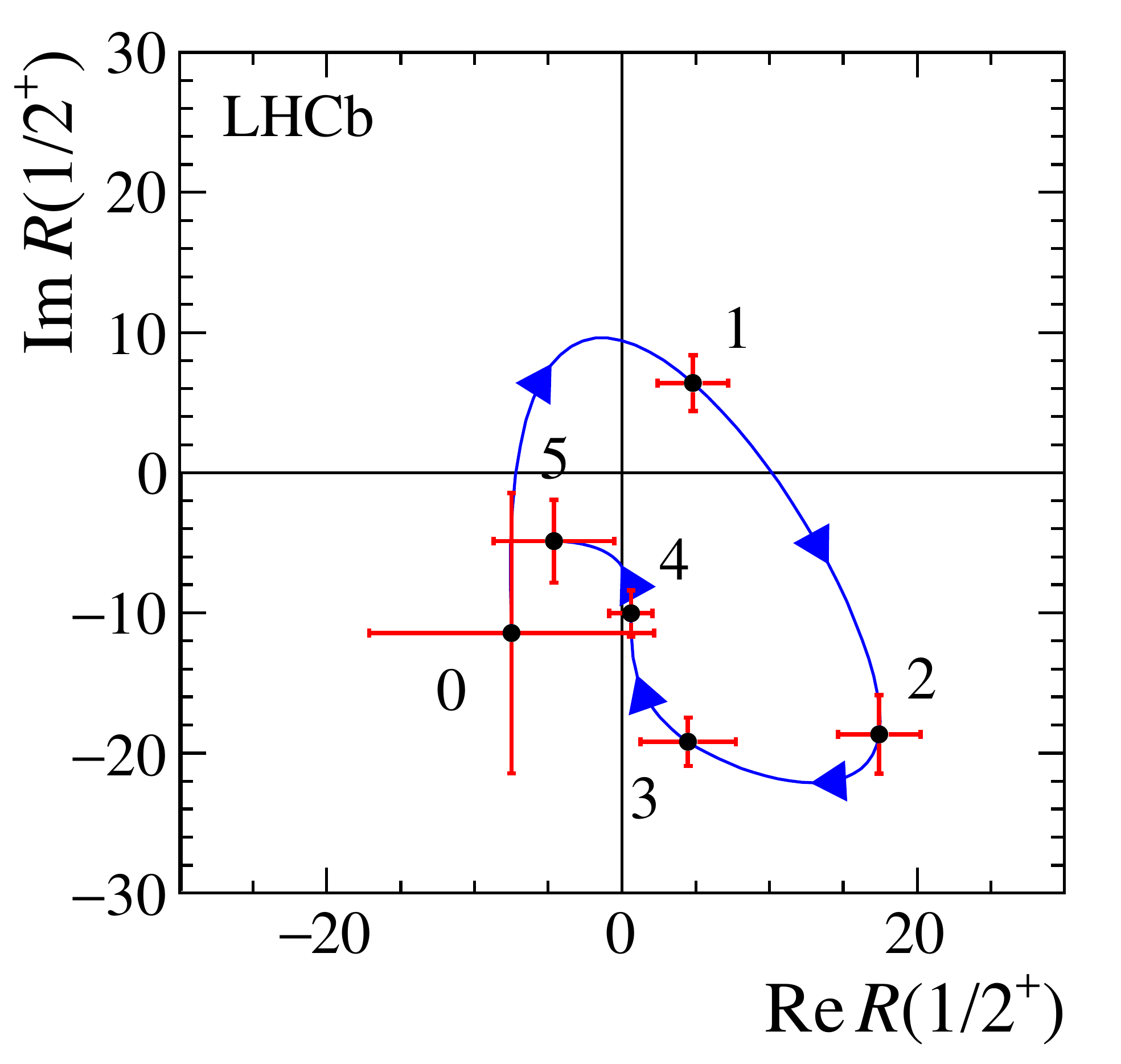}
  \put(-30,145){(b)}
\end{center}
\caption{
Argand diagrams for the complex spline components used in two fits, represented by blue lines with arrows
indicating the phase motion with increasing $M(\Dz\proton)$. 
For subfigure~(a), the $J^P=3/2^+$ partial wave is modelled as a spline and the other 
components in the fit ($1/2^{+}$, $1/2^-$ and $3/2^-$) are described with exponential amplitudes. 
For comparison, results from a separate fit in which the $3/2^+$ partial wave is described 
with a Breit--Wigner function are superimposed: the green line represents its phase motion, 
and the green dots correspond to the $\Dz\proton$ masses at the spline knots. 
For subfigure~(b), the $J^P=1/2^+$ component is modelled as a spline and $1/2^{-}$, $3/2^+$ and $3/2^-$
components as exponential amplitudes.
}
\label{fig:lowerdp_argand}
\end{figure}

One model-independent test for the presence of structure in the broad component is to describe the 
real and imaginary parts with spline-interpolated shapes. 
Cubic splines with six knots at $\Dz\proton$ masses of 2800, 2820, 2840, 2860, 2880 and 2900\mev are used. 
Of the models where only one partial wave is described by a spline while the others remain exponential, 
the best fit is again given by the model where the spline-interpolated amplitude has $J^P=3/2^+$. 
The Argand diagram for the $3/2^+$ amplitude in this fit is shown in Fig.~\ref{fig:lowerdp_argand}(a). 
Each of the points numbered from 0 to 5 corresponds to one spline knot at increasing values of $M(\Dz\proton)$. 
Note that knots 3 and 5 at masses 2860 and 2900\mev correspond to the boundaries 
of the region 2 where the nonresonant amplitude is described by a linear function (Sec.~\ref{sec:lc2880})
and that the amplitudes and phases in those two knots can be compared 
directly to Fig.~\ref{fig:lc2880_argand}, since the convention is the 
same in both fits. The Argand diagram demonstrates resonance-like phase rotation of the $3/2^+$ partial wave 
with respect to the other broad components in the $\Dz\proton$ amplitude, which are assumed to be 
constant in phase. Note that the absolute phase motion cannot be obtained from this fit 
since there are no reference amplitudes covering the entire $\Dz\proton$ mass range used in the fit. 

As seen in Table~\ref{tab:lowerdp_lh}, inclusion of a spline-interpolated shape in the $1/2^+$ component 
instead of $3/2^+$ also gives a reasonable fit quality. The Argand diagram for the $1/2^+$ wave in this fit 
is shown in Fig.~\ref{fig:lowerdp_argand}(b). Since the phase rotates clockwise, this solution 
cannot be described by a single resonance. 

\begin{figure}[b]
\centering
\includegraphics[width=0.35\textwidth]{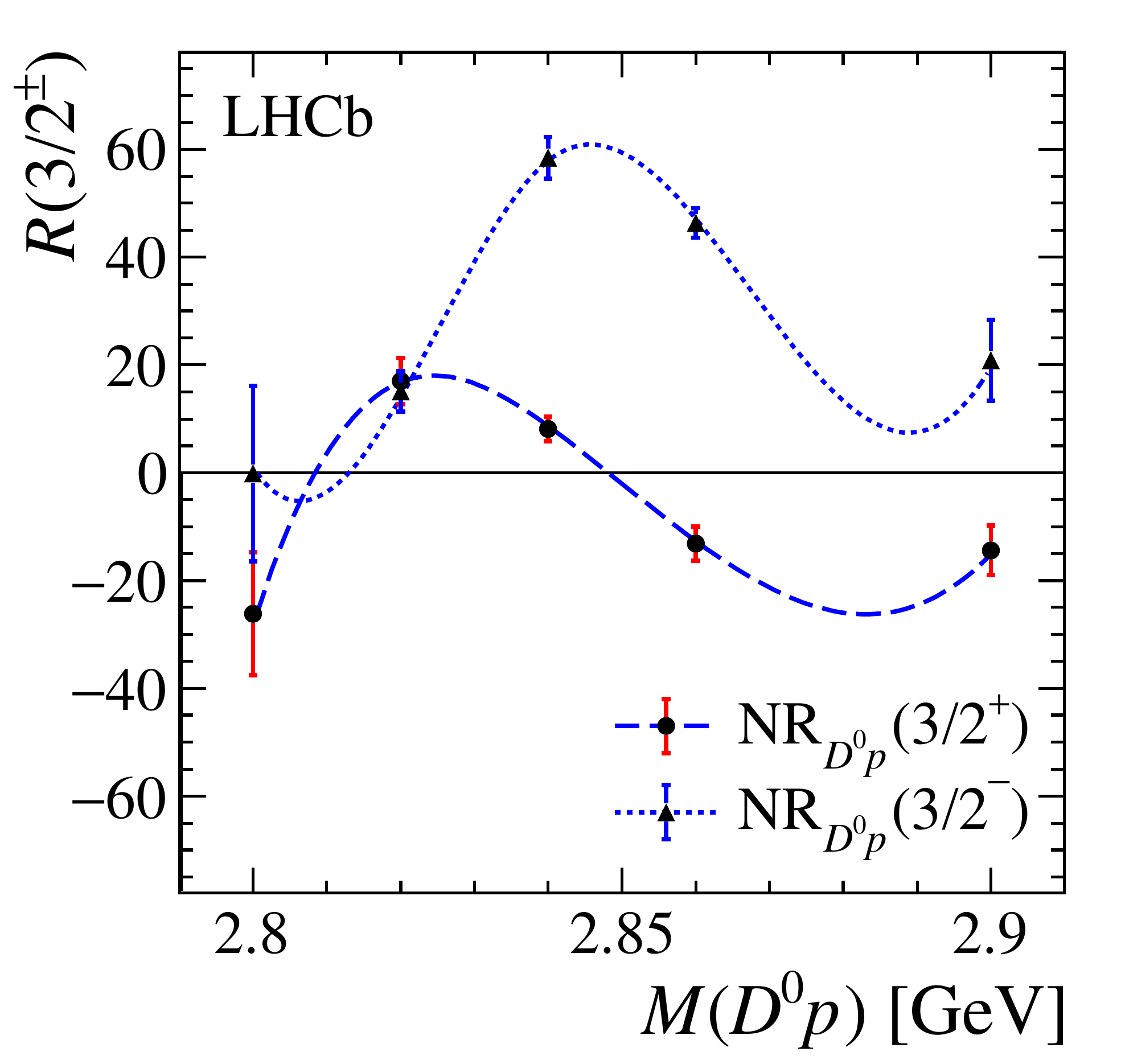} 
  \put(-125,116){(a)}
\hspace{0.01\textwidth}
\includegraphics[width=0.35\textwidth]{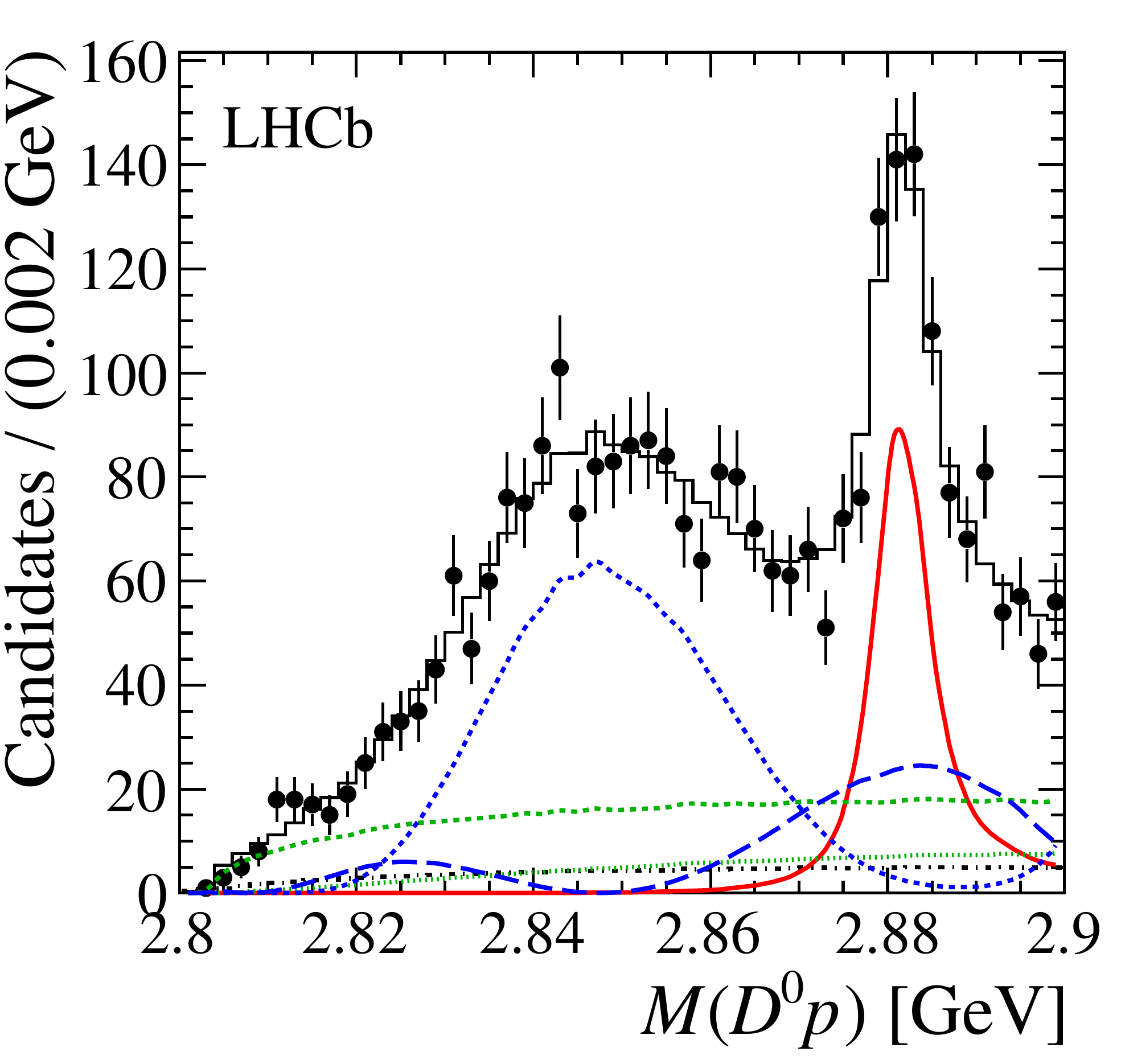} 
  \put(-125,116){(b)}
\hspace{0.01\textwidth}
\includegraphics[width=0.22\textwidth]{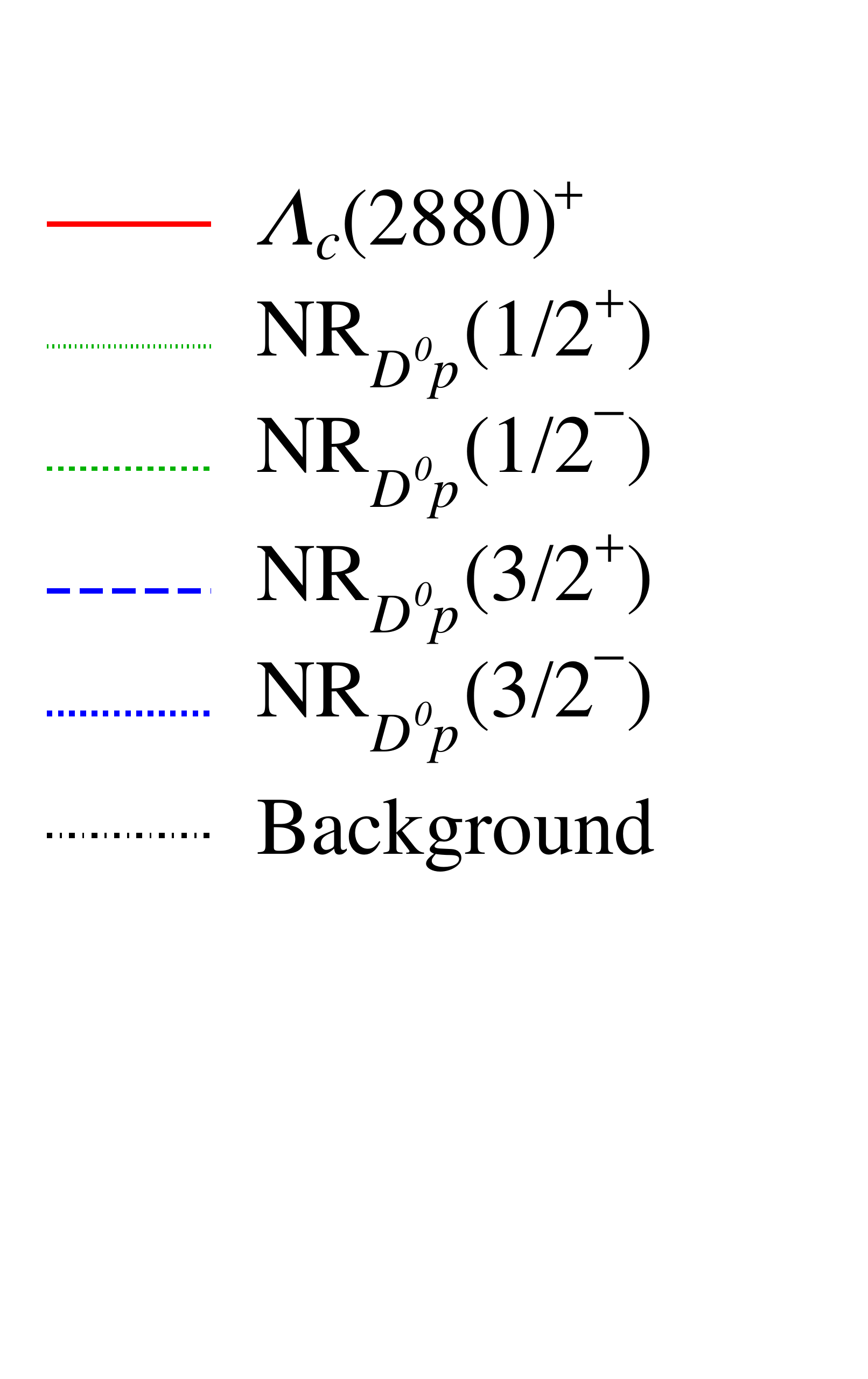} 
\caption{Results of the fit including the \Lcst state, two exponential nonresonant amplitudes with $J^P=1/2^{\pm}$
         and two real splines in $J^P=3/2^{\pm}$ partial waves. (a)
         Spline amplitudes for $J^P=3/2^{\pm}$ partial waves as functions of $M(\Dz\proton)$. 
         Points with the error bars are fitted values of the amplitude in the spline knots, 
         smooth curves are the interpolated amplitude shapes.  
         (b) The $M(\Dz\proton)$ projection of the decay density and the components of the fit model.
}
\label{fig:lowerdp_real_fit}
\end{figure}

A genuine resonance has characteristic phase motion as a function of $M(\Dz\proton)$. 
As a null test, the fits are repeated with a spline function with 
no phase motion. This is implemented as a real spline function multiplied by a constant phase. The fits where 
only one partial wave is replaced by a real spline give poor fits. If both spin-$3/2$
amplitudes are represented by real splines, the fit quality is good, but the 
resulting amplitudes oscillate as functions of $M(\Dz\proton)$, 
which is not physical. Figure \ref{fig:lowerdp_real_fit}(a) shows the real spline amplitudes 
without the contribution of the phase space term, which exhibit oscillating behaviour, 
while Fig.~\ref{fig:lowerdp_real_fit}(b) shows the $M(\Dz\proton)$
projection of the decay density for this solution.

As in the case of the amplitude fit in the \Lcst region, pseudoexperiments are used to validate the 
fit procedure, obtain uncertainties on the fit fractions, and determine values of ${\rm ndf}_{\rm eff}$ for the 
binned fit quality test. Pseudoexperiments 
are also used to obtain the  $\Delta\ln\mathcal{L}$ distributions for fits with various spin-parity hypotheses.  
After correcting for fit bias, the mass and width of the broad \Lcx resonance are found to be 
$m(\Lcx)=\lcxmassstat\mev$ and $\Gamma(\Lcx)=\lcxwidthstat\mev$, where the uncertainties are statistical only. 

Systematic uncertainties are obtained following the same procedure as for the amplitude fit in 
the \Lcst region (Sec.~\ref{sec:lc2880_amplitude}) and are summarised in Table~\ref{tab:lowerdp_syst}. 
An additional contribution to the list of systematic uncertainties is the uncertainty in the knowledge of the 
mass and width of the \Lcst resonance, which are fixed in the fit. It is estimated by varying these 
parameters within their uncertainties.
The model uncertainty associated with the parametrisation of the nonresonant components is estimated by performing 
fits with an additional exponential $3/2^+$ amplitude component and with the $3/2^-$ component removed, as well as 
by adding the $\proton\pim$ amplitude and using the covariant amplitude formalism in the same way as in 
Sec.~\ref{sec:lc2880_amplitude}. 

The $J^P=3/2^+$ hypothesis is preferred for the \Lcx state, since its fit likelihood, 
as measured by $\Delta\ln\mathcal{L}$, is substantially better than those of 
the other $J^P$ values tested. 
The significance of this difference is assessed with pseudoexperiments and corresponds to 
$8.8\sigma$, $6.3\sigma$, and $6.6\sigma$ for the $1/2^+$, $1/2^-$, and $3/2^-$ hypotheses, 
respectively. 
When systematic uncertainties are included, these reduce to $8.4\sigma$, $6.2\sigma$ and $6.4\sigma$.
For $J^P=3/2^+$, the following parameters are obtained for the near-threshold resonant state:
\begin{equation*}
  \begin{split}
    m(\Lcx) & = \lcxmass\mev, \\
    \Gamma(\Lcx) & = \lcxwidth\mev.\\
  \end{split}
\end{equation*}
The largest uncertainties are associated with the modelling of the nonresonant 
components of the $\Dz\proton$ amplitude. 

\begin{table}
  \caption{Systematic uncertainties on the \Lcx parameters and on $\Delta\ln\mathcal{L}$ between 
           the baseline $3/2^+$ and alternative spin-parity assignments. 
           The uncertainty due to the nonresonant model 
           includes a component associated with the helicity formalism, 
           which for comparison is given explicitly in the table, too.}
  \label{tab:lowerdp_syst}
  \centering
  \begin{tabular}{l|ccccc}
         & \multicolumn{5}{c}{Uncertainty} \\
  \cline{2-6}
  Source & $m(\Lcx)$ & $\Gamma(\Lcx)$ & \multicolumn{3}{c}{$\Delta\ln\mathcal{L}$} \\
  \cline{4-6}
         & $[\mev\,]$     & $[\mev\,]$            & $1/2^+$ & $1/2^-$ & $3/2^+$ \\
  \hline
       Background fraction & $0.22$ & $0.54$ & $2.3$ & $1.1$ & $1.8$  \\
      Efficiency profile & $0.20$ & $0.61$ & $0.5$ & $0.8$ & $0.4$  \\
        Background shape & $0.29$ & $0.77$ & $1.0$ & $0.4$ & $0.3$  \\
     Momentum resolution & $0.10$ & $0.49$ & $-$ & $-$ & $-$  \\
              Mass scale & $0.05$ & $-$ & $-$ & $-$ & $-$  \\
           Fit procedure & $0.17$ & $0.67$ & $-$ & $-$ & $-$  \\
      $\Lcst$ parameters & $0.02$ & $0.22$ & $0.7$ & $0.4$ & $0.5$  \\
\hline
        Total systematic & $0.46$ & $1.41$ & $2.7$ & $1.4$ & $2.0$  \\
\hline
     Breit--Wigner model & $${\small $+1.11$/$-1.65$}$$ & $${\small $+5.92$/$-8.02$\phantom{1}}$$ & $0.2$ & $0.0$ & $0.2$  \\
       Nonresonant model & $${\small $+0.00$/$-5.35$}$$ & $${\small $+0.15$/$-18.29$}$$ & $2.4$ & $0.1$ & $0.5$  \\
--- of which helicity formalism & $${\small $+0.00$/$-1.23$}$$ & $${\small $+0.00$/$-5.67$\phantom{1}}$$ & $1.6$ & $0.1$ & $0.0$  \\
\hline
             Total model & $${\small $+1.11$/$-5.59$}$$ & $${\small $+5.93$/$-19.97$}$$ & $2.9$ & $0.2$ & $0.5$  \\

  \end{tabular}
\end{table}

\subsection{\boldmath Fit including \Lcstst}

Finally, the $\Dz\proton$ mass region in the amplitude fit is extended up to $M(\Dz\proton)=3.0\gev$ 
to include the $\Lcstst$ state (region 4). 
Since the behaviour of the slowly-varying $\Dz\proton$ amplitude is consistent with the presence of a resonance 
in the $J^P=3/2^+$ wave and nonresonant amplitudes in the $1/2^+$, $1/2^-$, and $3/2^-$ waves, 
the same model is used to describe those parts of the amplitude in the extended fit region. 
The $\Lcstst$ resonance is modelled by a Breit--Wigner lineshape. 
The masses and widths of the \Lcstst and \Lcx states are floated in the fit, while those 
of the \Lcst resonance are fixed to their nominal values~\cite{PDG2016}.
Several variants of the fit are performed in which the spin of $\Lcstst$ is assigned 
to be $1/2$, $3/2$, $5/2$ or $7/2$, with both 
positive and negative parities considered. 
Two different parametrisations of the nonresonant components are considered: the exponential model 
(taken as the baseline) and a second-order polynomial (Eq.~(\ref{eq:nr_poly})). 

\begin{figure}[ht!]
  \centering
  \vspace{-10pt}
  \includegraphics[width=0.33\textwidth]{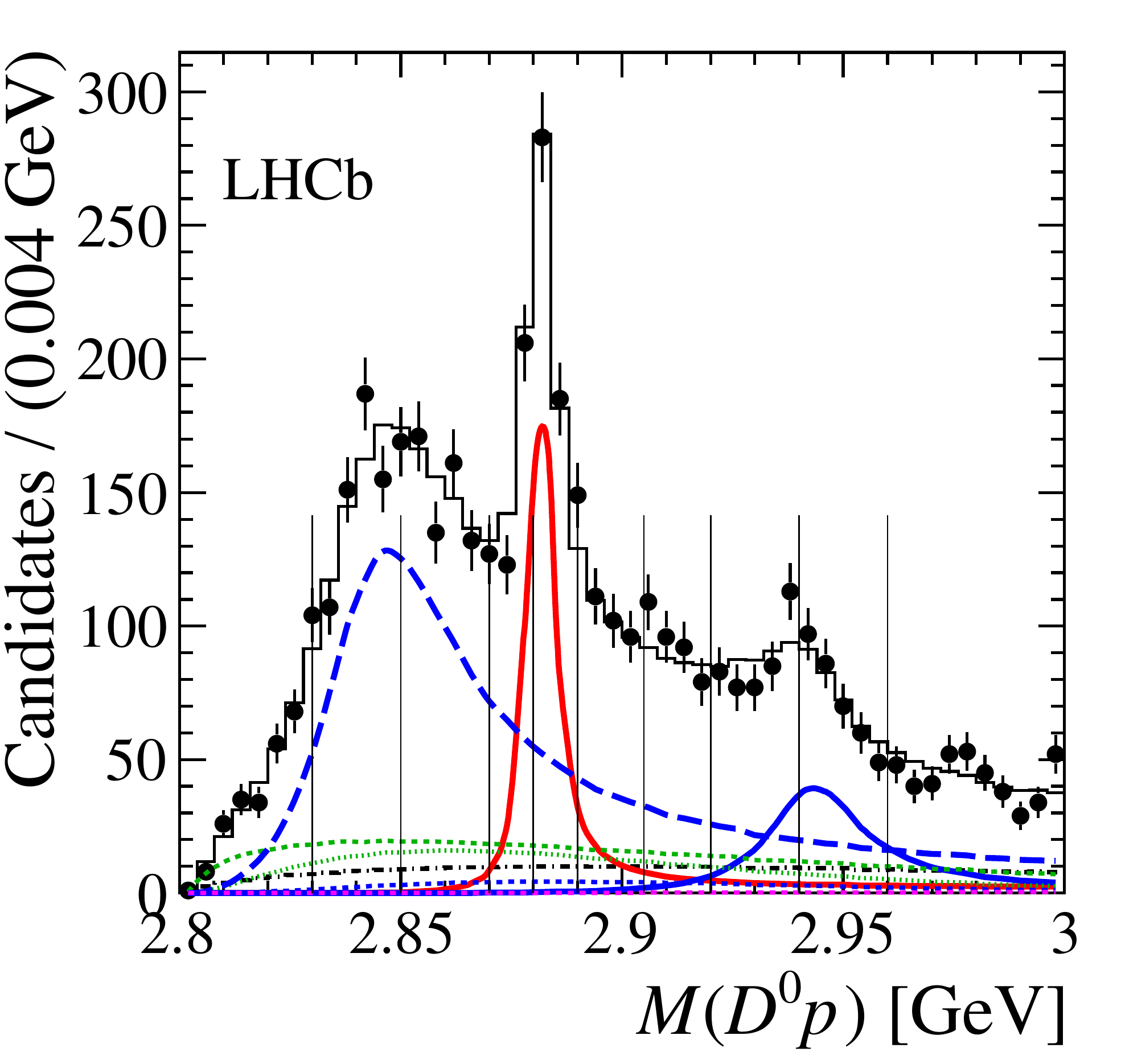} 
  \put(-113,104){(a)}
  \includegraphics[width=0.33\textwidth]{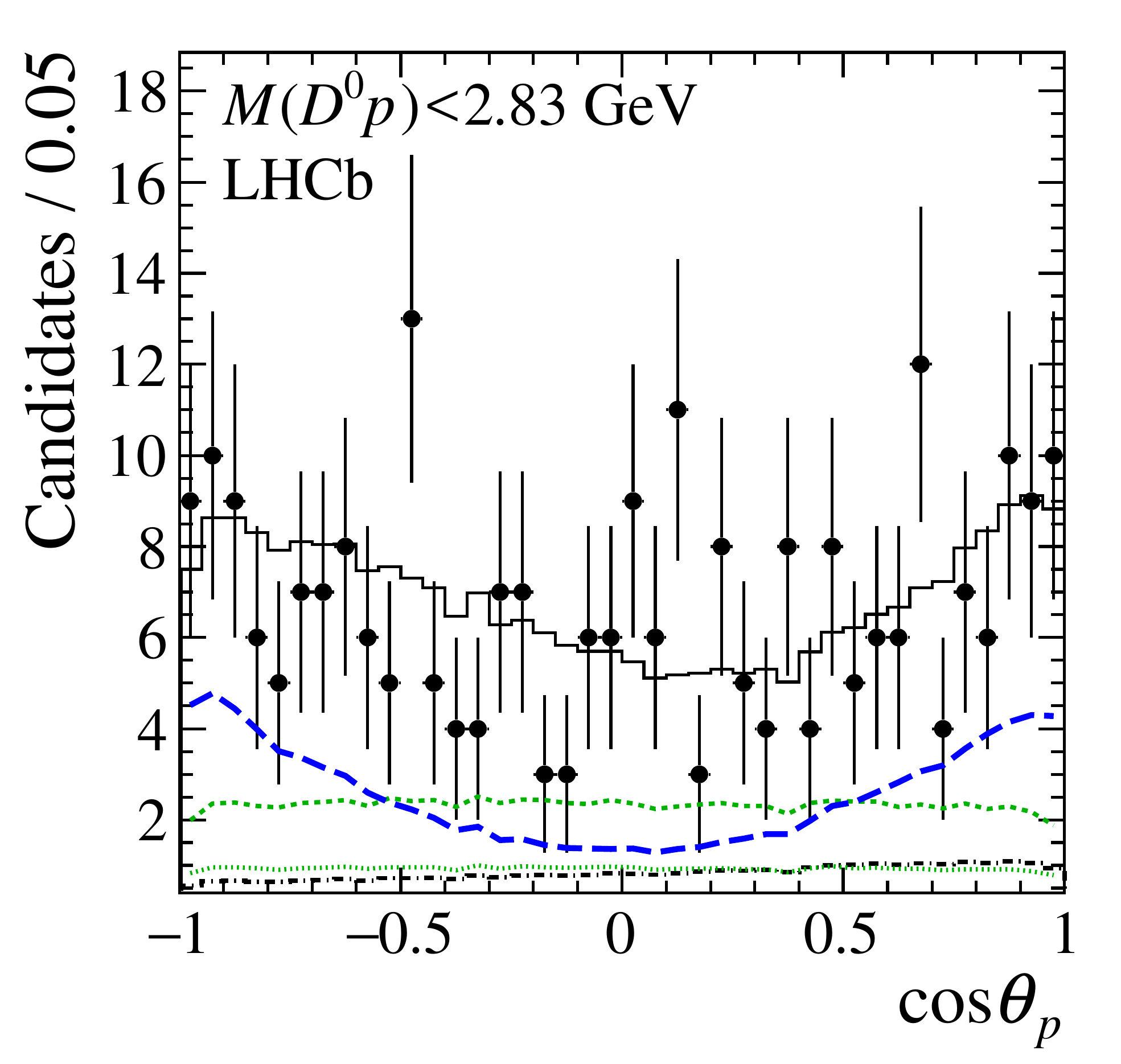} 
  \put(-113,104){(b)}
  \includegraphics[width=0.33\textwidth]{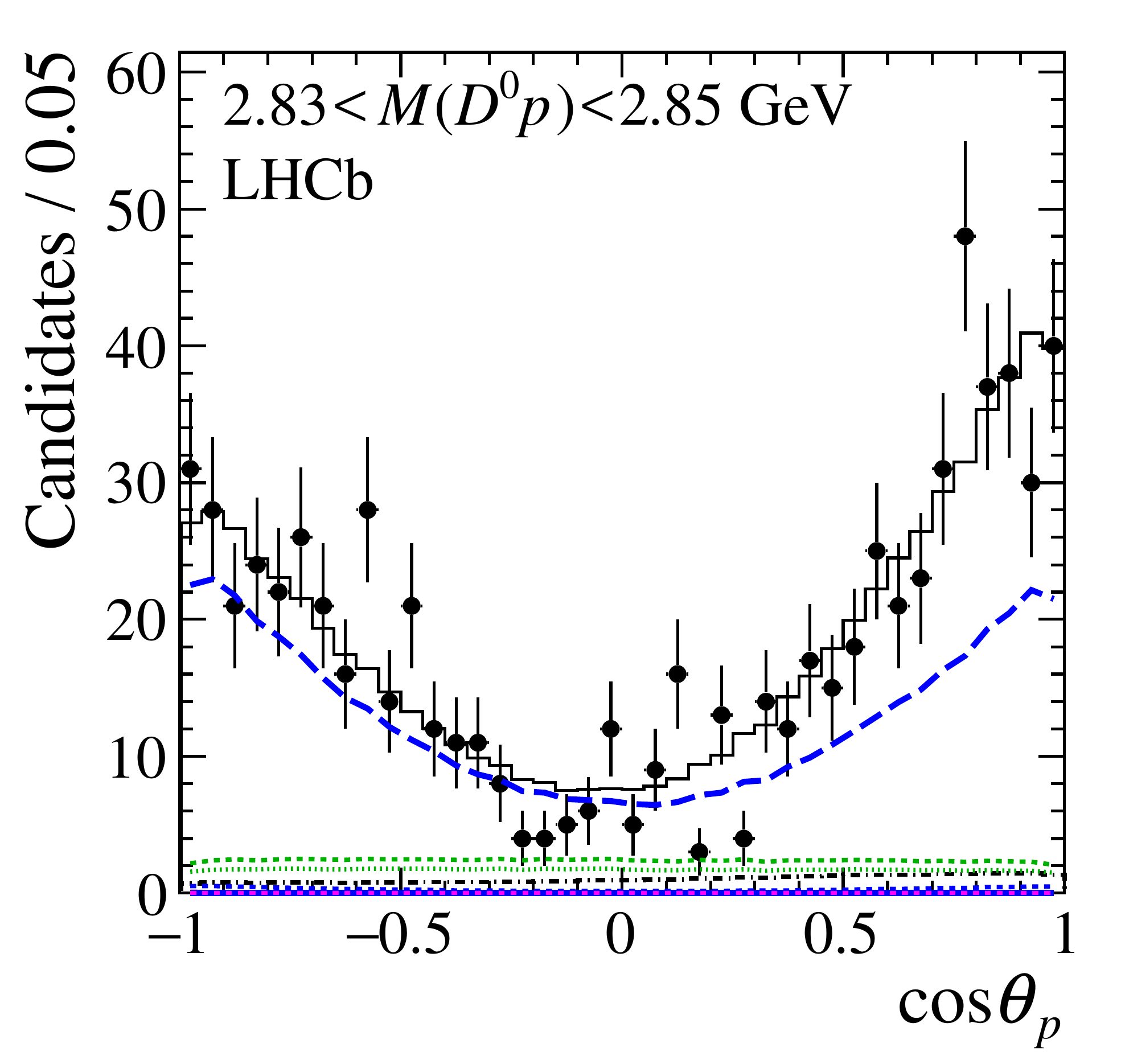} 
  \put(-113,104){(c)}

  \includegraphics[width=0.33\textwidth]{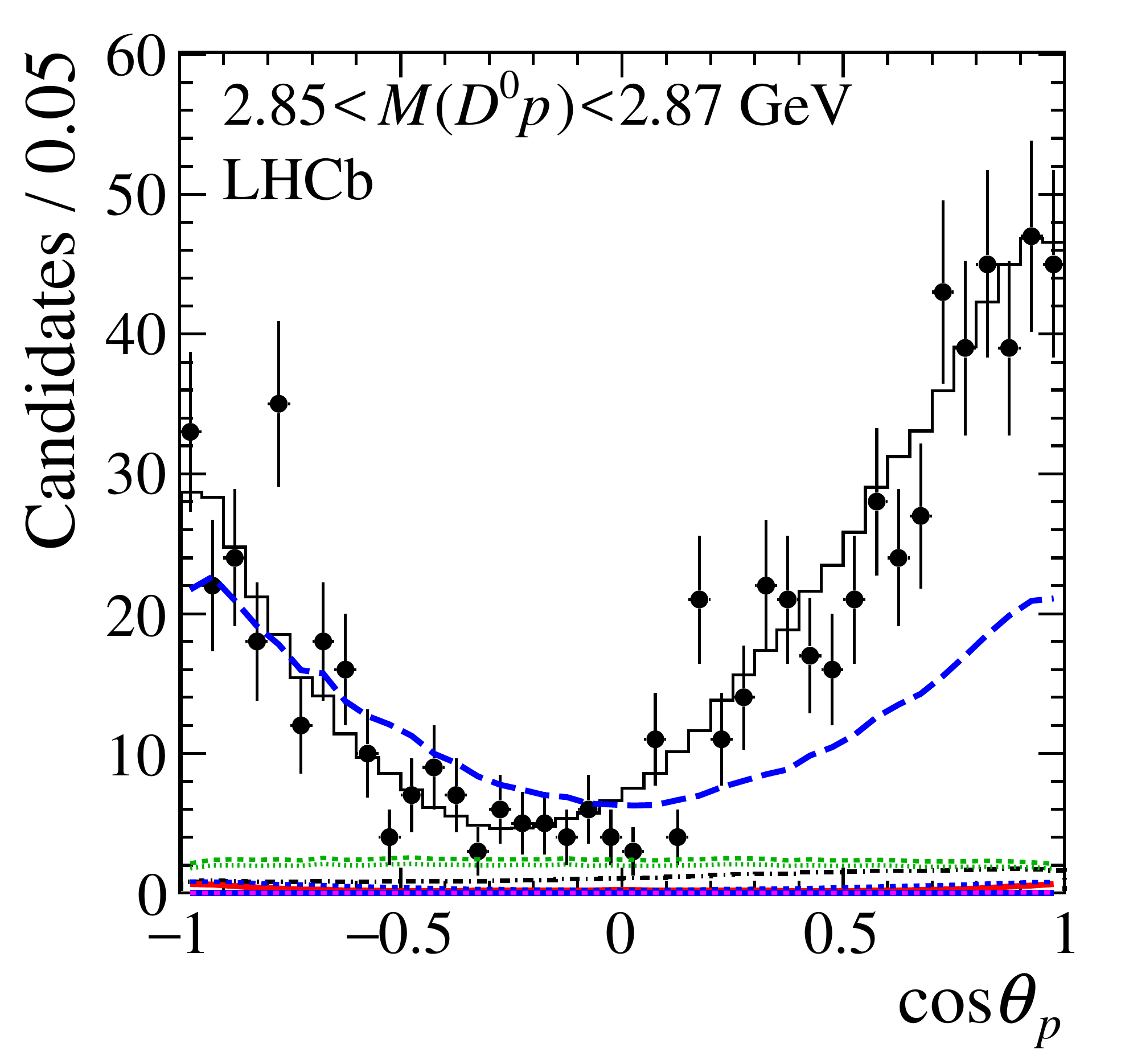} 
  \put(-113,104){(d)}
  \includegraphics[width=0.33\textwidth]{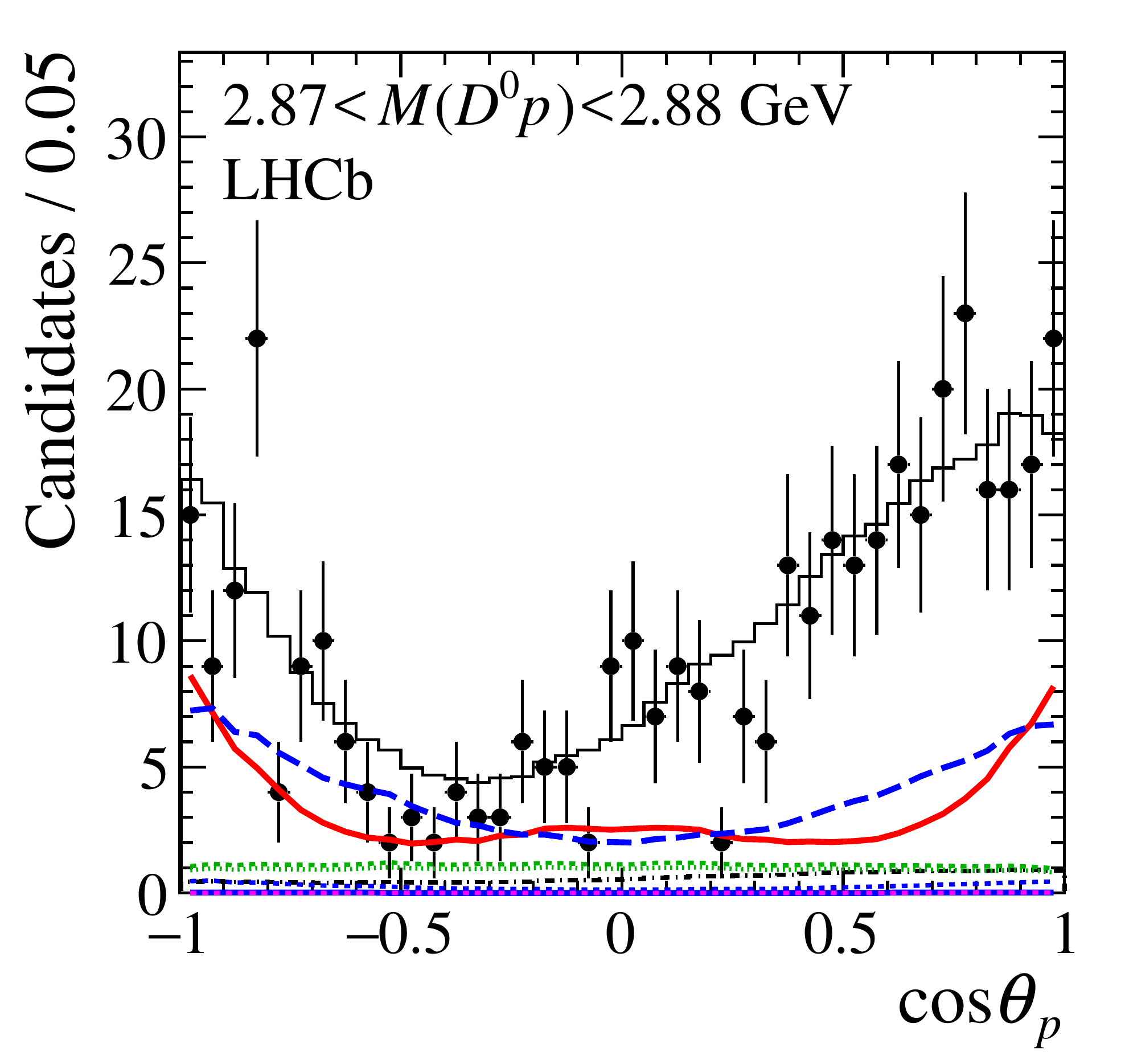} 
  \put(-113,104){(e)}
  \includegraphics[width=0.33\textwidth]{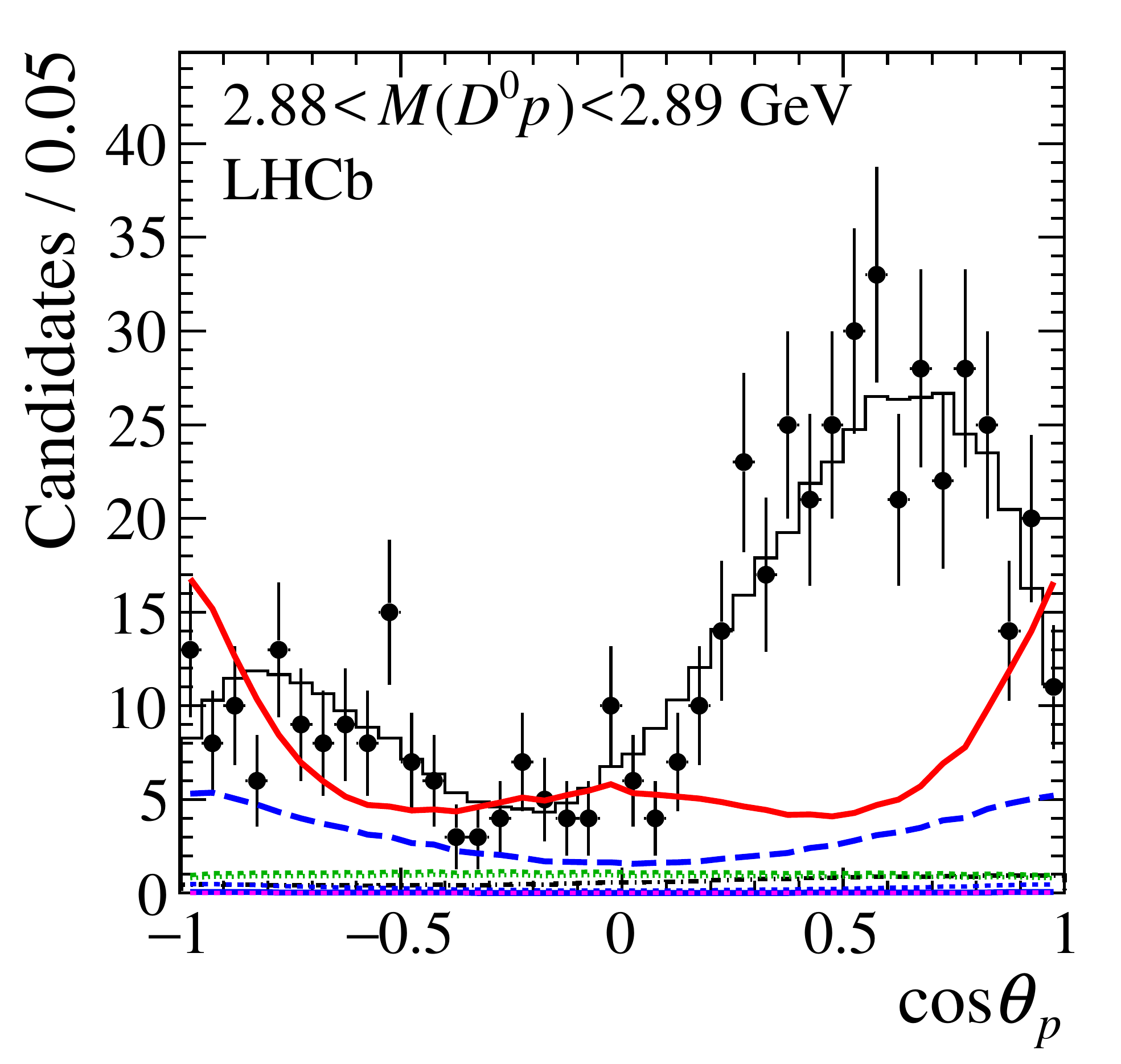} 
  \put(-113,104){(f)}

  \includegraphics[width=0.33\textwidth]{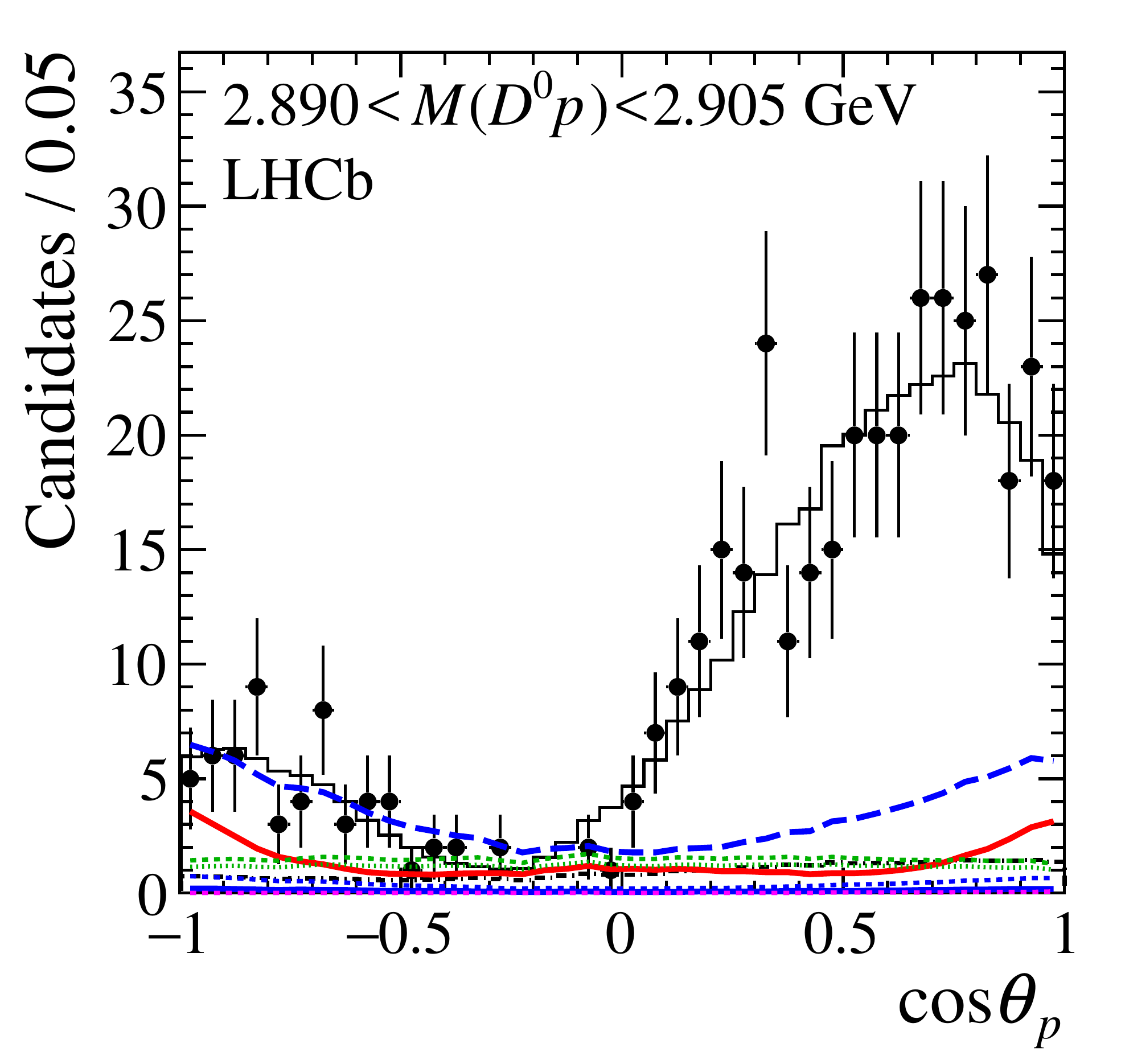} 
  \put(-113,104){(g)}
  \includegraphics[width=0.33\textwidth]{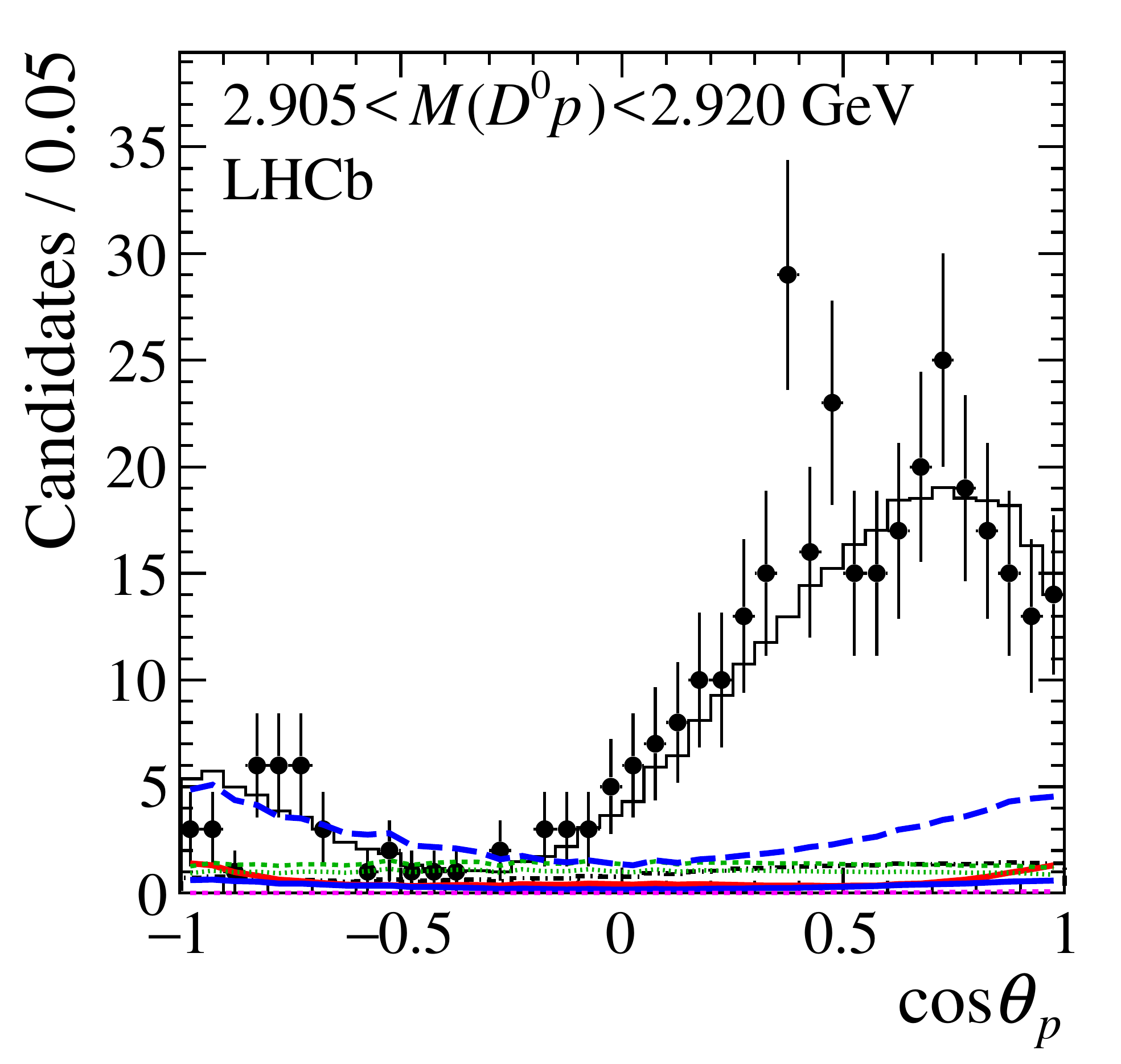} 
  \put(-113,104){(h)}
  \includegraphics[width=0.33\textwidth]{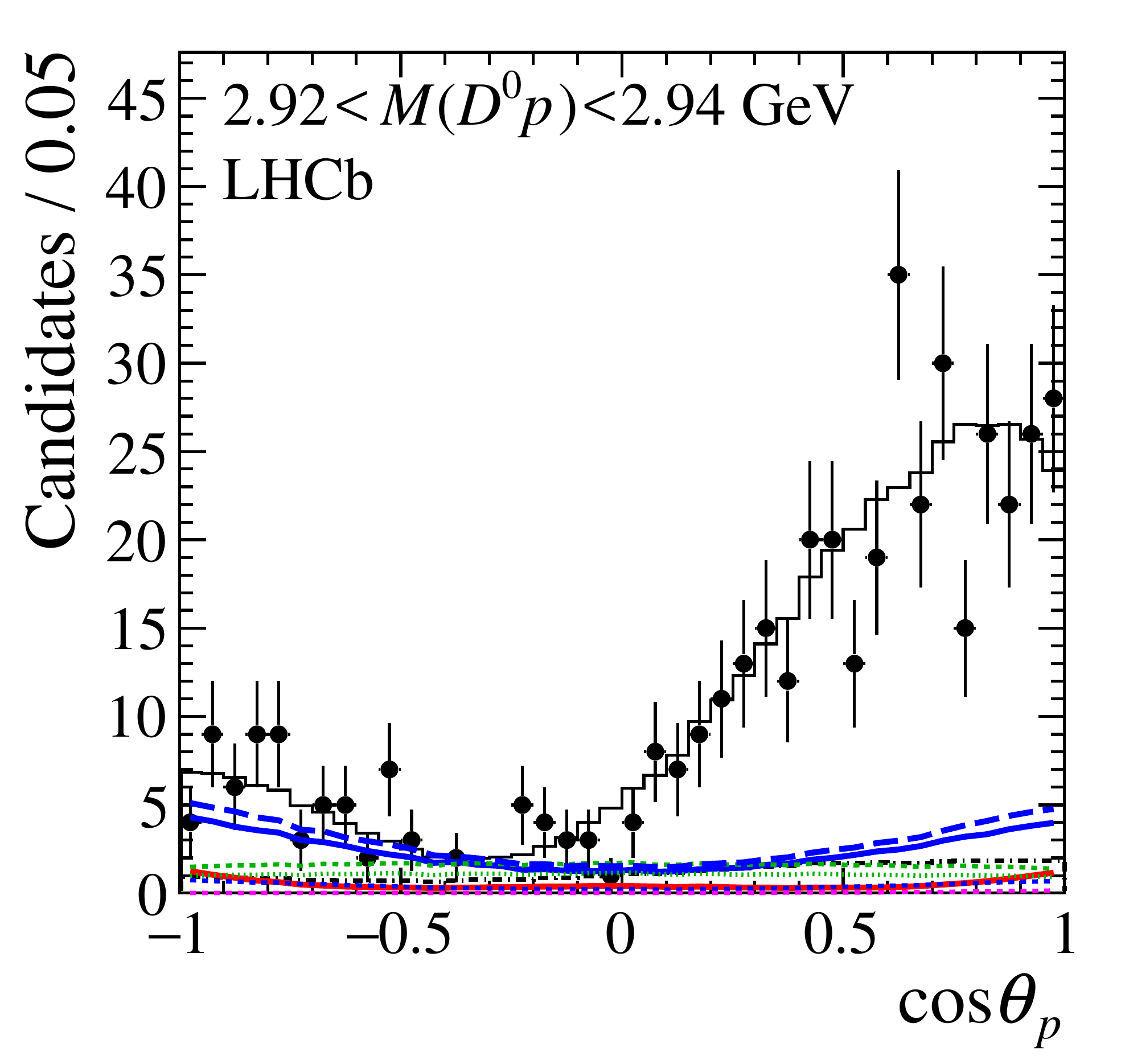} 
  \put(-113,104){(i)}
  
  \includegraphics[width=0.33\textwidth]{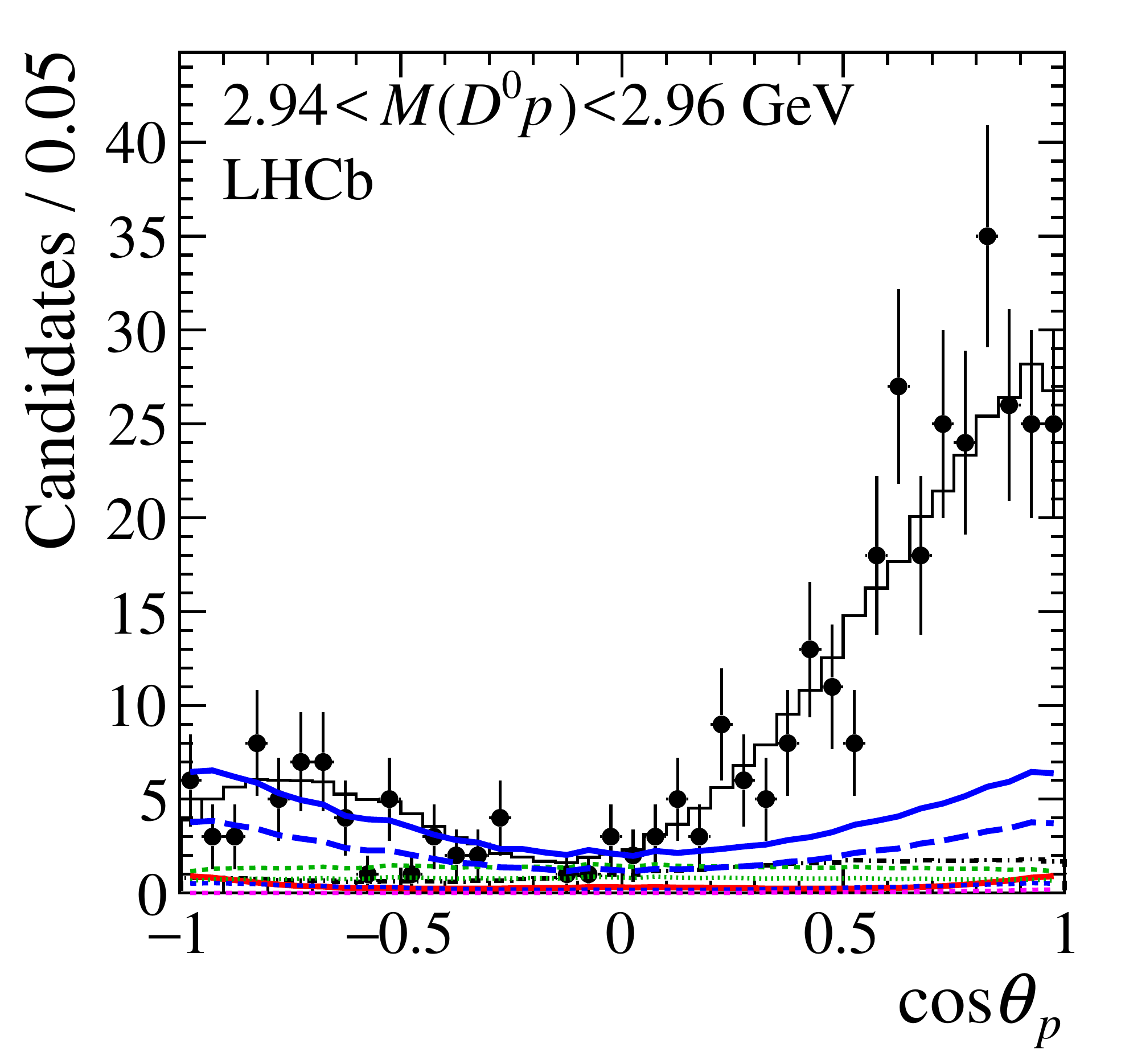} 
  \put(-113,104){(j)}
  \includegraphics[width=0.33\textwidth]{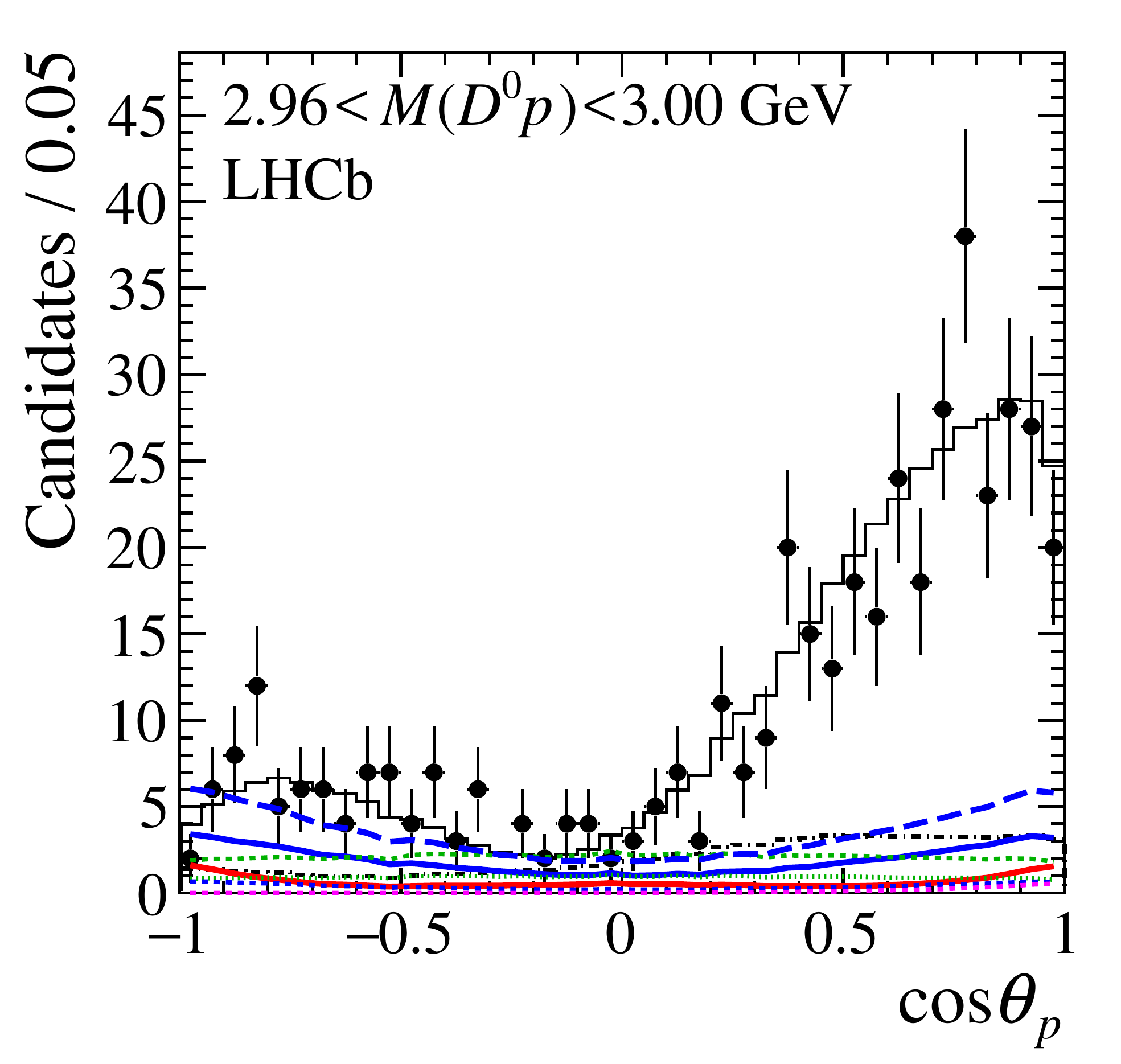} 
  \put(-113,104){(k)}
  \includegraphics[width=0.33\textwidth]{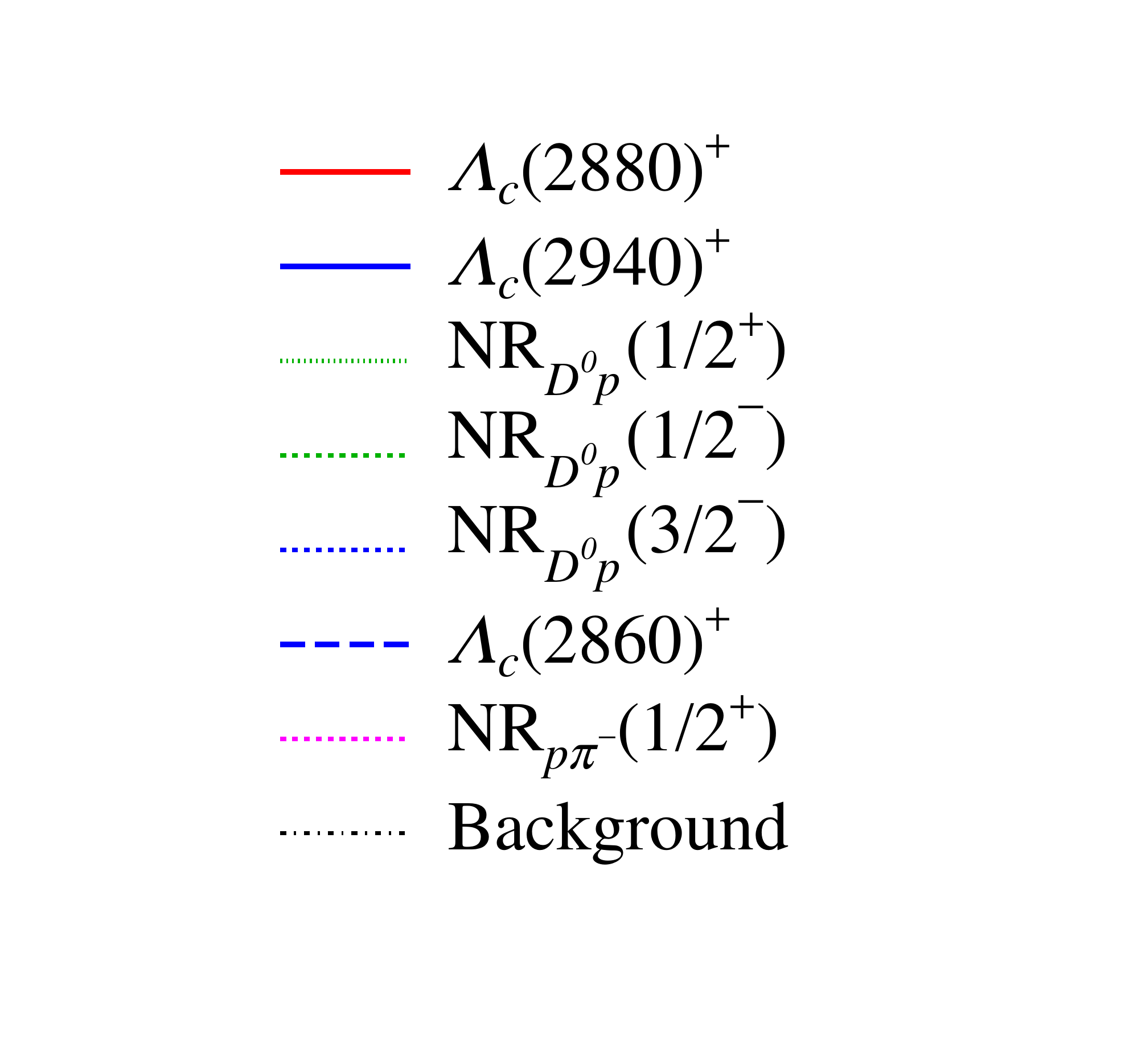} 
  \vspace{-8pt}
  \caption{Results of the fit of the \lbdnppi data in the $\Dz\proton$ 
   mass region including the $\Lcst$ and $\Lcstst$ resonances (region 4): 
   (a) $m(\Dz\proton)$ projection and (b--k) $\cos\theta_{\proton}$ projections for slices 
   of $\Dz\proton$ invariant mass. An exponential model is used for the nonresonant partial 
   waves, and the $J^P=3/2^-$ hypothesis is used for the \Lcstst state.
   Vertical lines in (a) indicate the boundaries of the $\Dz\proton$ invariant mass slices. 
   Due to interference effects the total is not necessarily equal to the sum of the components. 
  }
  \label{fig:dp_ampl_fit}
\end{figure}

The results of the fits are given in Table~\ref{tab:dp_lh}. For both nonresonant parametrisations, 
the best fit has a $\Lcstst$ spin-parity assignment of 
$3/2^-$. The results of the fit with 
this hypothesis and an exponential model for the nonresonant amplitudes, which is taken as the baseline for fit region 4, 
are shown in Fig.~\ref{fig:dp_ampl_fit}. 
Although the $3/2^-$ hypothesis describes the data significantly better than all others in fits using an exponential 
nonresonant model, this is not the case for the more flexible polynomial model: the assignment $J^P=5/2^-$ is only slightly 
worse ($\Delta\ln\mathcal{L}=3.6$) and a number of other spin-parity assignments are not excluded either. 

In the baseline model, the mass of the \Lcstst state is measured to be
$m(\Lcstst)=\lcststmassstat\mev$, and the width is 
$\Gamma(\Lcstst)=\lcststwidthstat\mev$. The fit fractions 
for the resonant components of the \Dz\proton amplitude are $\mathcal{F}(\Lcx)=(\lcxfracstat)\%$, 
$\mathcal{F}(\Lcst)=(\lcstfracstat)\%$, and $\mathcal{F}(\Lcstst)=(\lcststfracstat)\%$. 
All these uncertainties are statistical. Pseudoexperiments are used to correct for fit bias, 
which is small compared to the statistical uncertainties, and to determine the linear correlation 
coefficients for the statistical uncertainties between the measured masses, widths and fit fractions 
(Table~\ref{tab:dp_corr}). 

\begin{table}
  \caption{Correlation matrix associated to the statistical uncertainties
           of the fit results in the fit region 4.}
  \label{tab:dp_corr}
  \centering
  \begin{tabular}{r|ccccccc}
    & \rotatebox{90}{$  \mathcal{F}(\Lcst)$}  & \rotatebox{90}{$   \mathcal{F}(\Lcx)$}  & \rotatebox{90}{$             M(\Lcx)$}  & \rotatebox{90}{$        \Gamma(\Lcx)$}  & \rotatebox{90}{$\mathcal{F}(\Lcstst)$}  & \rotatebox{90}{$          M(\Lcstst)$}  & \rotatebox{90}{$     \Gamma(\Lcstst)$}  \\ \hline 
 $  \mathcal{F}(\Lcst)$  & $+1.00$ &         &         &         &         &         &        \\ 
 $   \mathcal{F}(\Lcx)$  & $+0.02$ & $+1.00$ &         &         &         &         &        \\ 
 $             M(\Lcx)$  & $-0.14$ & $+0.24$ & $+1.00$ &         &         &         &        \\ 
 $        \Gamma(\Lcx)$  & $-0.14$ & $+0.34$ & $+0.61$ & $+1.00$ &         &         &        \\ 
 $\mathcal{F}(\Lcstst)$  & $+0.18$ & $+0.03$ & $-0.02$ & $-0.12$ & $+1.00$ &         &        \\ 
 $          M(\Lcstst)$  & $+0.02$ & $+0.13$ & $-0.08$ & $-0.09$ & $+0.45$ & $+1.00$ &        \\ 
 $     \Gamma(\Lcstst)$  & $+0.15$ & $+0.06$ & $-0.04$ & $-0.11$ & $+0.78$ & $+0.54$ & $+1.00$\\

  \end{tabular}
\end{table}

\begin{table}
  \caption{Fit quality for various 
           \Lcstst spin-parity assignments. Exponential and 
           polynomial parametrisations of the nonresonant lineshapes are considered. 
           The baseline model is shown in bold face. }
  \label{tab:dp_lh}
  \centering
  \begin{tabular}{l|c|rcc}
  Nonresonant model & $\Lcstst$ $J^P$ & $\Delta\ln\mathcal{L}$ & $\chi^2/$ndf & $P(\chi^2, \mbox{ndf})$ [\%] \\
  \hline
  Exponential  & No $\Lambda_{c}^+(2940)$ & $54.6$ & 337.3/230 &  \phantom{0}0.0 \\ 
     & $1/2^{-}$ & $25.5$ & 293.1/228 &  \phantom{0}0.2 \\ 
     & $1/2^{+}$ & $34.2$ & 306.4/228 &  \phantom{0}0.0 \\ 
     & \boldmath{$3/2^{-}$} &  \boldmath{$0.0$} & {\bf 246.9/228} & {\bf 18.6} \\ 
     & $3/2^{+}$ & $14.8$ & 269.1/228 &  \phantom{0}3.2 \\ 
     & $5/2^{-}$ & $14.5$ & 269.9/228 &  \phantom{0}3.0 \\ 
     & $5/2^{+}$ & $15.6$ & 271.7/228 &  \phantom{0}2.5 \\ 
     & $7/2^{-}$ & $23.0$ & 276.4/228 &  \phantom{0}1.6 \\ 
     & $7/2^{+}$ & $29.0$ & 300.2/228 &  \phantom{0}0.1 \\ 
\hline 
Polynomial & No $\Lambda_{c}^+(2940)$ & 25.5 & 296.0/228 &  \phantom{0}0.2 \\ 
     & $1/2^{-}$ & $ 8.9$ & 270.0/226 &  \phantom{0}2.4 \\ 
     & $1/2^{+}$ & $ 7.2$ & 266.1/226 &  \phantom{0}3.5 \\ 
     & $3/2^{-}$ & $-4.2$ & 238.0/226 & 27.9 \\ 
     & $3/2^{+}$ & $ 4.9$ & 253.4/226 & 10.2 \\ 
     & $5/2^{-}$ & $-0.6$ & 249.0/226 & 14.0 \\ 
     & $5/2^{+}$ & $ 4.9$ & 250.5/226 & 12.6 \\ 
     & $7/2^{-}$ & $10.6$ & 270.0/226 &  \phantom{0}2.4 \\ 
     & $7/2^{+}$ & $11.7$ & 273.0/226 &  \phantom{0}1.8 \\ 

  \end{tabular}
\end{table}

The systematic and model uncertainties for the parameters given above, 
obtained following the procedure described in Sections~\ref{sec:lc2880_amplitude} 
and \ref{sec:lowerdp_amplitude}, are presented in Table~\ref{tab:dp_syst}. 
The part of the model uncertainty associated with the nonresonant amplitude is estimated 
from fits that use the polynomial nonresonant parametrisation instead of the default 
exponential form, by adding a $3/2^+$ nonresonant amplitude or removing the $3/2^-$ or $\proton\pim$ amplitudes, and by using the 
covariant formalism instead of the baseline helicity formalism.
The uncertainty due to the unknown quantum numbers of the $\Lcstst$ state is estimated 
from the variation among the fits with spin-parity assignments that give reasonable fit quality ($P(\chi^2, {\rm ndf})>5\%$): 
$3/2^+$, $3/2^-$, $5/2^+$, $5/2^-$. 

\begin{table}
  \caption{Systematic and model uncertainties of the 
           \Lcstst parameters and the resonance fit fractions.
           The uncertainty due to the nonresonant model 
           includes a component associated with the helicity formalism, 
           which for comparison is given explicitly in the table, too.
          }
  \label{tab:dp_syst}
  \centering
  \scalebox{0.82}{
  \begin{tabular}{l|ccccc}
         & \multicolumn{5}{c}{Uncertainty} \\
  \cline{2-6}
  Source & $m(\Lcstst)$ & $\Gamma(\Lcstst)$ & $\mathcal{F}(\Lcx)$ & $\mathcal{F}(\Lcst)$ & $\mathcal{F}(\Lcstst)$ \\
         & $[\mev\,]$     & $[\mev\,]$            & [\%] & [\%] & [\%] \\
  \hline
       Background fraction & $0.09$ & $0.23$ & $0.29$ & $0.12$ & $0.19$  \\
      Efficiency profile & $0.12$ & $0.34$ & $0.50$ & $0.24$ & $0.11$  \\
        Background shape & $0.15$ & $0.68$ & $1.13$ & $0.09$ & $0.48$  \\
     Momentum resolution & $0.07$ & $0.09$ & $0.03$ & $0.07$ & $0.02$  \\
              Mass scale & $0.05$ & $-$ & $-$ & $-$ & $-$  \\
           Fit procedure & $0.30$ & $0.45$ & $0.25$ & $0.08$ & $0.15$  \\
      $\Lcst$ parameters & $0.01$ & $0.16$ & $0.17$ & $0.03$ & $0.03$  \\
\hline
        Total systematic & $0.38$ & $0.92$ & $1.30$ & $0.30$ & $0.55$  \\
\hline
     Breit--Wigner model & $${\small $+0.10$/$-0.16$}$$ & $${\small $+0.00$/$-0.34$\phantom{1}}$$ & $${\small $+0.00$/$-0.59$}$$ & $${\small $+0.01$/$-0.16$}$$ & $${\small $+0.17$/$-0.31$}$$  \\
       Nonresonant model & $${\small $+0.00$/$-1.43$}$$ & $${\small $+5.21$/$-7.39$\phantom{1}}$$ & $${\small $+8.77$/$-1.60$}$$ & $${\small $+0.86$/$-0.41$}$$ & $${\small $+2.06$/$-2.38$}$$  \\
--- of which hel. form.  & $${\small $+0.00$/$-0.38$}$$ & $${\small $+2.18$/$-0.00$\phantom{1}}$$ & $${\small $+1.15$/$-0.00$}$$ & $${\small $+0.00$/$-0.23$}$$ & $${\small $+0.38$/$-0.00$}$$  \\
         $\Lcstst$ $J^P$ & $${\small $+0.00$/$-4.32$}$$ & $${\small $+0.00$/$-7.25$\phantom{1}}$$ & $${\small $+0.00$/$-5.79$}$$ & $${\small $+0.00$/$-0.67$}$$ & $${\small $+0.00$/$-3.29$}$$  \\
\hline
             Total model & $${\small $+0.10$/$-4.58$}$$ & $${\small $+5.22$/$-10.36$}$$ & $${\small $+8.82$/$-6.04$}$$ & $${\small $+0.86$/$-0.80$}$$ & $${\small $+2.07$/$-4.08$}$$  \\

  \end{tabular}
  }
\end{table}

The systematic uncertainties on $\Delta\ln\mathcal{L}$ between the various \Lcstst 
spin-parity hypotheses and the baseline hypothesis, $J^P=3/2^-$, are shown in Table~\ref{tab:dp_lh_syst}
(for the exponential nonresonant model) and Table~\ref{tab:dp_lh_syst_poly} (for the polynomial model). 
Only those systematic variations from Table~\ref{tab:dp_syst} that can affect the 
significance of the quantum number assignment are considered. 
Since the cases with exponential and polynomial nonresonant amplitudes
are treated separately, the model uncertainty associated with the nonresonant amplitudes 
does not include the difference between these two models. 

For each $J^P$ hypothesis, the significance with respect to the baseline is obtained from ensembles of 
pseudoexperiments and shown in Table~\ref{tab:dp_twofits}. The column marked ``Statistical'' includes 
only statistical uncertainties on $\Delta\ln\mathcal{L}$, while that marked ``Total'' is the sum in 
quadrature of the statistical, systematic, and model uncertainties. 

\begin{table}
  \caption{Systematic and model uncertainties on $\Delta\ln\mathcal{L}$
           between the baseline fit with $J^P=3/2^-$ for the \Lcstst state 
           and other fits without a \Lcstst contribution or 
           with other spin-parity assignments, for the exponential nonresonant model. }
  \label{tab:dp_lh_syst}
  \centering
  \scalebox{0.95}{
  \begin{tabular}{l|cccccccc}
         & \multicolumn{8}{c}{$\Delta\ln\mathcal{L}$ uncertainty for \Lcstst $J^P$} \\
         \cline{2-9}
  Source & No \Lcstst & $1/2^+$ & $1/2^-$ & $3/2^+$ & $5/2^+$ & $5/2^-$ & $7/2^+$ & $7/2^-$ \\
  \hline
       Background fraction & $0.3$ & $0.7$ & $0.3$ & $0.9$ & $0.7$ & $0.6$ & $0.7$ & $0.8$  \\
      Efficiency profile & $0.3$ & $0.2$ & $0.6$ & $0.6$ & $0.6$ & $0.6$ & $0.9$ & $1.1$  \\
        Background shape & $3.6$ & $3.4$ & $3.3$ & $2.6$ & $1.4$ & $2.0$ & $2.4$ & $4.0$  \\
     Momentum resolution & $0.1$ & $0.0$ & $0.1$ & $0.1$ & $0.1$ & $0.1$ & $0.1$ & $0.1$  \\
      $\Lcst$ parameters & $0.2$ & $0.2$ & $0.9$ & $0.2$ & $0.3$ & $0.1$ & $0.5$ & $0.4$  \\
\hline
        Total systematic & $3.6$ & $3.5$ & $3.4$ & $2.8$ & $1.7$ & $2.2$ & $2.6$ & $4.2$  \\
\hline
     Breit--Wigner model & $2.1$ & $1.2$ & $1.9$ & $1.6$ & $2.3$ & $0.4$ & $1.4$ & $1.4$  \\
       Nonresonant model & $3.7$ & $2.4$ & $0.4$ & $1.5$ & $1.0$ & $1.9$ & $1.4$ & $0.1$  \\
\hline
             Total model & $4.3$ & $2.7$ & $1.9$ & $2.1$ & $2.5$ & $1.9$ & $2.0$ & $1.4$  \\

  \end{tabular}
  }
\end{table}

\begin{table}
  \caption{Systematic and model uncertainties on $\Delta\ln\mathcal{L}$
           between the baseline fit with $J^P=3/2^-$ for the \Lcstst state 
           and other fits without a \Lcstst contribution or 
           with other spin-parity assignments, for the polynomial nonresonant model. }
  \label{tab:dp_lh_syst_poly}
  \centering
  \scalebox{0.95}{
  \begin{tabular}{l|cccccccc}
         & \multicolumn{8}{c}{$\Delta\ln\mathcal{L}$ uncertainty for \Lcstst $J^P$} \\
         \cline{2-9}
  Source & No \Lcstst & $1/2^+$ & $1/2^-$ & $3/2^+$ & $5/2^+$ & $5/2^-$ & $7/2^+$ & $7/2^-$ \\
  \hline
       Background fraction & $0.6$ & $0.1$ & $0.2$ & $0.3$ & $0.3$ & $0.4$ & $0.1$ & $0.6$  \\
      Efficiency profile & $0.6$ & $0.5$ & $0.5$ & $0.3$ & $0.2$ & $0.6$ & $0.7$ & $0.7$  \\
        Background shape & $1.2$ & $0.5$ & $0.6$ & $1.4$ & $1.6$ & $0.7$ & $1.5$ & $1.3$  \\
     Momentum resolution & $0.5$ & $0.2$ & $0.1$ & $0.1$ & $0.1$ & $0.1$ & $0.1$ & $0.1$  \\
      $\Lcst$ parameters & $0.2$ & $0.6$ & $0.2$ & $0.2$ & $0.1$ & $0.4$ & $0.3$ & $0.5$  \\
\hline
        Total systematic & $1.6$ & $0.9$ & $0.8$ & $1.5$ & $1.6$ & $1.1$ & $1.7$ & $1.7$  \\
\hline
     Breit--Wigner model & $1.1$ & $0.7$ & $0.4$ & $0.6$ & $1.1$ & $0.5$ & $0.9$ & $0.3$  \\
       Nonresonant model & $3.7$ & $2.2$ & $2.2$ & $1.6$ & $0.8$ & $1.3$ & $2.1$ & $3.2$  \\
\hline
             Total model & $3.8$ & $2.3$ & $2.3$ & $1.7$ & $1.3$ & $1.4$ & $2.3$ & $3.2$  \\

  \end{tabular}
  }
\end{table}

\begin{table}
  \caption{Significances of the $J^P=3/2^-$ spin-parity assignment for \Lcstst state 
           with respect to the alternative models without a \Lcstst contribution or 
           with other spin-parity assignments. }
  \label{tab:dp_twofits}
  \centering
  \begin{tabular}{l|c|cc}
           
  Nonresonant model & \Lcstst & \multicolumn{2}{c}{Significance, $\sigma$}\\
                     \cline{3-4}
           &   $J^P$ & Statistical & Total \\
  \hline
  Exponential  & No \Lcstst & 19.0 &  8.2 \\
     & $1/2^{+}$  & 18.3 &  7.9 \\
     & $1/2^{-}$  & 10.6 &  5.6 \\
     & $3/2^{+}$  & \phantom{1}7.5 &  3.7 \\
     & $5/2^{+}$  & \phantom{1}7.5 &  4.4 \\
     & $5/2^{-}$  & \phantom{1}7.4 &  4.5 \\
     & $7/2^{+}$  & 13.0 &  6.1 \\
     & $7/2^{-}$  & \phantom{1}9.9 &  6.1 \\
\hline
Polynomial & No \Lcstst & 11.8 &  5.6 \\
     & $1/2^{+}$  & \phantom{1}7.3 &  4.1 \\
     & $1/2^{-}$  & \phantom{1}7.8 &  4.5 \\
     & $3/2^{+}$  & \phantom{1}5.5 &  3.6 \\
     & $5/2^{+}$  & \phantom{1}4.8 &  3.1 \\
     & $5/2^{-}$  & \phantom{1}3.3 &  2.2 \\
     & $7/2^{+}$  & \phantom{1}8.0 &  6.2 \\
     & $7/2^{-}$  & \phantom{1}7.9 &  4.0 \\

  \end{tabular}
\end{table}

Including the systematic and model uncertainties, the mass and width of the \Lcstst resonance are 
  \begin{equation*}
  \begin{split}
    m(\Lcstst) & = \lcststmass\mev \\
    \Gamma(\Lcstst) & = \lcststwidth\mev. \\
  \end{split}
  \end{equation*}
The largest uncertainties in the measurement of these parameters, apart from those of statistical origin, 
are related to the model of the nonresonant amplitude 
and the uncertainties for the $\Lcstst$ quantum numbers. 
The fit fractions of the resonances in the region of the \lbdnppi phase space 
used in the fit, $M(\Dz\proton)<3$\gev, are 
\begin{equation*}
  \begin{split}
  \mathcal{F}(\Lcx) & =(\lcxfrac)\%, \\
  \mathcal{F}(\Lcst) & =(\lcstfrac)\%, \\
  \mathcal{F}(\Lcstst) & =(\lcststfrac)\%. \\
  \end{split}
\end{equation*}

The contributions of individual resonant components, 
integrated over the entire phase space of the \lbdnppi decay, 
can be used to extract the ratios of branching fractions
\begin{equation*}
  \small
  \begin{split}
  \frac{\BR(\Lb\to\Lcx\pim)\times \BR(\Lcx\to\Dz\proton)}{\BR(\Lb\to\Lcst\pim)\times \BR(\Lcst\to\Dz\proton)} & =\lcxratio, \\
  \frac{\BR(\Lb\to\Lcstst\pim)\times \BR(\Lcstst\to\Dz\proton)}{\BR(\Lb\to\Lcst\pim)\times \BR(\Lcst\to\Dz\proton)} & = \lcststratio, \\
  \end{split}
\end{equation*}
which assumes the ratios of the branching fractions 
to be equal to the ratios of the fit fractions. 

The constraints on the \Lcstst quantum numbers depend on the description of the 
nonresonant amplitudes. If an exponential model is used for the nonresonant components, the single 
best spin-parity assignment is $J^P=3/2^-$, and the $3/2^+$, $5/2^+$ and $5/2^-$
assignments are excluded at the levels of $3.7$, $4.4$ and $4.5$ standard deviations, respectively
(including systematic uncertainties), while spins of $1/2$ or $7/2$
are excluded by more than $5\sigma$. 
If a polynomial nonresonant parametrisation is used, 
the solution with $3/2^-$ is again the most likely one, 
though the data are consistent with the $5/2^-$ hypothesis at $2.2\sigma$.
Several $J^P$ assignments 
($5/2^+$, $3/2^+$, $7/2^-$, $1/2^+$ and $1/2^-$) are disfavoured with respect to the $3/2^-$ 
hypothesis with significances between $3.1$ and $4.5\sigma$, 
and only the $7/2^+$ hypothesis is excluded by more than $5\sigma$. 
Since the data are consistent with both the exponential and polynomial nonresonant models, 
only weak constraints on the spin and parity are obtained, 
with $J^P=3/2^-$ favoured and with positive parity excluded at the $3\sigma$ level.

\section{Conclusion}

\label{sec:conclusion}

An amplitude analysis of the decay \lbdnppi is performed in the region of the phase space containing $\Dz\proton$ resonant 
contributions. 
This study provides important information about the structure of the $\Dz\proton$ amplitude 
for future studies of CP violation in $\lbdpk$ decays, as well as on the spectroscopy of excited 
$\Lc$ states.

The preferred spin of the \Lcst state is found to be $J=5/2$, with the $J=7/2$ hypothesis disfavoured by $4.0$
standard deviations.
The solutions with $J=1/2$ and $3/2$ are excluded with a significance of more than $5$ standard deviations.  
The mass and width of the \Lcst state are found to be: 
  \begin{equation*}
   \begin{split}
    m(\Lcst) & = \lcstmass\mev, \\
    \Gamma(\Lcst) & = \lcstwidth\mev. \\
   \end{split}
  \end{equation*}
These results are consistent with and have comparable precision to the current world averages (WA), which are 
$m_{\rm WA}(\Lcst)=2881.53\pm 0.35\mev$, and $\Gamma_{\rm WA}(\Lcst)=5.8 \pm 1.1\mev$~\cite{PDG2016}. 

A near-threshold enhancement in the $\Dz\proton$ amplitude is studied. The enhancement is consistent with being a
resonant state (referred to here as the \Lcx) with mass and width 
  \begin{equation*}
   \begin{split}
    m(\Lcx) & =\lcxmass\mev, \\
    \Gamma(\Lcx) & =\lcxwidth\mev \\
   \end{split}
  \end{equation*}
and quantum numbers $J^P=3/2^+$, with the parity measured relative to that of the $\Lcst$ state. 
The other quantum numbers are excluded with a significance of 
more than $6$ standard deviations. The phase motion of the $3/2^+$ component with respect to the nonresonant 
amplitudes is obtained in a model-independent way and is consistent with resonant behaviour. 
With a larger dataset, it should be possible to constrain the phase motion of the $3/2^+$
partial wave using the \Lcst amplitude as a reference, without making assumptions on the nonresonant 
amplitude behaviour. The mass of the \Lcx state is consistent with recent predictions 
for an orbital $D$-wave $\Lc$ excitation with quantum numbers $3/2^+$ 
based on the nonrelativistic heavy quark-light diquark model~\cite{Chen:2016iyi} and from 
QCD sum rules in the HQET framework~\cite{Chen:2016phw}. 

First constraints on the spin and parity of the \Lcstst state are obtained in this analysis, 
and its mass and width are measured. 
The most likely spin-parity assignment for \Lcstst is $J^P=3/2^-$ but the other solutions 
with spins $1/2$ to $7/2$ cannot be excluded. 
The mass and width of the \Lcstst state are measured to be 
  \begin{equation*}
  \begin{split}
  m(\Lcstst) & =\lcststmass\mev, \\
  \Gamma(\Lcstst) & =\lcststwidth\mev. \\
  \end{split}
  \end{equation*}
The $J^P=3/2^-$ assignment for \Lcstst state is consistent with its interpretations as 
a $D^*N$ molecule~\cite{Ortega:2012cx, Zhang:2012jk, Zhang:2014ska}
or a radial $2P$ excitation~\cite{Chen:2014nyo}.

\section*{Acknowledgements}
 
\noindent We express our gratitude to our colleagues in the CERN
accelerator departments for the excellent performance of the LHC. We
thank the technical and administrative staff at the LHCb
institutes. We acknowledge support from CERN and from the national
agencies: CAPES, CNPq, FAPERJ and FINEP (Brazil); NSFC (China);
CNRS/IN2P3 (France); BMBF, DFG and MPG (Germany); INFN (Italy); 
FOM and NWO (The Netherlands); MNiSW and NCN (Poland); MEN/IFA (Romania); 
MinES and FASO (Russia); MinECo (Spain); SNSF and SER (Switzerland); 
NASU (Ukraine); STFC (United Kingdom); NSF (USA).
We acknowledge the computing resources that are provided by CERN, IN2P3 (France), KIT and DESY (Germany), INFN (Italy), SURF (The Netherlands), PIC (Spain), GridPP (United Kingdom), RRCKI and Yandex LLC (Russia), CSCS (Switzerland), IFIN-HH (Romania), CBPF (Brazil), PL-GRID (Poland) and OSC (USA). We are indebted to the communities behind the multiple open 
source software packages on which we depend.
Individual groups or members have received support from AvH Foundation (Germany),
EPLANET, Marie Sk\l{}odowska-Curie Actions and ERC (European Union), 
Conseil G\'{e}n\'{e}ral de Haute-Savoie, Labex ENIGMASS and OCEVU, 
R\'{e}gion Auvergne (France), RFBR and Yandex LLC (Russia), GVA, XuntaGal and GENCAT (Spain), Herchel Smith Fund, The Royal Society, Royal Commission for the Exhibition of 1851 and the Leverhulme Trust (United Kingdom).


\addcontentsline{toc}{section}{References}
\setboolean{inbibliography}{true}
\bibliographystyle{LHCb}
\bibliography{main,LHCb-PAPER,LHCb-DP}

\newpage


 
\newpage
\centerline{\large\bf LHCb collaboration}
\begin{flushleft}
\small
R.~Aaij$^{40}$,
B.~Adeva$^{39}$,
M.~Adinolfi$^{48}$,
Z.~Ajaltouni$^{5}$,
S.~Akar$^{59}$,
J.~Albrecht$^{10}$,
F.~Alessio$^{40}$,
M.~Alexander$^{53}$,
S.~Ali$^{43}$,
G.~Alkhazov$^{31}$
% ...

{\footnotesize \it
% ...
$ ^{5}$Clermont Universit{\'e}, Universit{\'e} Blaise Pascal, CNRS/IN2P3, LPC, Clermont-Ferrand, France\\
% ...
$ ^{10}$Fakult{\"a}t Physik, Technische Universit{\"a}t Dortmund, Dortmund, Germany\\
% ...
$ ^{31}$Petersburg Nuclear Physics Institute (PNPI), Gatchina, Russia\\
% ...
$ ^{39}$Universidad de Santiago de Compostela, Santiago de Compostela, Spain\\
$ ^{40}$European Organization for Nuclear Research (CERN), Geneva, Switzerland\\
% ...
$ ^{43}$Nikhef National Institute for Subatomic Physics, Amsterdam, The Netherlands\\
% ...
$ ^{48}$H.H. Wills Physics Laboratory, University of Bristol, Bristol, United Kingdom\\
% ...
$ ^{53}$School of Physics and Astronomy, University of Glasgow, Glasgow, United Kingdom\\
% ...
$ ^{59}$University of Cincinnati, Cincinnati, OH, United States\\
% ...
}
\end{flushleft}

\end{document}